\pdfoutput=1
\RequirePackage{fix-cm}
\documentclass[twocolumn,a4paper]{svjour3}          

\def\makeheadbox{\relax}
\smartqed  
\usepackage{graphicx}
\usepackage{pslatex,indentfirst,balance,parskip,graphicx,tikz}
\usepackage{pgfplots, booktabs, multirow, subfigure, amsmath, hyperref}
\usepackage{xargs, xcolor, mathtools, xspace}
\usepackage{algorithm, algorithmic, amssymb}
\floatname{algorithm}{\small Algorithm}
\usepackage[margin=3cm]{geometry}
\usetikzlibrary{arrows, patterns, plotmarks}
\usepackage[flushleft]{threeparttable}
\usepackage{mathrsfs}
\usepackage{lineno}
\usepackage[misc,geometry]{ifsym} 
\modulolinenumbers[5]
\usepackage[english]{babel}
\usepackage{siunitx}
\definecolor{Darkgreen}{RGB}{0,100,0}
\definecolor{Darkblue}{RGB}{25,25,200}
\definecolor{Darkred}{RGB}{100,2,10}
\definecolor{Brombeer}{RGB}{160,80,160}
\definecolor{green1}{RGB}{25,151,93}

\definecolor{sand}{RGB}{255,165,79}
\definecolor{olive}{RGB}{107,142,35}
\definecolor{newgray}{RGB}{112,128,144}
\definecolor{hellesbrombeer}{rgb}{0.8,0.4,0.8}
\definecolor{aquamarine}{RGB}{102,205,170}
\definecolor{SteelBlue}{RGB}{70,130,180}
\definecolor{rwthblue}{RGB}{0,84,159} 
\definecolor{rwth-blue}{RGB}{0,84,159} 
\definecolor{rwth-orange}{RGB}{246,168,0}
\definecolor{rwth-red}{RGB}{204,7,30}
\definecolor{rwth-lgreen}{RGB}{189,205,0}
\definecolor{rwth-green}{RGB}{87,171,39}
\definecolor{red1}{RGB}{255,106,106}
\definecolor{blue1}{RGB}{100,149,237}

\newcommand{\SimulationModel}{SiM-Care\xspace}

\newcommand{\PIT}{point in time\xspace}
\newcommand{\PITS}{points in time\xspace}
\newcommand{\PointsInTime}{\mathcal{T}}
\newcommand{\pointInTime}{t}
\newcommand{\decimalTime}{\eta}
\newcommand{\DecimalTimes}{\mathcal{H}}
\newcommand{\timeHorizon}{T}

\newcommand{\Locations}{\mathcal{L}}
\newcommand{\location}{\ell}
\newcommand{\dist}{\text{dist}}
\newcommand{\travelTime}{\tau}

\newcommand{\Day}{\delta}
\newcommand{\TimeOfDay}{\gamma}
\newcommand{\session}{\lambda}
\newcommand{\SessionOfWeek}{\Sessions{/}{\sim}}
\newcommand{\Sessions}{\Lambda}
\newcommand{\sessionBegin}{\underline{\openingHours}}
\newcommand{\sessionEnd}{\overline{\openingHours}}

\newcommand{\SP}{\mathcal{P}}
\newcommand{\SPs}{\mathcal{P}^{\text{s}}}
\newcommand{\SPm}{\mathcal{P}^{\text{m}}}
\newcommand{\p}{\rho}
\newcommand{\availabilities}{\alpha}
\newcommand{\ScGP}{\mathcal{G}^\text{con}}
\newcommand{\fgp}{\gp^\text{fam}}
\newcommand{\ratingApppointment}{r_\p^{\text{app}}}
\newcommand{\rApp}{\ratingApppointment (\gp)}
\newcommand{\ratingWalkIn}{r_\p^{\text{walk}}}
\newcommand{\rWalk}{\ratingWalkIn (\gp, [\session])}
\newcommand{\condition}{c}
\newcommand{\Illnesses}{\mathcal{I}}
\newcommand{\chronicIllness}{\varsigma}
\newcommand{\tmpIllnesses}{\Illnesses^{\text{act}}}
\newcommand{\chroIllnesses}{\Illnesses^{\text{chro}}}
\newcommand{\emergencyFlag}{\varepsilon}

\newcommand{\SGP}{\mathcal{G}}
\newcommand{\SGPs}{\mathcal{G}^{\text{s}}}
\newcommand{\SGPm}{\mathcal{G}^{\text{m}}}
\newcommand{\gp}{\phi}

\newcommand{\speedup}{\zeta}
\newcommand{\openingHours}{o}
\newcommand{\appointmentStrategy}{S}
\newcommand{\AppointmentStrategies}{\mathcal{S}^{\text{app}}}
\newcommand{\admissionStrategy}{S}
\newcommand{\AdmissionStrategies}{\mathcal{S}^{\text{adm}}}
\newcommand{\treatmentStrategy}{S}
\newcommand{\TreatmentStrategies}{\mathcal{S}^{\text{tmt}}}

\newcommand{\illness}{i}
\newcommand{\AllIllnesses}{\mathscr{I}}
\newcommand{\AllTmpIllnesses}{\AllIllnesses^{\text{act}}}
\newcommand{\AllChroIllnesses}{\AllIllnesses^{\text{chro}}}
\newcommand{\expAnnualIllnesses}{I_\age}
\newcommand{\illnessAgeclassDist}{\pi^{\text{act}}}
\newcommand{\chroIllnessAgeclassDist}{\pi^{\text{chro}}}
\newcommand{\SFamiliesIllnesses}{\mathcal{F}}
\newcommand{\ChronicFamiliesIllnesses}{\SFamiliesIllnesses^{\text{chro}}}
\newcommand{\TemporaryFamiliesIllnesses}{\SFamiliesIllnesses^{\text{act}}}
\newcommand{\familyOfIllnesses}{f_{\illness}}

\newcommand{\seriousness}{s_\illness}
\newcommand{\nominalDuration}[1]{D_{#1}}
\newcommand{\nominalWTW}[1]{W_{#1}}
\newcommand{\treatmentFrequency}[1]{N_{#1}}
\newcommand{\duration}{d_\illness}
\newcommand{\tf}{\nu_\illness}
\newcommand{\wtw}{\omega}
\newcommand{\wtwIllness}{\omega_\illness}
\newcommand{\ageAdjustedExpWTW}{\mathbb{E}^\wtw_\age(\familyOfIllnesses,\seriousness)}
\newcommand{\ageAdjustedExpDuration}{\mathbb{E}^d_\age(\familyOfIllnesses,\seriousness)}
\newcommand{\chronicAttribute}[1]{\kappa_{#1}}

\newcommand{\age}{a}
\newcommand{\SAgeClass}{\mathcal{A}}
\newcommand{\changeInDuration}{\Delta^d_{\age}}
\newcommand{\changeInWTW}{\Delta^{\wtw}_{\age}}
\newcommand{\probToCancel}{p_{\age}}
\newcommand{\youngClass}{16-24}
\newcommand{\midClass}{25-65}
\newcommand{\oldClass}{\textgreater 65}

\newcommand{\arr}{\text{arv}}
\newcommand{\arrivalEvent}{\event^\arr}
\newcommand{\rel}{\text{rel}}
\newcommand{\releaseEvent}{\event^\rel}
\newcommand{\fol}{\text{fol}}
\newcommand{\followUpEvent}{\event^\fol}
\newcommand{\rec}{\text{rec}}
\newcommand{\recoveryEvent}{\event^\rec}
\newcommand{\ill}{\text{ill}}
\newcommand{\illnessEvent}{\event^\ill}
\newcommand{\ope}{\text{opn}}
\newcommand{\openEvent}{\event^\ope}
\newcommand{\clo}{\text{clo}}
\newcommand{\closeEvent}{\event^\clo}
\newcommand{\Events}{\mathcal{E}}

\newcommand{\eventQueue}{\mathcal{Q}}
\newcommand{\event}{e}
\newcommand{\feasibleWalkInVisits}{W}
\newcommand{\timeBetween}{w_\pointInTime}
\newcommand{\matches}{m}
\newcommand{\stdApp}{acute\xspace}
\newcommand{\regApp}{regular\xspace}
\newcommand{\Appointments}{\mathcal{B}}
\newcommand{\appointment}{b}
\newcommand{\user}{modeler\xspace}
\newcommand{\users}{modelers\xspace}
\newcommand{\stdappointment}{\appointment^{\text{act}}}
\newcommand{\regappointment}{\appointment^{\text{reg}}}
\newcommand{\N}{\mathbb{N}}
\newcommand{\Rplus}{\mathbb{R}_{\geq 0}}
\newcommand{\act}{acute\xspace}

\DeclareMathOperator{\accessTime}{ac-time}
\DeclareMathOperator{\travelDist}{tr-dist}
\DeclareMathOperator{\waitingTime}{wait-time}
\DeclareMathOperator{\utilization}{util}
\DeclareMathOperator{\overtime}{over}

\begin{document}

\title{Patients, Primary Care, and Policy: Simulation Modeling for Health Care Decision Support \thanks{This work is supported by the Freigeist-Fellowship of the Volkswagen Stiftung. \newline
		This work is supported by the German research council (DFG) Research Training Group 2236 UnRAVeL.}
}

\author{Martin Comis 
        \and Catherine Cleophas
        \and Christina B\"using 
}

\institute{Christina B\"using \and  Martin Comis  (\begin{scriptsize}\Letter
	\end{scriptsize}) \at
              Lehrstuhl II f\"ur Mathematik, RWTH Aachen University, Pontdriesch 10--12, 52062 Aachen, Germany\\
              \email{\{buesing, comis\}@math2.rwth-aachen.de}           
           \and
           Catherine Cleophas \at
              Working Group Service Analytics, Christian-Albrechts-Universit\"at zu Kiel, Westring 425, 24118 Kiel, Germany\\
              \email{cleophas@bwl.uni-kiel.de}
}

\date{Received: date / Accepted: date}

\maketitle
\begin{abstract}
Demand for health care is constantly increasing due to the ongoing demographic change, while at the same time health service providers face difficulties in finding skilled personnel.
This creates pressure on health care systems around the world, such that the efficient, nationwide provision of primary health care has become one of society's greatest challenges.
Due to the complexity of health care systems, unforeseen future events, and a frequent lack of data, analyzing and optimizing the performance of health care systems means tackling a wicked problem.
To support this task for primary care, this paper introduces the hybrid agent-based simulation model \SimulationModel. \SimulationModel models the interactions of patients and primary care physicians on an individual level.
By tracking agent interactions, it enables \users to assess multiple key indicators such as patient waiting times and physician utilization.
Based on these indicators, primary care systems can be assessed and compared.
Moreover, changes in the infrastructure, patient behavior, and service design can be directly evaluated.
To showcase the opportunities offered by \SimulationModel and aid model validation, we present a case study for a primary care system in Germany.
Specifically, we investigate the effects of an aging population, a decrease in the number of physicians, as well as the combined effects.
\end{abstract}

\section{Introduction}
\label{sec:intro}
Health is one of the most important factors for the prosperity and well-being of a society.
Therefore, all member states of the World Health Organization (WHO) made the commitment to ensure everyone's access to health services~\cite{Dye199ed13}.
The foundation of universally accessible health services is usually laid by a primary care system. 
Following the definition of the American Academy of Family Physicians~\cite{AAFP}, primary care systems ``serve as the patient's first point of entry into the health care system and the continuing focal point for all needed health services''.
To that end, they feature a set of primary care physicians (PCPs) who provide ``primary care services to a defined population of patients''.
Such primary care services include ``health promotion, disease prevention, health maintenance, counseling, patient education, diagnosis and treatment of acute and chronic illnesses''.

Demographic change poses serious challenges to primary care systems:
Medical and technological progress paired with improved living conditions and reduced birth rates lead to a rise in the share of elderly citizens.
In the United States, the percentage of individuals aged 65 and older is predicted to exceed $\SI{21}{\si{\percent}}$ of the total population by 2030~\cite{USCB17}.
In Germany, this demographic shift is even more severe with elderly citizens (aged 65 and above) being expected to account for more than $\SI{26}{\si{\percent}}$ of the total population by 2030~\cite{POE19}.
As populations age, their demand for primary care services tends to increase.
This is primarily due to the prevalence of chronic illnesses which disproportionately affect older adults~\cite{MAN10,doi:10.1111}.
On the supply side, primary care physicians are also aging, e.g., $\SI{34.1}{\si{\percent}}$ of all primary care physicians in Germany were $\num{60}$ years or older by the end of 2017 \cite{KBV} and thus about to retire.
Moreover, fewer medical students are willing to practice primary care~\cite{MAN10}, let alone open a private primary care practice~\cite{MONITOR14}.
This leads to reduced treatment capacities and exacerbates the risk for supply disruptions.

In the United States, the ``confluence of a rising demand for primary care services and a decreasing supply of professionals providing these services'' is considered a ``crisis in primary care''~\cite{MAN10}.
In order to manage this crisis, existing primary care systems have to be fundamentally adjusted~\cite{pfaff2017lehrbuch}.
Various new concepts and policies to maintain the standard of health care provision are discussed by the statutory health insurances, the governments, and the Associations of Statutory Health Insurance Physicians~\cite{GP14,MAN10}.
However, all such concepts call for validation and evaluation prior to their potentially costly implementation \cite{pfaff2017lehrbuch}.
Naturally, this leads to the pressing question: How can the quality of primary care systems and the effect of changes to them be quantified?

In German legislation, this question is answered by the 2012 GKV-Versorgungsstrukturgesetz \cite{GKV12} which defines adequate health care supply on the basis of profession-specific ratios.
The law subdivides the country into zones and specifies the required population-to-provider ratio for each medical specialization.
For example, the predefined nominal ratio of primary care physicians is one PCP per $\num{1671}$ inhabitants \cite[\S 11(4)]{GKV12}. This base indicator can be adjusted to account for a zone's individual demographic and geographic characteristics \cite[\S 2]{GKV12}.
If the actual ratio of a zone is significantly higher than the nominal ratio, closing practices will not be replaced.
If it is significantly lower, new practices are permitted to be opened.

Beyond Germany, we can find similar ratio-based measures in other European countries like Bulgaria, Estonia, Italy, and Spain~\cite{KRI15}.
But also in the United States,~the Health Resources and Services Administration (HRSA) defines adequate health care supply based on profession- and region-specific population-to-provider ratios.
If the predefined popu\-lation-to-provider ratio (for primary care $\num{3500}$ to $1$; or $\num{3000}$ to $1$ for unusually high needs~\cite{HRSA}) of a geographic area is exceeded, HRSA designates it a health professional shortage area to which National Health Service Corps personnel is directed with priority.

Obviously all ratio-based assessments have several shortcomings: Even after adjustment, population-to-provider ratios can only provide a very rough estimate for the actual demand. Furthermore, adjustment rules are highly dependent on the definition of the underlying zones or geographic areas. Factors such as the accessibility of practices and the PCP's individual workload are completely neglected.
Finally, ratio-based assessments cannot account for new health care concepts such as tele\-medicine,  mobile medical units, or centralized appointment scheduling.

To overcome these limitations, a new approach to model dynamic effects in primary care systems is required.
However, analyzing and evaluating health care systems is a complex and complicated task due to the large number of involved individuals and uncertain nature of health care processes, e.g., fluctuating demand, arrival time of patients, emergency patients, and durations of treatments.
To model such  ``wicked problems''~\cite{Rittel1973}, researchers have achieved promising results using agent-based modeling (ABM), which account for individual agents and their interactions on the micro-level. A general introduction to the concept of ABM is provided in~\cite{gilbert2008agent}.
Existing studies implementing ABM have considered diverse social systems.
Examples include matters such as tobacco control~\cite{doi:10.7326/M15-1567} and educational policy research~\cite{maroulis2010complex}.
But there are also numerous applications of ABM in field of health care~\cite{Barnes2013}, e.g., for accountable care organizations~\cite{alibrahim2018agent,liu2016agent}, for medical workforce forecasting~\cite{Lopes2018}, and in epidemiology~\cite{10.1007/11553762_21,Meng2010}.

\begin{figure}
	\centering
	\begin{tikzpicture}
	\node at (0,0) {\includegraphics[width=0.95\linewidth]{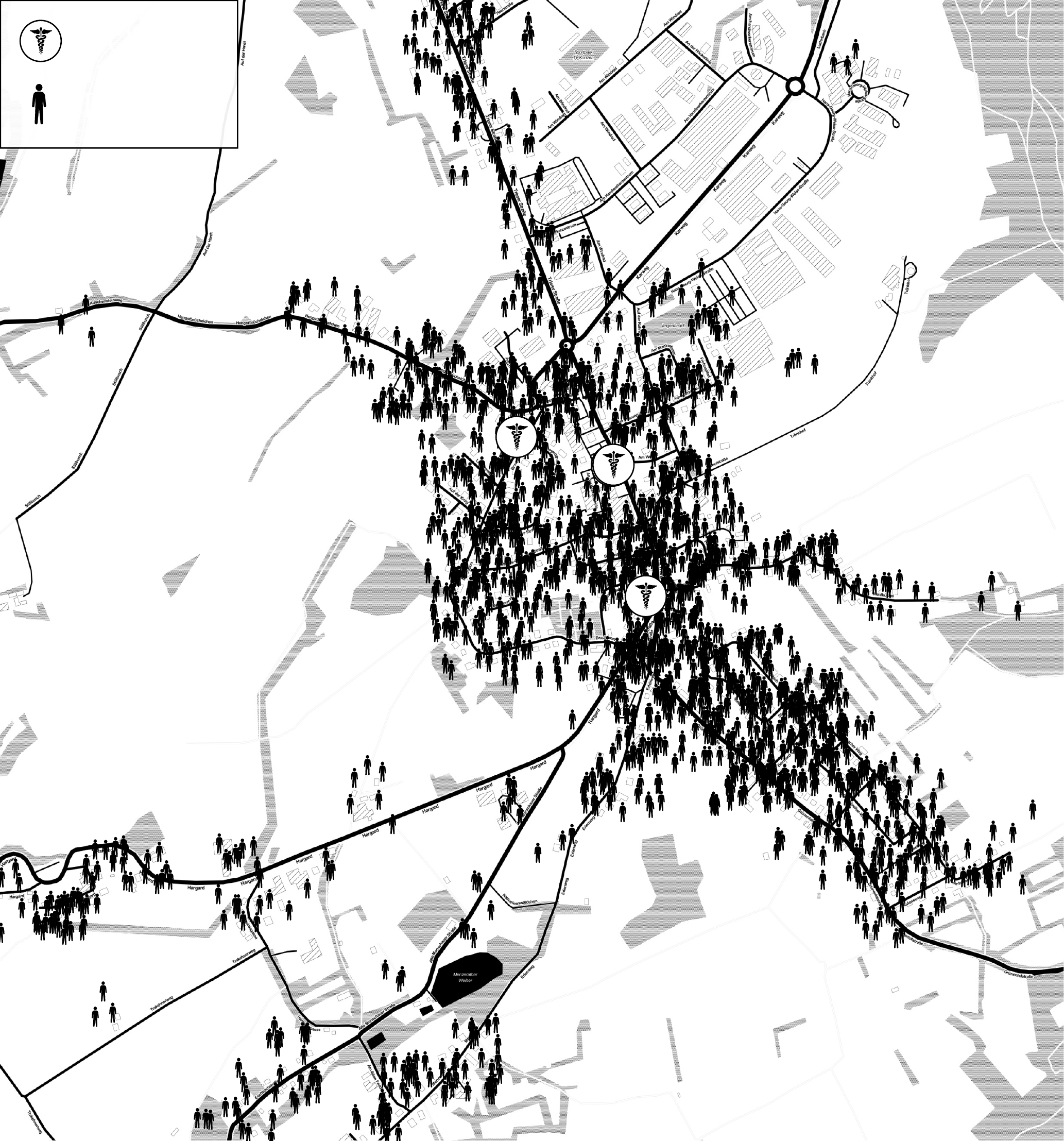}};
		\node[right] at (-3.05,3.4) {\footnotesize PCPs};
		\node[right] at (-3.05,2.98) {\footnotesize Patients};
\end{tikzpicture}
	\caption[theCaption]{Geo-social system of patients and physicians.\footnotemark{}}
	\label{fig:domain}
\end{figure}

This paper introduces the hybrid agent-based simulation tool \SimulationModel (\textbf{Si}mulation \textbf{M}odel for primary \textbf{Care}) to model the dynamics of primary care systems. \SimulationModel models patients and PCPs on an individual level as illustrated by Figure~\ref{fig:domain}:
Patients and primary care physicians are modeled via a geo-social system, in which patients decide whether and where to request an appointment based on their state of health and PCPs handle appointment requests, manage patient admission, and treat patients.
By tracking the resulting interactions in \SimulationModel, planners can identify dependencies of different subproblems, evaluate new planning approaches, and quantify the effects of interventions on the basis of multiple meaningful performance measurements.
From empirical data, we develop realistic test scenarios including a controllable degree of uncertainty realized via stochastic simulation experiments.

The main contribution of this paper can be summarized as follows: 
We introduce the simulation model \SimulationModel, which provides decision makers with a versatile decision support tool for primary care planning.
\SimulationModel is very generic and can be easily modified and extended to meet each \user's needs.
Patients and physicians are modeled as individuals who follow their own objectives, learn, and adapt. 
To ensure computational tractability, the model incorporates a global event queue at its core.
As such, \SimulationModel can be considered an integrated hybrid simulation model that combines paradigms from agent-based modeling and discrete event simulation.
Based on empirical data for a German primary care system, we illustrate how scenarios for the simulation model can be generated.
Finally, we showcase the opportunities of \SimulationModel through a case study.
To the best of our knowledge, \SimulationModel is the first model of its kind that captures entire primary care system with all physicians and patients as individual agents and allows for the simultaneous consideration of microsystem improvements as well as macrosystem reforms.
\footnotetext{Map tiles by Stamen Design, under CC BY 3.0. Data by OpenStreetMap, under ODbL.}

The remainder of this paper is structured as follows:
Section~\ref{sec:relatedWork} discusses related work on agent-based simulations with a focus on health care applications.
Section~\ref{sec:simulationModel} introduces \SimulationModel based on the ODD framework by Grimm et al.~\cite{Grimm2010ODD}.
Section~\ref{sec:computationalExamples} presents a case study based on real-world data to aid model validation and showcase how \SimulationModel can be applied to support health care planning.
Section~\ref{sec:discussion} discusses the potential applications and entry requirements of \SimulationModel.
Finally, Section~\ref{sec:conclusion} concludes and provides directions for future work.
\section{Related Work}
\label{sec:relatedWork}
Decision support for health care planning is an area with increasing importance~\cite{hamrock2013discrete}.
To analyze health care systems, decision support tools have to deal with the detail complexity that is inherent to the health care sector~\cite{Fone2003}: Patients schedule appointments based on their preferences and state of health, while PCPs offer appointments and treat patients.
Thereby, micro-interactions affect macro-level indicators as agents observe, learn and adapt, decide and act, and - as a group - determine the system's behavior~\cite{gilbert2008agent}.
When a system's status depends on such micro-interactions, its behavior becomes difficult to predict and a ``wide range of possible outcomes may arise from any policy change''~\cite{Fone2003}.
Simulation modeling can deal with this kind of complexity by ``simulating the life histories of individuals and then estimating the population effect from the sum of the individual effects''~\cite{Fone2003}.
As such, simulation models represent a powerful tool to inform policy makers:
They can provide valuable insights into the dependencies within health care systems and allow for the prediction of the outcome of a change in strategy ahead of a  potentially costly and risky  real-world intervention~\cite{Fone2003,hamrock2013discrete}.

Given these potentials, the use of computer simulation to study health care delivery systems has significantly increased over the recent years~\cite{zhong2016discrete}.
The resulting body of literature on applications of simulation in health care is rich and can, from a methodological point of view, be classified into four main groups~\cite{Brailsford2009}: System dynamics~\cite{homer2006system,brailsford2008system},  discrete-event simulations~\cite{hamrock2013discrete,Jacobson2006}, agent-based models~\cite{Barnes2013,doi:10.1146/annurev-publhealth-040617-014317}, and hybrid simulations that combine two or more of the former paradigms~\cite{BRAILSFORD2019721,Brailsford:2010:THG:2433508.2433790}.
To categorize related work on the study of primary care systems even further, we distinguish in the following models that focus on microsystem improvements and models that aim at macrosystem reforms.

Simulation models aimed at studying microsystem improvements in primary care systems mostly include a detailed model of a single (specific) outpatient practice and focus on a predefined subset of potential improvements.
Zong et al.~\cite{zhong2016discrete} present a discrete event simulation for a pediatric clinic at the University of Wisconsin Health.
Their model includes a very detailed representation of the sequential stages during a patient's visit.
In a set of ``what-if`'' scenarios, the authors investigate how the overall performance of the clinic is impacted by different scheduling templates, a change in the medical assistant to physician ratio, and the pairing of resident doctors with clinicians.
Shi et al.~\cite{SHI2014165} developed a discrete event simulation model for a primary care clinic of the Department of Veteran Affairs.
Within the model, the different patient flow routes for appointment patients, walk-ins, and nurse-only patients are distinguished.
In a scenario analysis, the authors investigate how the clinic's performance is affected by  six distinct factors that include walk-in and no-show rates as well as the double booking of appointments. 
Cayirli et al.~\cite{Cayirli2006} used empirical data collected at a primary care clinic in New York to devise a discrete event simulation of a generic single-server primary care practice.
The model distinguishes new and returning patients and accounts for walk-ins, no-shows, patient punctuality, and service time variations.
In a simulation study, the authors evaluate $42$ appointment systems that vary in the implemented sequencing- and appointment rules.
A similar discrete event simulation of a generic single-server primary care practice is introduced by Schacht~\cite{SCHACHT2018119}. 
In his model, all arriving patients have a stochastic willingness to wait and always request an appointment.
If the indirect waiting time for this appointment exceeds a patient's willingness to wait, they become walk-ins.
The arrival rate of patients depends on the session, day, and month to model seasonality.
In a case study, the author evaluates a class of appointment systems that can account for seasonal variations in demand through reconfigurations.
Further simulation models aimed at the study of microsystem improvements in primary care practices can be found in \cite{Wiesche2017,7019975,giachetti2005assessing}.
In contrast to \SimulationModel, all of the models above include only one single primary care practice out of the many providers that make up a primary care system.
Moreover, all of these models adopt a different approach to the representation of patients:
While \SimulationModel models a persistent patient population that is shared by all providers, the previous models perceive patients as non-persistent, i.e., patients are generated as they arrive at the practice and cease to exist as soon as they are discharged.
As a result, the previous models cannot account for the effects of individual microsytem improvements on the entire primary care system itself.

Simulation models aimed at investigating macrosystem reforms of primary care systems mostly include an entire primary care system, however they are usually much more high level.
Matchar et al.~\cite{7822255} use the methodology of system dynamics to develop a simulation model to aid primary care planning in Singapore.
The model captures the causal relationships between the stakeholders' aims and the provision of services in an analytical framework.
The authors evaluate three policy changes that constitute in reducing the service gap, reducing the out-of-pocket costs, and increasing the number of physicians.
Through the use of system dynamics, the model is much more high level than \SimulationModel and does not model patients or physicians as individuals.
Consequently, the model cannot account for the objectives and satisfaction of individual stakeholders which limits the possibilities for evaluation.
Homa et.~al~\cite{Homa2015} present an agent-based model to investigate the paradox of primary care.
Their model features patients, PCPs, and specialists as individual agents.
Every patient has a health status that changes over time:
The contraction of illnesses leads to a (temporary) decrease in the patients' health;
the treatment of acute illnesses by PCPs and specialists as well as regular check-ups (performed exclusively by PCPs) lead to an increase in the patients' health.
Tracking the evolution of the patients' average health status over time, the authors investigate how public health is positively affected by the interplay of different mechanisms in primary care.
As such, the model of Homa et.~al has a different objective than \SimulationModel:
While Homa et al.~investigate the external effects of treatments in primary care on the entire health care system, \SimulationModel focuses on the processes within primary care systems.
To that end, \SimulationModel models the scheduling of appointments, the patients' actual practice visits that result in waiting times through the interaction of patients, and the physicians' treatments of patients with variable service times which are not part of the model by Homa et al.

To the best of our knowledge, there is no previous work on simulation models for the evaluation of primary care systems that allow for the simultaneous consideration of microsystem improvements and macrosystem reforms as in \SimulationModel.
\section{Simulation Model}
\label{sec:simulationModel}
\begin{figure*}
	\centering
	\input{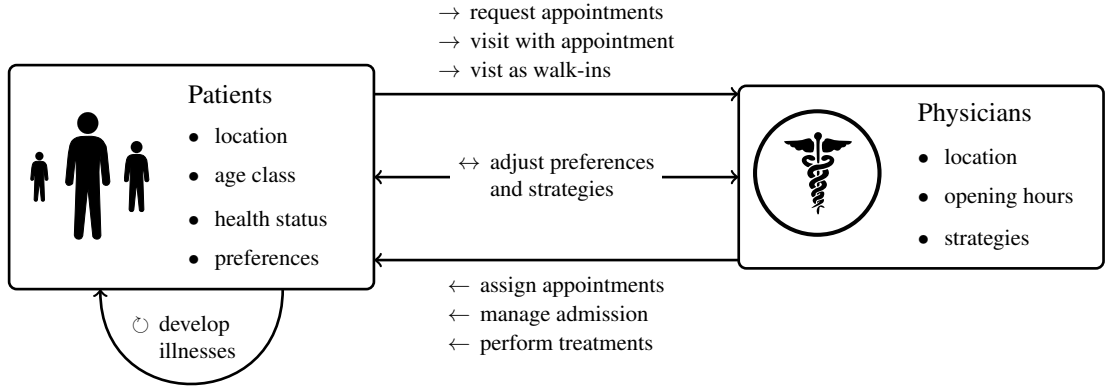}
	\caption{Concept of \SimulationModel showing both types of agents with their main attributes as well as interactions between agents.}
	\label{overview}
\end{figure*}

\SimulationModel models the interaction of patients and primary care physicians in a restricted geographical area over a given period of time; compare Figure~\ref{fig:domain}.
The model features two types of agents: a population of potential patients $\SP$ and a population of primary care physicians $\SGP$.
Every patient $\p \in \SP$ resides at a specific location, belongs to a certain age group, has an individual health status, as well as individual treatment preferences; compare Figure~\ref{overview}.
Patients continuously develop acute illnesses that depend on their age and health status and require treatment until they subside.
Additionally, patients may suffer from long term chronic illnesses which need to be monitored by a physician on a regular basis.
To receive medical attention, patients either schedule an appointment or visit a PCP's practice without prior notice.
Thereby, the patients' decision making process depends on their individual preferences and health status which determine the choice of physician, the type of the visit (walk-in/ appointment), as well as the time of the visit.
Physicians $\gp \in \SGP$ practice at a certain location and have weekly opening hours during which they admit patients for treatment; see Figure~\ref{overview}.
Moreover, every physician $\gp \in \SGP$ follows individual strategies that govern how they manage appointments, admit patients, and perform treatments.
As patients and physicians interact, they influence each other and adapt by adjusting their preferences and strategies.

The simulation's purpose is to model trade-offs between the objectives pursued by three stakeholder groups: patients, primary care physicians, and policy makers.
These objectives are assumed by the model as follows: While PCPs strive to efficiently utilize their time,  patients strive for a quick response to their health concerns.
Thereby, the model illustrates the trade-off between efficiency and patient-centered care. 
The objectives pursued by policy makers can range from minimizing the cost of health care to maximizing the degree of patient-centered care.
Policy makers are not represented by agents within \SimulationModel.
Instead, policy decisions set relevant model parameters such as the number of physicians in the area, treatment standards, and the financial reimbursement system.

To describe the simulation model in detail, we rely on guidance from the ODD framework  described by Grimm et al.~\cite{Grimm2010ODD}.
To concisely highlight the interaction of relevant components in that framework, we order and group the design questions given by this framework as follows:
First, Section~\ref{scales} defines the temporal and geographical scales within \SimulationModel.
Second, Section~\ref{epp} describes the representation of the relevant entities and state variables, including patients and physicians with their sensing, predicting, adapting, interacting, and learning actions.
Third, Section~\ref{poas}  provides process overviews and describes matters of scheduling.
Fourth, Section~\ref{sec:stochasticity} explains how and where the model captures the uncertain nature of health care systems through stochastic parameters.
Fifth, Section~\ref{emergenceObs} briefly reviews the indicators that result from running the simulation and explains their emergent properties.
Sixth, Section~\ref{sec:init} discusses the initialization of a simulation experiment.
Seventh, Section~\ref{submodels} documents the submodels that implement, e.g., the PCP's strategies to handle appointments.
Finally, Section~\ref{sec:structuralValiVeri} describes our structural validation as well as our approach to verification taken when implementing the model.

\subsection{Simulation Environment}
\label{scales}

\SimulationModel's environment entails the geographical and temporal structure as well as policy effects. Within the model, locations $\location\in \Locations \coloneqq [-90, 90] \times [-180,180]$ are represented using the geographic coordinates latitude and longitude indicating the north-south and east-west position, respectively.

The modeled time period is considered as a continuum structured by \PITS and durations.
For any time object (\PIT or duration) $\pointInTime=(\Day, \decimalTime) \in \PointsInTime \coloneqq \N \times [0,1) $, $\Day \in \mathbb{N}$ indicates the day and  \mbox{$\decimalTime \in [0,1) \eqqcolon \DecimalTimes$} specifies the time as an increment of day known as \textit{decimal time}.
That is, we use the same encoding for \PITS and durations as context uniquely defines which of the former a time object refers to. 
For example, $(38,0.55) \in \PointsInTime$ corresponds to day $38$ and $24\cdot60\cdot0.55 = 792$ minutes, i.e., $1$:$12$ p.m. as a \PIT or, analogously, to a duration of 38 days, $13$ hours, and $12$ minutes.
To ease notation, we associate every \PIT and duration  $(\Day, \decimalTime) \in \PointsInTime$ with the non-negative value $\Day+\decimalTime \in \Rplus$ which yields a bijection between $\PointsInTime$ and $\Rplus$.

In addition to the continuous representation of time, we structure each day into a morning and an afternoon session as it is common practice in primary care~\cite{KLASSEN199683}.
Each \emph{session} $\session=(\Day,\TimeOfDay) \in \Sessions \coloneqq \N \times \{0,1\}$ is uniquely defined by a day $\Day \in \mathbb{N}$ and a binary indicator $\TimeOfDay\in \{0,1\}$.
Thereby, the binary indicator $\TimeOfDay$ defines whether it is the morning ($\TimeOfDay=0$) or the afternoon ($\TimeOfDay=1$) session.
Sessions reoccur on a weekly basis which yields an equivalence relation $\sim$ on the set of sessions $\Sessions$ via 
$$(\Day_1, \TimeOfDay_1) \sim (\Day_2, \TimeOfDay_2) :\Leftrightarrow \Day_1\equiv \Day_2 \,\text{mod}\, 7 \;\land\; \TimeOfDay_1= \TimeOfDay_2. $$ 
The resulting equivalence class for a session $\session \in \Sessions$ defined as $[\session] \coloneqq \{ \session^\prime \in \Sessions : \session^\prime \sim \session\}$
contains all sessions sharing the same day of the week and time of the day, e.g., all Thursday afternoon sessions.
Thus, we can associate the set of all equivalence classes $\SessionOfWeek \coloneqq \{[\session] : \session \in \Sessions\}$ with the $14$ sessions of the week, i.e., Monday to Sunday with a respective morning and afternoon session.

\subsection{Entities and State Variables}
\label{epp}

Modeled as interacting agents, patients $\p  \in \SP$ and PCPs $\gp \in \SGP$ are the active entities in the simulation.
Their interaction is motivated by patients' suffering from illnesses and therefore seeking treatment with PCPs via appointments or walk-in visits.
Both patients and PCPs are complex individuals featuring characteristics that represent entities themselves.
Going from simple to more elaborated, we begin by describing the self-containing entities of \SimulationModel and end with the description of the agents representing patients and physicians.

\subsubsection{Objectives}
When patients suffer from an acute illness, they want to be treated as soon as possible, ideally by their preferred physician.
For the continuous treatment of chronic illnesses and the follow-up care of acute illnesses, patients prefer treatment by the same physician through appointments in regular intervals. 
Physicians, on the other hand, aim at efficiently utilizing their available time while minimizing overtime.
Thus, patients' and physicians' objectives are in conflict as it is ineffective for physicians to fully comply with patient demands:
To ensure that all short-notice appointment requests can be accommodated, PCPs would have to withhold too much treatment time.
Providing follow-up appointments in strict intervals would prevent PCPs from reacting to demand fluctuations.

Policy makers, while not explicitly modeled, follow a multitude of conflicting objectives.
On the one hand, they need to ensure a certain minimum standard in health care quality to guarantee patients are treated when necessary.
On the other hand, they cannot afford to subsidize an excessive number of physicians.
Thus, policy makers necessarily aim at a trade-off:  A purely patient-based system that disregards efficiency is likely to turn out to be unaffordable, a health system optimized only for efficiency might lead to unacceptable waiting and access times.
\SimulationModel represents policy decisions through their resulting parameter values, e.g., the number of physicians and their distribution.

\subsubsection{Illnesses and Families of Illnesses}
\label{illnesses}

Illnesses are health concerns that cause discomfort to patients and require treatment.
They belong to a certain illness family (e.g. cold or heartburn), have a certain seriousness (e.g. mild or severe), persist over a certain period of time, and require an initial treatment within an acceptable time frame as well as subsequent follow-up visits in regular time intervals.
In \SimulationModel, we formalize illnesses as tuples $\illness=(\seriousness, \familyOfIllnesses,\duration, \wtwIllness, \tf) \in \AllIllnesses$ with attributes as shown in Table~\ref{tab:illness}.
Thereby, $\seriousness\in [0,1]$ defines the seriousness of the illness, $\familyOfIllnesses \in \SFamiliesIllnesses$ defines the illness family of the illness, and $\duration \in \PointsInTime$ defines the duration of the illness.
The parameter $\wtwIllness \in \PointsInTime$ defines the illness' willingness to wait, which is the patient's maximum accepted waiting time for the initial treatment.
The parameter $\tf \in \PointsInTime$ defines the illness' follow-up interval, which specifies the frequency of the required aftercare that follows the initial treatment.
When we use this representation to model health concerns that are not strictly illnesses like the need for vaccination, the characteristics duration and follow-up interval may not apply.
In such cases, setting parameter values $\duration= \emptyset$ and $\tf= \emptyset$  indicates that the  respective characteristic is not applicable for $\illness\in \AllIllnesses$.

Families of illnesses serve as the classification system of illnesses within \SimulationModel.
While emerging illnesses vary in their manifestation, families of illnesses define the common constant traits of all illnesses belonging to the same illness family.
In our model, the common constant traits of all illnesses $\illness\in \AllIllnesses$ with seriousness $\seriousness\in [0,1]$ belonging to illness family $\familyOfIllnesses\in \SFamiliesIllnesses$ are the expected duration $\nominalDuration{\familyOfIllnesses}(\seriousness) \in \PointsInTime$,  the expected willingness to wait $\nominalWTW{\familyOfIllnesses}(\seriousness) \in \PointsInTime$, and the follow-up interval $\treatmentFrequency{\familyOfIllnesses}(\seriousness) \in \PointsInTime$.
The expected duration $\nominalDuration{\familyOfIllnesses}(\seriousness)$ and expected willingness to wait $\nominalWTW{\familyOfIllnesses}(\seriousness)$ are exclusively used during the generation of new emerging illnesses and serve as the means for distributions from which we sample each illness' stochastic duration $\duration$ and stochastic willingness to wait $\wtwIllness$.
Thus for all emerged illnesses $\illness\in \AllIllnesses$, it generally holds that $\duration \neq \nominalDuration{\familyOfIllnesses}(\seriousness)$ and $\wtwIllness \neq \nominalWTW{\familyOfIllnesses}(\seriousness)$.
Only the follow-up interval of emerged illnesses $\illness\in \AllIllnesses$ derives from the illness family in a deterministic way, i.e., $\tf=\treatmentFrequency{\familyOfIllnesses}(\seriousness)$.

In order to define the common traits of emerging illnesses, families of illnesses $f \in \SFamiliesIllnesses$ are formally specified by three functions:
A linear function $\nominalDuration{f}\colon [0,1] \to \PointsInTime$ that defines the expected duration $\nominalDuration{f}(s)$ in days for all emerging illnesses with seriousness $s\in[0,1]$ that derive from illness family $f\in \SFamiliesIllnesses$.
Moreover, linear functions $\nominalWTW{f}\colon [0,1] \to \PointsInTime$ and $\treatmentFrequency{f}\colon [0,1] \to \PointsInTime$ that analogously define the expected willingness to wait in days and follow-up interval in days; see Table~\ref{coi}.
As above, we indicate the inapplicability of the characteristics duration or follow-up interval to families of illnesses by setting $\nominalDuration{f} = \emptyset$ and $\treatmentFrequency{f}=\emptyset$, respectively.

\begin{table}[tb]
	\centering
	\caption{Attributes of illnesses $\illness\in \AllIllnesses$.}
	\label{tab:illness}
	\begin{tabular*}{\linewidth}{@{}p{2.8cm}p{2.3cm}l@{}}
		\toprule
		Attribute                    & Type & Unit                   \\ \midrule
		seriousness	     & $\seriousness \in [0,1]$ & \\
		illness family   & $\familyOfIllnesses \in \SFamiliesIllnesses$ & \\
		duration         & $\duration\in \PointsInTime $ & [days]   \\
		willingness to wait      & $\wtwIllness \in \PointsInTime$ & [days]  \\
		follow-up interval & $\tf\in \PointsInTime $ & [days]  \\
		\bottomrule
	\end{tabular*}
\end{table}

\begin{table}[tb]
	\centering
	\caption{Attributes of families of illnesses $f \in \SFamiliesIllnesses$.}
	\label{coi}
	\begin{tabular*}{\linewidth}{@{}ll@{}}
		\toprule
		Attribute                    & Type                   \\ \midrule
		linear function for expected duration       &$\nominalDuration{f}\colon [0,1] \to \PointsInTime$ \\
		linear function for expected willingness      &$\nominalWTW{f}\colon [0,1] \to \PointsInTime$ \\
		linear function for  follow-up interval  	  &$\treatmentFrequency{f}\,\colon [0,1] \to \PointsInTime$ \\
		chronic attribute	 					  & $\chronicAttribute{f}\in \{0,1\}$\\  \bottomrule
	\end{tabular*}
\end{table}

To illustrate the concept of illnesses and families of illnesses, consider the illness family ``common cold'' with expected illness duration defined by $\nominalDuration{f}(s)=10\,s +3$, expected  willingness to wait defined by $\nominalWTW{f}(s)=-3\,s +3$, and follow-up interval defined by $\treatmentFrequency{f}(s)=-2\,s +7$. When a patient develops a mild ($\seriousness=0.2$) ``common cold'', the illness family ``common cold'' defines the expected duration, expected willingness to wait, and follow-up interval of the mild cold as $\nominalDuration{f}(\seriousness)=5$ days, $\nominalWTW{f}(\seriousness)=2.4$ days, and $\treatmentFrequency{f}(\seriousness)=6.6$ days. The actual duration and willingness to wait of the developed mild ``common cold'' are stochastic and vary around their expected counterparts, e.g., $\duration=5.5$ days and $\wtwIllness=2.7$ days.
The illness' follow-up interval is deterministic and derives from the illness family via $\tf=\treatmentFrequency{\familyOfIllnesses}(\seriousness)=6.6$ days.

To model chronic health concerns such as diabetes that persist over an extended period of time, a chronic attribute $\chronicAttribute{f}\in \{0,1\}$ identifies families of chronic illnesses.
Thereby, $\chronicAttribute{f}$ partitions $\SFamiliesIllnesses$ into the set of acute families of illnesses  $\TemporaryFamiliesIllnesses\coloneqq \{f\in\SFamiliesIllnesses: \chronicAttribute{f}=0\}$ and the set of chronic families of illnesses $\ChronicFamiliesIllnesses\coloneqq \{f\in\SFamiliesIllnesses: \chronicAttribute{f}=1\}$.
This directly induces a partition of the set of illnesses $\AllIllnesses$ into the set of acute $\AllTmpIllnesses$ and the set of chronic illnesses $\AllChroIllnesses$.

Acute illnesses $\illness \in \AllTmpIllnesses$ develop and subside over time and patients can simultaneously suffer from an arbitrary number of acute illnesses $\tmpIllnesses \subseteq \AllTmpIllnesses$.
Chronic illnesses $\chronicIllness\in\AllChroIllnesses$ are conceived as static by the model -- they neither develop nor heal in the modeled time period.
Instead, each patient $\p\in\SP$ either suffers from exactly one chronic illness $\chronicIllness_\p\in \AllChroIllnesses$ throughout the modeled time period, i.e., $\chroIllnesses_\p = \{\chronicIllness_\p\} \subseteq \AllChroIllnesses$, or no chronic illness at all, i.e., $\chroIllnesses_\p = \emptyset$.
To distinguish patients suffering from a chronic illness from those who do not, we refer to the former as \textit{chronic patients}.

\subsubsection{Appointments}
Appointments specify the \PIT at which the treatment of a specific patient is scheduled to take place.
To that end, appointments $\appointment\in \Appointments$ are defined by the time of the appointment $\pointInTime_\appointment \in \PointsInTime$, the attending primary care physician $\gp_\appointment \in \SGP$, and the  patient $\p_\appointment \in \SP$ receiving treatment.
At any point in time, non-chronic patients can have at most one scheduled appointment $\stdappointment\in \Appointments$, called the \emph{\stdApp} appointment.
Acute appointments are intended for the initial treatment of acute illnesses, the follow-up treatment of acute illnesses, or both.
Chronic patients, may have a \emph{\regApp} appointment $\regappointment\in \Appointments$ to treat their chronic illness in addition to the \stdApp appointment to treat their acute illnesses.
While chronic illnesses are only treated during \regApp appointments, acute illness are treated during any appointment.
Thus, all of a patients' acute illnesses $\tmpIllnesses$ are treated during every appointment.

\subsubsection{Age Classes}
\label{ageClass}
Age classes group the modeled set of patients and serve the purpose of defining the common characteristics of patients within the respective classes.
For patients of age class $a\in \SAgeClass$, these characteristics are the deviation from the expected illness duration $\changeInDuration > 0$, the deviation from the expected willingness to wait $\changeInWTW \geq 0$, the probability to cancel an appointment after full recovery $p_a \in [0,1]$, and the expected number of annual acute illnesses defined through the linear function $\expAnnualIllnesses\colon [0,1] \to \Rplus$; see Table~\ref{ageclass}.
\begin{table}[t]
	\centering
	\caption{Attributes of age classes $\age \in \SAgeClass$.}
	\label{ageclass}
	\begin{tabular*}{\linewidth}{@{}ll@{}}
		\toprule
		Attribute                      & Type              \\ \midrule
		lin.~function exp.~annual \act illnesses     & $\expAnnualIllnesses\colon [0,1] \to \Rplus$ \\
		dev.~from exp.~illness duration    & $\changeInDuration>0$   \\ 
		dev.~from exp.~willingness to wait & $\changeInWTW \geq 0$   \\ 
		probability to cancel appointments          & $\probToCancel \in [0,1]$        \\ \bottomrule
	\end{tabular*}
\end{table} 
The deviation from the expected illness duration $\changeInDuration$ is a multiplicative factor, that determines whether the expected illness duration $\nominalDuration{\familyOfIllnesses}(\seriousness)\in \PointsInTime$ extends ($\changeInDuration>1$) or shortens ($\changeInDuration<1$) for patients of age class $\age\in \SAgeClass$. 
Analogously, the deviation from the expected willingness to wait $\changeInWTW$, determines how the expected willingness to wait $\nominalWTW{\familyOfIllnesses}(\seriousness)\in \PointsInTime$ of an illnesses changes for patients of age class $\age\in \SAgeClass$.
The linear function $\expAnnualIllnesses\colon [0,1] \to \Rplus$ defines the expected number of annual acute illnesses $\expAnnualIllnesses(\condition)\in \Rplus$ for patients in age class $\age \in \SAgeClass$ which depends on  the patient's individual health condition $\condition\in [0,1]$ which can range from perfectly healthy ($\condition = 0$) to extremely delicate ($\condition = 1$).

\subsubsection{Age Class-Illness Distribution}
\label{illnessAgeDist}
The age class-illness distribution $\illnessAgeclassDist \colon \SAgeClass \times \TemporaryFamiliesIllnesses \to [0,1]$ builds the connection between the set of age classes $\SAgeClass$ and the set of acute families of illnesses $\TemporaryFamiliesIllnesses$.
To that end, $\illnessAgeclassDist$ defines the expected distribution of acute illness families for each age class, i.e., among all developed acute illnesses by patients of age class $\age \in \SAgeClass$, a fraction $\illnessAgeclassDist(\age,\familyOfIllnesses)\in [0,1]$ is expected to belong to illness family $\familyOfIllnesses\in \TemporaryFamiliesIllnesses$.
As a result, $\illnessAgeclassDist$ defines a discrete probability distribution on the set of acute families of illnesses $\TemporaryFamiliesIllnesses$ for fixed age class $\age\in \SAgeClass$, i.e., $\sum_{\familyOfIllnesses\in \TemporaryFamiliesIllnesses}\, \illnessAgeclassDist(\age, \familyOfIllnesses) = 1$. 

\subsubsection{Patients}
\label{patients}
Patients are the driving force of the simulation, as their health concerns trigger the events that underly most of the simulation's processes.
All non-chronic patients $\p\in\SP$ are characterized by their geographical location $\location\in \Locations$, health condition $\condition\in [0,1]$, acute illnesses $\tmpIllnesses \subseteq \AllTmpIllnesses$, age class $\age\in \SAgeClass$, \stdApp appointment $\stdappointment \in \Appointments$, and preferences.
While the location, health condition, and age class of each patient remain constant throughout a simulation experiment, a patient's acute illnesses, \stdApp appointment and preferences are variable and change over time.
Chronic patients possess all the characteristics of non-chronic patients, but are additionally identified by a constant chronic illness $\chronicIllness \in \AllChroIllnesses$ and a variable \regApp appointment $\regappointment\in \Appointments$.

Patients' preferences determine when, where and how they pursue treatment.
Specifically, each patient considers a set of PCPs $\ScGP \subseteq \SGP$ and never seeks treatment with PCPs outside the consideration set.
Since continuity in the treatment of chronic illnesses is particularly important, chronic patients select a distinguished family physician $\fgp \in \ScGP$ with whom all \regApp appointments $\regappointment\in \Appointments$ are exclusively arranged.
While every patients' consideration set $\ScGP$ remains constant throughout the modeled time period, patients reevaluate and vary their family physician.
Naturally, patients have personal schedules and cannot attend all weekly sessions.
Thus, the model assumes that each patient has a constant set of weekly-reoccurring session availabilities given by $\availabilities\colon \SessionOfWeek \to \{0,1\}$, where $0$ encodes unavailability.
Finally, patients maintain individual appointment ratings $\rApp \geq 0$ as well as session-specific walk-in ratings $\rWalk \geq 0$ for every weekly session  $[\session] \in \SessionOfWeek$ and every considered physician $\gp \in \ScGP$.

Ratings are the means by which patients express their satisfaction with a physician's services.
Whenever a patient seeks consultation, the choice of physician is determined by the patient's current ratings.
To that end, ratings incorporate patients' sense of geographic distance, matching of opening hours with availabilities, and previous positive and negative experiences.
As patients adjust their ratings over time, they adjust their choice of PCP.
If a physician is unable to meet an appointment request, incurs excessive waiting time, or rejects patients due to capacity overruns, patients reduce their rating.
Positive experiences such as successful appointment arrangements or short waiting times increase ratings.
In other terms, through their sensing of the quality of treatment and the adaptation of their ratings, patients learn about the quality of PCPs throughout the simulation cycle.

When patients begin to suffer from a new illness, they always seek treatment. To that end, patients first request an appointment from the set of considered PCPs $\ScGP$. Appointment requests are one of the ways in which patients and PCPs interact. Patients attempt up to two appointment requests in order of the appointment rating $\rApp \geq 0$ they assign to the considered primary care physicians $\gp \in \ScGP$.
If both requested PCPs fail to offer a feasible appointment within the patient's willingness to wait, patients resort to their second way of interacting with physicians: They forgo an appointment and visit a PCP as a walk-in patient. The selection of the PCP for the walk-in visit is based on the corresponding  walk-in rating $\rWalk$ of the targeted session $\session\in\Sessions$. 

Upon arrival, patients may be rejected by physicians due to, e.g., capacity overloads.
Following a rejection, patients update their rating of the rejecting PCP and attempt a new visit as walk-in patient at the then-highest-rated PCP.
Rejected patients are flagged as emergencies ($\emergencyFlag=1$) for as long as they unsuccessfully continue to seek treatment.
In our model, this emergency state does not enforce a particular PCP behavior.
Instead, PCPs may include the emergency state in their decision making.

Until an illness $\illness\in \tmpIllnesses$ subsides, patients continuously try to arrange follow-up appointments to the initial treatment with the attending physician in the follow-up interval $\tf\in \PointsInTime$.
Analogously, chronic patients continuously try to arrange regular appointments with their family physician $\fgp \in \ScGP$ in the follow-up interval $\nu_\chronicIllness\in \PointsInTime$ of their unique chronic illness $\chronicIllness\in\chroIllnesses$.
Only if the arrangement of a follow-up or regular appointment fails and the aftercare of the patients is endangered, do patients seek follow-up treatment as walk-in patients.
As a result, a chronic patient's chronic illness $\chronicIllness\in \chroIllnesses$ can be treated by a physician other than the family physician $\fgp\in \ScGP$, but only through a walk-in visit triggered by the unavailability of a \regApp appointment.

In \SimulationModel, patients do not directly interact with other patients. However, an indirect form interaction emerges as patients compete with each other for timely treatment by their preferred PCP.

The attributes shared by all patients as well as the attributes specific to chronic patients are summarized in Table~\ref{patientAttr}.

\begin{table}[t]
	\centering
	\caption{Attributes of (chronic) patients $\p \in \SP$.}
	\label{patientAttr}
	\begin{tabular*}{\linewidth}{@{}lll@{}}
		\toprule
		Attribute                      & Domain    &Type             \\ \midrule
		location    & $\location\in \Locations$ & constant \\
		health condition    & $\condition\in [0,1]$  & constant \\ 
		age class & $\age \in \SAgeClass$  & constant \\ 
		acute illnesses & $\tmpIllnesses \subseteq \AllTmpIllnesses$ & variable\\
		emergency flag			& $\emergencyFlag \in \{0,1\}$ & variable \\
		\stdApp appointment			& $\stdappointment \in \Appointments$ & variable\\
		considered PCPs			& $\ScGP \subseteq \SGP$ & constant\\
		availabilities          & $\availabilities\colon \SessionOfWeek \to \{0,1\}$ & constant\\
		appointment ratings		& $\rApp \geq 0, \,\forall\gp {\in} \ScGP$ & variable\\
		\multirow{1}{*}{walk-in ratings}			& $\rWalk \geq 0,$ & \multirow{1}{*}{variable}\\
		& $\forall\gp {\in} \ScGP,\, \forall[\session]{\in}\SessionOfWeek $\\
		\midrule
		chronic illness			& $\chroIllnesses= \{\chronicIllness\}  \subseteq \AllChroIllnesses$ & constant\\
		\regApp appointment			& $\regappointment \in \Appointments$ & variable\\
		family physician		& $\fgp \in \ScGP$ & variable     
		 \\ \bottomrule
	\end{tabular*}
\end{table}

\subsubsection{Primary Care Physicians}
\label{PCPs}
Primary care physicians operate practices featuring an uncapacitated waiting room to offer medical services to patients in need.
The model characterizes physicians $\gp \in \SGP$ by their geographic location $\location\in\Locations$, opening hours, as well as an individual set of strategies to schedule appointments, manage patient admission, and organize treatments.

\begin{figure*}
	\begin{tikzpicture}[xscale=0.8, line width=1pt, yscale=1.2]

\draw[white, pattern= vertical lines, pattern color=blue1]  (5.5,-0.25) rectangle (6.5,0.25);
\draw[white, pattern=north east lines, pattern color=red1]  (6.5,-0.25) rectangle (8,0.25);
\draw[white, pattern=vertical lines, pattern color=blue1]  (11.5,-0.25) rectangle (12.5,0.25);
\draw[white, pattern=north east lines, pattern color=red1]  (12.5,-0.25) rectangle (16.7,0.25);
\draw[white, pattern=north east lines, pattern color=red1]  (1.5,-0.25) rectangle (0.0,0.25);

\draw [->,-latex] (0,0) -- (17,0) node[right=5pt, align=center]{Working\\Day};

\node[align=center] at (1.5, -0.5){ $\sessionBegin(\session_0)$};
\node[align=center] at (5.5, -0.5){ $\sessionEnd(\session_0)$};

\node[align=center] at (8, -0.5){ $\sessionBegin(\session_1)$};
\node[align=center] at (11.5, -0.5){$\sessionEnd(\session_1)$};

\begin{scope}[shift={(0,2.1)}]
\draw[white, pattern=vertical lines, pattern color=blue1] (0.0, -1.3) rectangle    (0.5,-1.6) ;
\draw[white, pattern=north east lines, pattern color=red1] (3.5, -1.3) rectangle   (4,-1.6);
\node[right] at (0.5,-1.45){Buffer time};
\node[right] at (4,-1.45){Closed};
\end{scope}

\begin{scope}[shift={(0,0)}]
\foreach \x in {1.5, 1.75, 2, 2.5, 2.75, 3.75, 4, 4.25, 4.5, 4.75, 5, 5.25, 5.5, 5.75, 6, 6.25}{
\draw[green1, opacity=0.5, line width=.3cm] (\x,0.0)--(\x+0.25,0.0);
}

\begin{scope}[shift={(0,0)}]
\draw[dashed,gray, opacity=0.5, line width=.3cm] (3,0)--(3.75,0);
\draw[dashed ,gray, opacity=0.5, line width=.3cm] (2.25,0)--(2.5,0);
\draw[red1, opacity=0.5, line width=.3cm] (6.5,0)--(7,0);
\end{scope}

\foreach \x in {8, 8.25, 8.5, 8.75, 9, 9.25, 10, 10.25, 10.5, 10.75, 11, 11.25, 11.5, 11.75, 12, 12.25}{
\draw[ green1, opacity=0.5, line width=.3cm] (\x,0.0)--(\x+0.25,0.0);
}

\begin{scope}[shift={(0,0)}]
\draw[ dashed,gray, opacity=0.5, line width=.3cm] (9.5,0)--(10,0);
\draw[red1,opacity=0.5, line width=.3cm] (12.5,0)--(14,0);
\end{scope}

\begin{scope}[shift={(0,-.9)}]
\draw[ green1,opacity=0.5, line width=.3cm] (7, 1.55) --  node[opacity=1.0, right=7pt, black]{Service time} (7.5,1.55);
\draw[dashed,gray, opacity=0.5, line width=.3cm] (10.5, 1.55) --  node[opacity=1.0,right=7pt,  black]{Idle time} (11,1.55);
\draw[red1, opacity=0.5, line width=.3cm] (14, 1.55) --  node[opacity=1.0,right=7pt, black]{Overtime} (14.5,1.55);
\end{scope}

\end{scope}

\foreach \x in {1.5, 5.5, 6.5,8,11.5,12.5}{
	\draw (\x,0.2) -- (\x, -0.2);
}
\end{tikzpicture}
	\caption{Schematic representation of a PCP's morning ($\session_0$) and afternoon ($\session_1$) session visualizing service-, idle- and overtime.}
	\label{timeline}
\end{figure*}
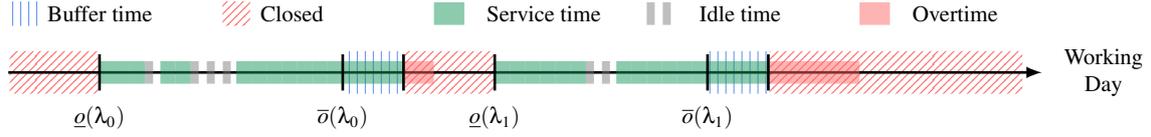

\SimulationModel assumes that all physicians operate in clinical sessions. 
Opening hours for these sessions are weekly-reoccurring and therefore defined over the session of the week via  $\openingHours\colon \SessionOfWeek \to \DecimalTimes \times \DecimalTimes$ where $\DecimalTimes$ denotes the set of decimal times defined in Section~\ref{scales}. Opening hours specify for each session $\session \in \Sessions$ the time window $\openingHours([\session])\coloneqq [\sessionBegin([\session]), \sessionEnd([\session])]$ during which a physician generally admits patients for treatment. The beginning of session $\session=(\Day, \TimeOfDay) \in \Sessions$ is defined as $\sessionBegin(\session)\coloneqq (\Day,\sessionBegin([\session]) \in \PointsInTime$,
the session's end as $\sessionEnd(\session)\coloneqq (\Day,\sessionEnd([\session]) \in \PointsInTime$.
To encode that a PCP is closed for a weekly session $[\session]\in \SessionOfWeek$, we set $\openingHours([\session])= \emptyset$. Physicians utilize the first hour after the end of each session as time buffer to compensate for possible delays and walk-in patients.
Buffers are considered anticipated working time so that only service time that extends beyond the buffer constitutes overtime.
Figure~\ref{timeline} provides a schematic visualization of a PCP's working day.

Primary care physicians implement a set of \emph{strategies} to schedule appointments, decide on patient admissions, and organize the treatment of patients.
These strategies govern the physicians' interactions with patients and incorporate all of their sensing, predicting, adapting, and learning.

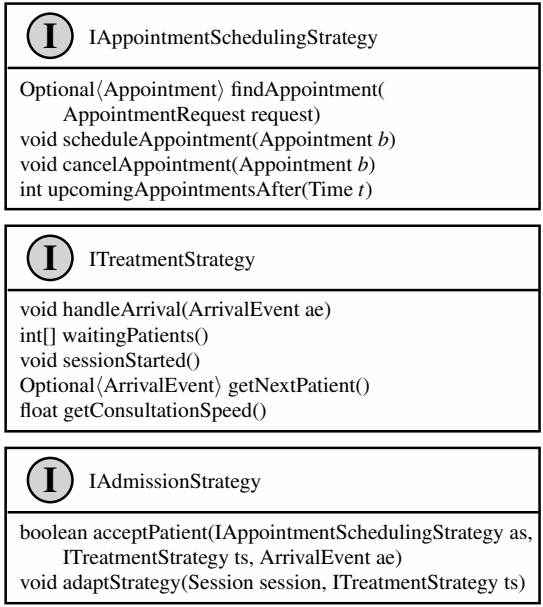
\begin{figure}
\definecolor{plantucolor0000}{RGB}{254,254,254}
\definecolor{plantucolor0001}{RGB}{0,0,0}
\definecolor{plantucolor0002}{RGB}{211,211,211}
\definecolor{plantucolor0003}{RGB}{0,0,0}
\resizebox*{\linewidth}{!}{
\begin{tikzpicture}[yscale=-.8, xscale = .8
,pstyle2/.style={color=plantucolor0001,line width=1pt}
]
\draw[color=plantucolor0001,fill=plantucolor0000,line width=1.0pt] (6pt,8pt) rectangle (269.4602pt,110.2188pt);
\begin{scope}[xshift=-1.3cm]
\draw[color=plantucolor0001,fill=plantucolor0002,line width=1.0pt] (64.9676pt,24pt) ellipse (11pt and 11pt);
\node at (64.9676pt,24pt)[]{\textbf{\Large I}};
\node at (80.4676pt,17.0156pt)[below right,color=black]{{IAppointmentSchedulingStrategy}};
\end{scope}
\draw[pstyle2] (7pt,40pt) -- (269.4602pt,40pt);
\begin{scope}[yshift=-.35cm, xshift=-.1cm]
\node[align=left] at (12pt,52pt)[below right,color=black]{Optional$\langle$Appointment$\rangle$  findAppointment( \\ \qquad AppointmentRequest request)\\
void scheduleAppointment(Appointment $\appointment$)\\
void cancelAppointment(Appointment $\appointment$)\\
int upcomingAppointmentsAfter(Time $t$)
};
\end{scope}
\end{tikzpicture}
}
		
		\vspace*{-.2cm}
\definecolor{plantucolor0000}{RGB}{254,254,254}
\definecolor{plantucolor0001}{RGB}{0,0,0}
\definecolor{plantucolor0002}{RGB}{211,211,211}
\definecolor{plantucolor0003}{RGB}{0,0,0}
\resizebox*{\linewidth}{!}{
\begin{tikzpicture}[yscale=-.8, xscale = .8
,pstyle2/.style={color=plantucolor0001,line width=1pt}
]
\draw[color=plantucolor0001,fill=plantucolor0000,line width=1.0pt] (6pt,8pt) rectangle (269.4602pt,110.2188pt);
\begin{scope}[xshift=-1.3cm]
\draw[color=plantucolor0001,fill=plantucolor0002,line width=1.0pt] (64.9676pt,24pt) ellipse (11pt and 11pt);
\node at (64.9676pt,24pt)[]{\textbf{\Large I}};
\node at (80.4676pt,17.0156pt)[below right,color=black]{{ITreatmentStrategy}};
\end{scope}
\draw[pstyle2] (7pt,40pt) -- (269.4602pt,40pt);
\begin{scope}[yshift=-.35cm, xshift=-.1cm]
\node[align=left] at (12pt,52pt)[below right,color=black]{void handleArrival(ArrivalEvent ae) \\
int[] waitingPatients() \\
void sessionStarted()\\
Optional$\langle$ArrivalEvent$\rangle$ getNextPatient()\\
float getConsultationSpeed()
};
\end{scope}
\end{tikzpicture}
}
		
		\vspace*{-.2cm}
\definecolor{plantucolor0000}{RGB}{254,254,254}
\definecolor{plantucolor0001}{RGB}{0,0,0}
\definecolor{plantucolor0002}{RGB}{211,211,211}
\definecolor{plantucolor0003}{RGB}{0,0,0}
\centering
\resizebox*{ \linewidth}{!}{
\begin{tikzpicture}[yscale=-.8, xscale = .8
,pstyle2/.style={color=plantucolor0001,line width=1.0pt}
]
\draw[color=plantucolor0001,fill=plantucolor0000,line width=1.0pt] (6pt,8pt) rectangle (269.4602pt,85.2188pt);
\begin{scope}[xshift=-1.3cm]
\draw[color=plantucolor0001,fill=plantucolor0002,line width=1.0pt] (64.9676pt,24pt) ellipse (11pt and 11pt);
\node at (64.9676pt,24pt)[]{\textbf{\Large I}};
\node at (80.4676pt,17.0156pt)[below right,color=black]{{IAdmissionStrategy}};
\end{scope}
\draw[pstyle2] (7pt,40pt) -- (269.4602pt,40pt);
\begin{scope}[yshift=-.35cm, xshift=-.1cm]
\node[align=left] at (12pt,52pt)[below right,color=black]{boolean acceptPatient(IAppointmentSchedulingStrategy as, \\
	\qquad ITreatmentStrategy ts, ArrivalEvent ae) \\
void adaptStrategy(Session session, ITreatmentStrategy ts)
};
\end{scope}
\end{tikzpicture}
}
	\caption{Interfaces implemented by strategies.}
	\label{fig:inter}
\end{figure}

The PCP's \emph{appointment scheduling strategy} $\appointmentStrategy\in \AppointmentStrategies$ defines how consultation time is allocated to appointment slots and how the resulting slots are assigned to requesting patients. The feasible set of appointment scheduling strategies $\AppointmentStrategies$ is defined via the interface shown in Figure~\ref{fig:inter}.
That is, every appointment scheduling strategy $\appointmentStrategy\in \AppointmentStrategies$ has to provide the functionality to answer appointment requests with an appointment suggestion (that can be empty).
Thereby, every appointment request specifies the requesting patient, earliest possible appointment time, willingness to wait, whether the request is for a \regApp appointment, and whether patient's availabilities have to be respected.
Furthermore, every appointment scheduling  strategy $\appointmentStrategy\in \AppointmentStrategies$ has to provide the functionality  to schedule previously offered appointments as well as the functionality to cancel previously scheduled appointments.
Finally, every appointment scheduling strategy $\appointmentStrategy\in \AppointmentStrategies$  has to be able to compute the number of upcoming appointments within a session that are scheduled to take place after specified \PIT.

The PCP's \emph{treatment strategy} $\treatmentStrategy\in \TreatmentStrategies$ defines the order of treatment among patients from the waiting room.
Physicians sense their patients' waiting times as input for their strategy.
To account for the observation that physicians consciously or unconsciously adjust service times depending on demand \cite{diwakar2008}, treatment policies define when and how physicians adjust their consultation speed and thereby service times.
The feasible set of treatment strategies $\TreatmentStrategies$ is defined via the interface shown in Figure~\ref{fig:inter}.
That is, every treatment strategy $\treatmentStrategy\in \TreatmentStrategies$ has to keep track of admitted patients, count the number of waiting patients with and without appointment, and define how the treatment strategy is affected by the beginning of a session.
Moreover,  every treatment strategy $\treatmentStrategy\in \TreatmentStrategies$ has to determine the next patient to be treated (that might not exist) as well as the PCP's current consultation speed which is thoroughly discussed in Section~\ref{sub:str}.

The PCP's \emph{admission strategy} $\admissionStrategy\in \AdmissionStrategies$ determines whether a physician admits or rejects an arriving patient based on the current workload.
Admitted patients await their treatment in the physician's waiting room.
In \SimulationModel, PCPs are required to treat all admitted patients.
Thus, physicians underestimating their workload due to faulty predictions might have to work overtime as they accept too many patients.
On the other hand, physicians that overestimate their workload reject too many patients and fail to fully utilize their available time.
At the end of every session's buffer, physicians learn by reevaluating their predictions and adapting their admission policy.
The feasible set of admission strategies $\AdmissionStrategies$ is defined via the interface shown in Figure~\ref{fig:inter}.
That is, every admission strategy $\admissionStrategy\in \AdmissionStrategies$ has to be able to decide whether an arriving patient is admitted or not given the PCP's treatment and appointment scheduling strategy. Moreover, every admission strategy $\admissionStrategy\in \AdmissionStrategies$ has to define the adaptive traits that are performed at the end of every session's buffer and depend on the PCP's treatment strategy.

Physicians do not directly interact with other physicians. However, an indirect form of interaction emerges as PCPs compete for the patients' favor while striving for optimal utilization.

The attributes of PCP's are summarized in Table~\ref{PCPAttr}.
\begin{table}
	\centering
	\caption{Attributes of PCPs $\gp \in \SGP$.}
	\label{PCPAttr}
	\begin{tabular*}{\linewidth}{@{}ll@{}}
		\toprule
		Attribute                      & Type                 \\ \midrule
		location    & $\location\in \Locations$ \\
		opening hours    & $\openingHours\colon \SessionOfWeek \to \DecimalTimes \times \DecimalTimes$   \\ 
		appointment scheduling strategy & $\appointmentStrategy\in \AppointmentStrategies$   \\ 
		admission strategy & $\admissionStrategy \in \AdmissionStrategies$\\
		treatment strategy			& $\treatmentStrategy \in \TreatmentStrategies$   \\
		 \bottomrule
	\end{tabular*}
\end{table}
\subsection{Process Overview and Scheduling }
\label{poas}

Within \SimulationModel, the progression of time is modeled via the discrete event paradigm.
That is time is a continuum which is traversed between discrete events at which the system state is updated.
The model stores events of the form $(\pointInTime,\event)$ in a sequential queue $\eventQueue$ where $\pointInTime\in \PointsInTime$ is the \PIT an event of type $\event\in \Events$ occurs.

\begin{figure*}
	\definecolor{dred}{RGB}{168,0,54}
	\centering
	\begin{tikzpicture}[>=latex,xscale=2.4, yscale=1.6, line width=1.0pt]
	\fill[gray!10] (0.5, 0.75) rectangle (3.7, -0.35);
	\fill[gray!70] (3.7, 0.75) rectangle (6.5, -0.35);
	\draw (0.7,0.65) node {Past};
	\draw (6.2,0.65) node {Future};
	\draw (3.7,0.65) node {Now};
	\draw[-|] (3.7,0.1) -- (6.5,0.1);
	\draw[|-,gray] (0.5,0.1) -- (3.7,0.1);
	
	\foreach \x/\y  in {4.5/5, 5/6, 6.2/7}
	{\draw (\x ,0) -- (\x,0.2) node[above, pos=1.0] {$e_\y$};		
	}
	
	\foreach \x/\y  in {3.7/4}
	{\draw[dred] (\x ,0) -- (\x,0.2) node[above, pos=1.0] {$e_\y$};		
	}
	
	\draw (3.45,1.4) node{$\mathcal{Q}=\{\textcolor{gray}{(t_1,e_1), (t_2,e_2), (t_3,e_3),} \textcolor{dred}{ (t_4,e_4)},(t_5,e_5),(t_6,e_6),(t_7,e_7)\}$};
	
	\draw[|->] (3.5,1.2) -- (3.5 ,.9) node[midway, above, sloped, yshift=2pt] {}; 
	
	\foreach \x [count=\xi] in {0.6,1.7,3.1}
	{\draw[dotted,gray] (\x ,0) -- (\x,0.2) node[above, pos=1.0] {$e_\xi$};
	}
	
	\draw [decorate,decoration={brace,mirror,amplitude=5pt},xshift=0,yshift=-10pt]
	(3.1,0.3) -- (3.7,0.3) node[below,yshift=-4pt, xshift=0, pos=0.5] {$t_4-t_3$};
	\end{tikzpicture}
	\caption{Progression of time induced by the processing of event queue $\eventQueue$ via  the discrete event paradigm.}
	\label{pov}
\end{figure*}
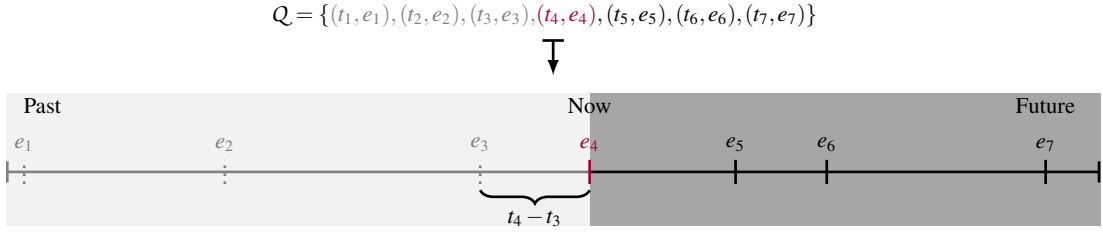

Events in $\eventQueue$ happen chronologically, i.e., $\eventQueue=\{(\pointInTime_1,\event_1),\dots, (\pointInTime_n,\event_n)\}$ with $\pointInTime_i \leq \pointInTime_{i+1}$ for $1 \leq i \leq n-1$.
As soon as an event $(\pointInTime_i,\event_i) \in \eventQueue$ occurs, the simulation advances from time $\pointInTime_{i-1}$ to time $\pointInTime_i$, compare Figure~\ref{pov}.
The simulation terminates at a specified \PIT $\timeHorizon \in \PointsInTime$, i.e., when the first element $(\pointInTime_i,\event_i)\in \eventQueue$ with $\pointInTime_i \geq \timeHorizon$ occurs.

Any event $(\pointInTime,\event)\in \eventQueue$ can generate new events or delete existing ones.
To be introduced or affected by event $(\pointInTime,\event)\in \eventQueue$, events $(\pointInTime^\prime, \event^\prime)\in \eventQueue$ must happen after time $\pointInTime$, i.e., we require $\pointInTime^\prime > \pointInTime$, so that time progresses in a consistent fashion.

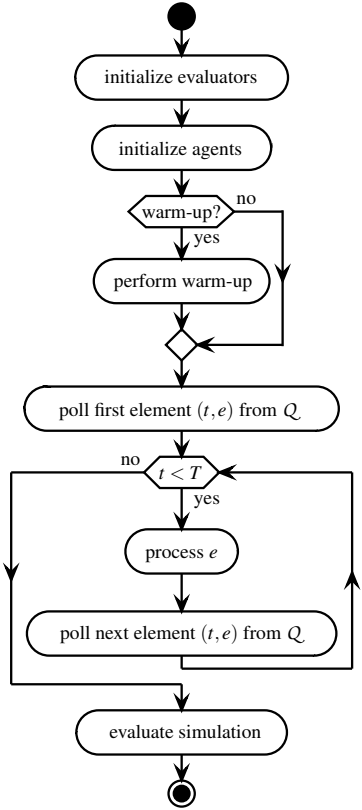
\begin{figure}
	\centering
	\resizebox{0.7\linewidth}{!}{
\definecolor{plantucolor0000}{RGB}{0,0,0}
\definecolor{plantucolor0001}{RGB}{255,255,255}
\definecolor{plantucolor0002}{RGB}{0,0,0}
\begin{tikzpicture}[yscale=-0.57, xscale=0.57
,pstyle0/.style={fill=black,line width=1.0pt}
,pstyle1/.style={color=plantucolor0002,fill=plantucolor0001,line width=1.0pt}
,pstyle3/.style={color=plantucolor0002,line width=1.0pt}
,pstyle4/.style={color=plantucolor0002,fill=plantucolor0002,line width=1.0pt}
,pstyle5/.style={color=plantucolor0002,fill=plantucolor0002,line width=1.0pt}
]
\draw[pstyle0] (153.9441pt,20pt) ellipse (10pt and 10pt);
\draw[pstyle1] (72.8608pt,66.9844pt) arc (180:270:16.9844pt) -- (89.8451pt,50pt) -- (218.043pt,50pt) arc (270:360:16.9844pt) -- (235.0274pt,66.9844pt) -- (235.0274pt,66.9844pt) arc (0:90:16.9844pt) -- (218.043pt,83.9688pt) -- (89.8451pt,83.9688pt) arc (90:180:16.9844pt) -- (72.8608pt,66.9844pt) -- cycle;
\node at (89.8608pt,56pt)[below right,color=black]{initialize evaluators};
\draw[pstyle1] (85.7155pt,120.9531pt) arc (180:270:16.9844pt) -- (102.6999pt,103.9688pt) -- (205.1883pt,103.9688pt) arc (270:360:16.9844pt) -- (222.1727pt,120.9531pt) -- (222.1727pt,120.9531pt) arc (0:90:16.9844pt) -- (205.1883pt,137.9375pt) -- (102.6999pt,137.9375pt) arc (90:180:16.9844pt) -- (85.7155pt,120.9531pt) -- cycle;
\node at (100.7155pt,110.9688pt)[below right,color=black]{initialize agents};
\draw[pstyle1] (86.9808pt,223.3242pt) arc (180:270:16.9844pt) -- (103.9652pt,206.3398pt) -- (203.923pt,206.3398pt) arc (270:360:16.9844pt) -- (220.9074pt,223.3242pt) -- (220.9074pt,223.3242pt) arc (0:90:16.9844pt) -- (203.923pt,240.3086pt) -- (103.9652pt,240.3086pt) arc (90:180:16.9844pt) -- (86.9808pt,223.3242pt) -- cycle;
\node at (96.9808pt,212.3398pt)[below right,color=black]{perform warm-up};
\begin{scope}[xscale=1.1, xshift=-0.5cm]
\draw[pstyle1] (129.3876pt,157.9375pt) -- (178.5006pt,157.9375pt) -- (190.5006pt,169.9375pt) -- (178.5006pt,181.9375pt) -- (129.3876pt,181.9375pt) -- (117.3876pt,169.9375pt) -- (129.3876pt,157.9375pt) -- cycle;
\end{scope}
\node at (157.9441pt,181.9375pt)[below right,color=black]{yes};
\node at (190.5006pt,152.1328pt)[below right,color=black]{no};
\draw[pstyle1] (153.9441pt,260.3086pt) -- (165.9441pt,272.3086pt) -- (153.9441pt,284.3086pt) -- (141.9441pt,272.3086pt) -- (153.9441pt,260.3086pt) -- cycle;
\draw[pstyle1] (34.142pt,321.293pt) arc (180:270:16.9844pt) -- (51.1263pt,304.3086pt) -- (256.7618pt,304.3086pt) arc (270:360:16.9844pt) -- (273.7462pt,321.293pt) -- (273.7462pt,321.293pt) arc (0:90:16.9844pt) -- (256.7618pt,338.2773pt) -- (51.1263pt,338.2773pt) arc (90:180:16.9844pt) -- (34.142pt,321.293pt) -- cycle;
\node at (55.142pt,310.3086pt)[below right,color=black]{ poll first element $(t,e)$ from $\mathcal{Q}$};
\draw[pstyle1] (111.0751pt,430.9722pt) arc (180:270:16.9844pt) -- (128.0595pt,413.9878pt) -- (179.8287pt,413.9878pt) arc (270:360:16.9844pt) -- (196.8131pt,430.9722pt) -- (196.8131pt,430.9722pt) arc (0:90:16.9844pt) -- (179.8287pt,447.9565pt) -- (128.0595pt,447.9565pt) arc (90:180:16.9844pt) -- (111.0751pt,430.9722pt) -- cycle;
\node at (121.0751pt,423.9878pt)[below right,color=black]{process $e$};
\draw[pstyle1] (36pt,494.0386pt) arc (180:270:16.9844pt) -- (52.9844pt,477.0542pt) -- (254.9038pt,477.0542pt) arc (270:360:16.9844pt) -- (271.8882pt,494.0386pt) -- (271.8882pt,494.0386pt) arc (0:90:16.9844pt) -- (254.9038pt,511.0229pt) -- (52.9844pt,511.0229pt) arc (90:180:16.9844pt) -- (36pt,494.0386pt) -- cycle;
\node at (56pt,482.0542pt)[below right,color=black]{poll next element $(t,e)$ from $\mathcal{Q}$};
\draw[pstyle1] (137.2774pt,358.2773pt) -- (170.6108pt,358.2773pt) -- (182.6108pt,370.2773pt) -- (170.6108pt,382.2773pt) -- (137.2774pt,382.2773pt) -- (125.2774pt,370.2773pt) -- (137.2774pt,358.2773pt) -- cycle;
\node at (157.9441pt,382.2773pt)[below right,color=black]{yes};

\node at (102.1917pt,352.4727pt)[below right,color=black]{no};
\draw[pstyle1] (73.6774pt,570.0073pt) arc (180:270:16.9844pt) -- (90.6618pt,553.0229pt) -- (217.2264pt,553.0229pt) arc (270:360:16.9844pt) -- (234.2108pt,570.0073pt) -- (234.2108pt,570.0073pt) arc (0:90:16.9844pt) -- (217.2264pt,586.9917pt) -- (90.6618pt,586.9917pt) arc (90:180:16.9844pt) -- (73.6774pt,570.0073pt) -- cycle;
\node at (93.6774pt,559.0229pt)[below right,color=black]{evaluate simulation};
\draw[color=black,line width=1.0pt] (153.9441pt,616.9917pt) ellipse (10pt and 10pt);
\draw[pstyle0] (153.9441pt,616.9917pt) ellipse (6pt and 6pt);
\draw[pstyle3] (153.9441pt,30pt) -- (153.9441pt,50pt);
\draw[pstyle4] (149.9441pt,40pt) -- (153.9441pt,50pt) -- (157.9441pt,40pt) -- (153.9441pt,44pt) -- cycle;
\draw[pstyle3] (153.9441pt,83.9688pt) -- (153.9441pt,103.9688pt);
\draw[pstyle4] (149.9441pt,93.9688pt) -- (153.9441pt,103.9688pt) -- (157.9441pt,93.9688pt) -- (153.9441pt,97.9688pt) -- cycle;
\draw[pstyle3] (153.9441pt,181.9375pt) -- (153.9441pt,206.3398pt);
\draw[pstyle4] (149.9441pt,196.3398pt) -- (153.9441pt,206.3398pt) -- (157.9441pt,196.3398pt) -- (153.9441pt,200.3398pt) -- cycle;
\draw[pstyle3] (192.5006pt,169.9375pt) -- (230.9074pt,169.9375pt);
\draw[pstyle5] (226.9074pt,213.3242pt) -- (230.9074pt,223.3242pt) -- (234.9074pt,213.3242pt) -- (230.9074pt,217.3242pt) -- cycle;
\draw[pstyle3] (230.9074pt,168.9375pt) -- (230.9074pt,273.3086pt);
\draw[pstyle3] (230.9074pt,272.3086pt) -- (165.9441pt,272.3086pt);
\draw[pstyle4] (175.9441pt,268.3086pt) -- (165.9441pt,272.3086pt) -- (175.9441pt,276.3086pt) -- (171.9441pt,272.3086pt) -- cycle;
\draw[pstyle3] (153.9441pt,240.3086pt) -- (153.9441pt,260.3086pt);
\draw[pstyle4] (149.9441pt,250.3086pt) -- (153.9441pt,260.3086pt) -- (157.9441pt,250.3086pt) -- (153.9441pt,254.3086pt) -- cycle;
\draw[pstyle3] (153.9441pt,137.9375pt) -- (153.9441pt,157.9375pt);
\draw[pstyle4] (149.9441pt,147.9375pt) -- (153.9441pt,157.9375pt) -- (157.9441pt,147.9375pt) -- (153.9441pt,151.9375pt) -- cycle;
\draw[pstyle3] (153.9441pt,284.3086pt) -- (153.9441pt,304.3086pt);
\draw[pstyle4] (149.9441pt,294.3086pt) -- (153.9441pt,304.3086pt) -- (157.9441pt,294.3086pt) -- (153.9441pt,298.3086pt) -- cycle;
\draw[pstyle3] (153.9441pt,447.9565pt) -- (153.9441pt,477.0542pt);
\draw[pstyle4] (149.9441pt,467.0542pt) -- (153.9441pt,477.0542pt) -- (157.9441pt,467.0542pt) -- (153.9441pt,471.0542pt) -- cycle;
\draw[pstyle3] (153.9441pt,382.2773pt) -- (153.9441pt,413.9878pt);
\draw[pstyle4] (149.9441pt,403.9878pt) -- (153.9441pt,413.9878pt) -- (157.9441pt,403.9878pt) -- (153.9441pt,407.9878pt) -- cycle;
\draw[pstyle3] (153.9441pt,511.0229pt) -- (153.9441pt,521.0229pt);
\draw[pstyle3] (153.9441pt,521.0229pt) -- (283.8882pt,521.0229pt);
\draw[pstyle5] (279.8882pt,457.0542pt) -- (283.8882pt,447.0542pt) -- (287.8882pt,457.0542pt) -- (283.8882pt,453.0542pt) -- cycle;
\draw[pstyle3] (283.8882pt,370.2773pt) -- (283.8882pt,521.0229pt);
\draw[pstyle3] (283.8882pt,370.2773pt) -- (182.6108pt,370.2773pt);
\draw[pstyle4] (192.6108pt,366.2773pt) -- (182.6108pt,370.2773pt) -- (192.6108pt,374.2773pt) -- (188.6108pt,370.2773pt) -- cycle;
\draw[pstyle3] (125.2774pt,370.2773pt) -- (24pt,370.2773pt);
\draw[pstyle5] (20pt,443.0542pt) -- (24pt,453.0542pt) -- (28pt,443.0542pt) -- (24pt,447.0542pt) -- cycle;
\draw[pstyle3] (24pt,370.2773pt) -- (24pt,533.0229pt);
\draw[pstyle3] (24pt,533.0229pt) -- (153.9441pt,533.0229pt);
\draw[pstyle3] (153.9441pt,533.0229pt) -- (153.9441pt,553.0229pt);
\draw[pstyle4] (149.9441pt,543.0229pt) -- (153.9441pt,553.0229pt) -- (157.9441pt,543.0229pt) -- (153.9441pt,547.0229pt) -- cycle;
\draw[pstyle3] (153.9441pt,338.2773pt) -- (153.9441pt,358.2773pt);
\draw[pstyle4] (149.9441pt,348.2773pt) -- (153.9441pt,358.2773pt) -- (157.9441pt,348.2773pt) -- (153.9441pt,352.2773pt) -- cycle;
\draw[pstyle3] (153.9441pt,586.9917pt) -- (153.9441pt,606.9917pt);
\draw[pstyle4] (149.9441pt,596.9917pt) -- (153.9441pt,606.9917pt) -- (157.9441pt,596.9917pt) -- (153.9441pt,600.9917pt) -- cycle;
\begin{scope}[xshift=-2mm, yshift=-1.2mm]
\node at (123.8876pt,162.5352pt)[below right,color=black]{warm-up?};
\node at (137.2774pt,363.875pt)[below right,color=black]{$t <  T$};
\end{scope}
\end{tikzpicture}
	}
	\caption{Structure of simulation run with time horizon $T$.}
	\label{sr}
\end{figure}

By construction, event queue $\eventQueue$ never runs empty.
Every simulation run follows the structure depicted in Figure~\ref{sr}, chronologically processing the events in $\eventQueue$ until time $\timeHorizon \in \PointsInTime$ is reached.
In this, the specific process depends on the event type $\event\in \Events$.
We now describe the different event types.

\paragraph{Arrival events} are indicated by $\arrivalEvent(\gp,\p)$. As illustrated in Figure~\ref{fig:arr}, they mark the event of patient $\p$ arriving at physician $\gp$'s practice,
either for an appointment or as a walk-in. The physician's decision to admit or reject arriving patients depends on $\gp$'s admission strategy.
Every admitted patient is guaranteed to receive treatment and enters the physician's waiting room. If the physician is currently idle, this triggers the physician's treatment strategy and treatment commences.

\begin{figure*}
	\centering
	
	\subfigure[]{
		\scalebox{.82}{
\definecolor{plantucolor0000}{RGB}{0,0,0}
\definecolor{plantucolor0001}{RGB}{255,255,255}
\definecolor{plantucolor0002}{RGB}{0,0,0}
\begin{tikzpicture}[yscale=-.59, xscale=0.59
,pstyle0/.style={fill=black,line width=1.0pt}
,pstyle1/.style={color=plantucolor0002,fill=plantucolor0001,line width=1.0pt}
,pstyle3/.style={color=plantucolor0002,line width=1.0pt}
,pstyle4/.style={color=plantucolor0002,fill=plantucolor0002,line width=1.0pt}
]

\begin{scope}[yscale=1.3, yshift=-0.75cm]
\draw[pstyle1] (65.4pt,50pt) -- (193.1115pt,50pt) -- (205.1115pt,62.8047pt) -- (193.1115pt,75.6094pt) -- (65.4pt,75.6094pt) -- (53.4pt,62.8047pt) -- (65.4pt,50pt) -- cycle;
\end{scope}
\begin{scope}[yshift=-.4cm,xshift=-.1cm]
\node at (65.4pt,46pt)[below right,color=black]{$\gp$ accepts $\p$ according };
\node at (65.4pt,61pt)[below right,color=black]{to admission policy };
\end{scope}
\begin{scope}[yshift=-.3cm]
\node at (28pt,46pt)[below right,color=black]{yes};
\node at (205.1115pt,46pt)[below right,color=black]{no};
\draw[pstyle3] (53.4pt,62.8047pt) -- (43.4pt,62.8047pt);
\draw[pstyle3] (43.4pt,62.8047pt) -- (43.4pt,90.6094pt);
\draw[pstyle3] (205.1115pt,62.8047pt) -- (215.1115pt,62.8047pt);
\draw[pstyle3] (215.1115pt,62.8047pt) -- (215.1115pt,90.6094pt);
\end{scope}

\begin{scope}[yshift=-.51cm]
\draw[pstyle0] (129.2557pt,20pt) ellipse (10pt and 10pt);
\draw[pstyle3] (129.2557pt,30pt) -- (129.2557pt,50pt);
\draw[pstyle4] (125.2557pt,40pt) -- (129.2557pt,50pt) -- (133.2557pt,40pt) -- (129.2557pt,44pt) -- cycle;
\end{scope}

\draw[pstyle1] (10pt,102.5938pt) arc (180:270:16.9844pt) -- (26.9844pt,85.6094pt) -- (59.8156pt,85.6094pt) arc (270:360:16.9844pt) -- (76.8pt,102.5938pt) -- (76.8pt,102.5938pt) arc (0:90:16.9844pt) -- (59.8156pt,119.5781pt) -- (26.9844pt,119.5781pt) arc (90:180:16.9844pt) -- (10pt,102.5938pt) -- cycle;
\node at (18pt,92.6094pt)[below right,color=black]{admit $\p$};
\draw[pstyle1] (181.5615pt,102.5938pt) arc (180:270:16.9844pt) -- (198.5459pt,85.6094pt) -- (231.6771pt,85.6094pt) arc (270:360:16.9844pt) -- (248.6615pt,102.5938pt) -- (248.6615pt,102.5938pt) arc (0:90:16.9844pt) -- (231.6771pt,119.5781pt) -- (198.5459pt,119.5781pt) arc (90:180:16.9844pt) -- (181.5615pt,102.5938pt) -- cycle;
\node at (189.5615pt,92.6094pt)[below right,color=black]{reject $\p$};
\draw[pstyle1] (129.2557pt,125.5781pt) -- (141.2557pt,137.5781pt) -- (129.2557pt,149.5781pt) -- (117.2557pt,137.5781pt) -- (129.2557pt,125.5781pt) -- cycle;
\draw[pstyle1] (23.9974pt,234.9648pt) arc (180:270:16.9844pt) -- (40.9817pt,217.9805pt) -- (217.5297pt,217.9805pt) arc (270:360:16.9844pt) -- (234.5141pt,234.9648pt) -- (234.5141pt,234.9648pt) arc (0:90:16.9844pt) -- (217.5297pt,251.9492pt) -- (40.9817pt,251.9492pt) arc (90:180:16.9844pt) -- (23.9974pt,234.9648pt) -- cycle;
\node at (25.9974pt,223.9805pt)[below right,color=black]{initiate treatment of next patient $\p$};

\node at (31.5186pt,291.9492pt)[below right,color=black]{ };
\draw[pstyle1] (79.3224pt,169.5781pt) -- (179.1891pt,169.5781pt) -- (191.1891pt,181.5781pt) -- (179.1891pt,193.5781pt) -- (79.3224pt,193.5781pt) -- (67.3224pt,181.5781pt) -- (79.3224pt,169.5781pt) -- cycle;
\node at (133.2557pt,193.5781pt)[below right,color=black]{yes};
\node at (190.2557pt,165.5781pt)[below right,color=black]{no};
\node at (79.3224pt,171.1758pt)[below right,color=black]{$\gp$ currently idle?};
\draw[pstyle4] (39.4pt,75.6094pt) -- (43.4pt,85.6094pt) -- (47.4pt,75.6094pt) -- (43.4pt,79.6094pt) -- cycle;
\draw[pstyle4] (211.1115pt,75.6094pt) -- (215.1115pt,85.6094pt) -- (219.1115pt,75.6094pt) -- (215.1115pt,79.6094pt) -- cycle;
\draw[pstyle3] (43.4pt,119.5781pt) -- (43.4pt,137.5781pt);
\draw[pstyle3] (43.4pt,137.5781pt) -- (117.2557pt,137.5781pt);
\draw[pstyle4] (107.2557pt,133.5781pt) -- (117.2557pt,137.5781pt) -- (107.2557pt,141.5781pt) -- (111.2557pt,137.5781pt) -- cycle;
\draw[pstyle3] (215.1115pt,119.5781pt) -- (215.1115pt,137.5781pt);
\draw[pstyle3] (215.1115pt,137.5781pt) -- (141.2557pt,137.5781pt);
\draw[pstyle4] (151.2557pt,133.5781pt) -- (141.2557pt,137.5781pt) -- (151.2557pt,141.5781pt) -- (147.2557pt,137.5781pt) -- cycle;

\draw[pstyle3] (129.2557pt,251.9492pt) -- (129.2557pt,286.9492pt);
\draw[pstyle3] (129.2557pt,193.5781pt) -- (129.2557pt,217.9805pt);
\draw[pstyle4] (125.2557pt,207.9805pt) -- (129.2557pt,217.9805pt) -- (133.2557pt,207.9805pt) -- (129.2557pt,211.9805pt) -- cycle;
\draw[pstyle3] (191.1891pt,181.5781pt) -- (251.9929pt,181.5781pt);

\begin{scope}[yshift=-1cm]
\draw[color=plantucolor0002,fill=plantucolor0002,line width=1.0pt] (247.9929pt,259.4492pt) -- (251.9929pt,269.4492pt) -- (255.9929pt,259.4492pt) -- (251.9929pt,263.4492pt) -- cycle;
\end{scope}

\draw[pstyle3] (251.9929pt,181.5781pt) -- (251.9929pt,288.5pt);

\draw[pstyle3] (129.2557pt,149.5781pt) -- (129.2557pt,169.5781pt);
\draw[pstyle4] (125.2557pt,159.5781pt) -- (129.2557pt,169.5781pt) -- (133.2557pt,159.5781pt) -- (129.2557pt,163.5781pt) -- cycle;

\begin{scope}[yshift=-2.3cm]
\draw[color=black,line width=1.0pt] (129.2557pt,394.918pt) ellipse (10pt and 10pt);
\draw[pstyle0] (129.2557pt,394.918pt) ellipse (6pt and 6pt);
\draw[pstyle1] (129.2557pt,340.918pt) -- (141.2557pt,352.918pt) -- (129.2557pt,364.918pt) -- (117.2557pt,352.918pt) -- (129.2557pt,340.918pt) -- cycle;

\draw[pstyle3] (251.9929pt,352.918pt) -- (141.2557pt,352.918pt);
\draw[pstyle4] (151.2557pt,348.918pt) -- (141.2557pt,352.918pt) -- (151.2557pt,356.918pt) -- (147.2557pt,352.918pt) -- cycle;

\draw[pstyle3] (129.2557pt,364.918pt) -- (129.2557pt,384.918pt);
\draw[pstyle4] (125.2557pt,374.918pt) -- (129.2557pt,384.918pt) -- (133.2557pt,374.918pt) -- (129.2557pt,378.918pt) -- cycle;
\end{scope}
\begin{scope}[yshift=-.39cm]
\draw[pstyle4] (125.2557pt,276.9492pt) -- (129.2557pt,286.9492pt) -- (133.2557pt,276.9492pt) -- (129.2557pt,280.9492pt) -- cycle;
\end{scope}
\end{tikzpicture}
		}
		\label{fig:arr}
	}
	\subfigure[]{
		\scalebox{.82}{
\definecolor{plantucolor0000}{RGB}{0,0,0}
\definecolor{plantucolor0001}{RGB}{255,255,255}
\definecolor{plantucolor0002}{RGB}{0,0,0}
\begin{tikzpicture}[yscale=-.59, xscale=0.59
,pstyle0/.style={fill=black,line width=1.0pt}
,pstyle1/.style={color=plantucolor0002,fill=plantucolor0001,line width=1.0pt}
,pstyle3/.style={color=plantucolor0002,line width=1.0pt}
,pstyle4/.style={color=plantucolor0002,fill=plantucolor0002,line width=1.0pt}
]
\draw[pstyle0] (132.8723pt,20pt) ellipse (10pt and 10pt);
\begin{scope}[xscale=.9, xshift=.53cm]

\draw[pstyle1] (50.6332pt,66.9844pt) arc (180:270:16.9844pt) -- (67.6176pt,50pt) -- (198.1269pt,50pt) arc (270:360:16.9844pt) -- (215.1113pt,66.9844pt) -- (215.1113pt,66.9844pt) arc (0:90:16.9844pt) -- (198.1269pt,83.9688pt) -- (67.6176pt,83.9688pt) arc (90:180:16.9844pt) -- (50.6332pt,66.9844pt) -- cycle;
\node at (62.6332pt,57pt)[below right,color=black]{$\gp$ treats illnesses of $\p$};
\draw[pstyle1] (48.416pt,120.9531pt) arc (180:270:16.9844pt) -- (65.4004pt,103.9688pt) -- (200.3441pt,103.9688pt) arc (270:360:16.9844pt) -- (217.3285pt,120.9531pt) -- (217.3285pt,120.9531pt) arc (0:90:16.9844pt) -- (200.3441pt,137.9375pt) -- (65.4004pt,137.9375pt) arc (90:180:16.9844pt) -- (48.416pt,120.9531pt) -- cycle;
\node at (60.416pt,110.9688pt)[below right,color=black]{$\p$ arranges follow-up};
\draw[pstyle1] (10pt,223.3242pt) arc (180:270:16.9844pt) -- (26.9844pt,206.3398pt) -- (238.7601pt,206.3398pt) arc (270:360:16.9844pt) -- (255.7445pt,223.3242pt) -- (255.7445pt,223.3242pt) arc (0:90:16.9844pt) -- (238.7601pt,240.3086pt) -- (26.9844pt,240.3086pt) arc (90:180:16.9844pt) -- (10pt,223.3242pt) -- cycle;
\node at (20pt,213.3398pt)[below right,color=black]{ initiate treatment of next patient $\p'$};
\node at (27.9037pt,280.3086pt)[below right,color=black]{ };
\draw[pstyle1] (67.791pt,157.9375pt) -- (197.9535pt,157.9375pt) -- (209.9535pt,169.9375pt) -- (197.9535pt,181.9375pt) -- (67.791pt,181.9375pt) -- (55.791pt,169.9375pt) -- (67.791pt,157.9375pt) -- cycle;
\node at (136.8723pt,181.9375pt)[below right,color=black]{yes};
\node at (67.791pt,159.5352pt)[below right,color=black]{ $\exists$ waiting patients? };
\node at (209.9535pt,154.1328pt)[below right,color=black]{no};

\draw[pstyle3] (132.8723pt,30pt) -- (132.8723pt,50pt);
\draw[pstyle4] (128.8723pt,40pt) -- (132.8723pt,50pt) -- (136.8723pt,40pt) -- (132.8723pt,44pt) -- cycle;
\draw[pstyle3] (132.8723pt,83.9688pt) -- (132.8723pt,103.9688pt);
\draw[pstyle4] (128.8723pt,93.9688pt) -- (132.8723pt,103.9688pt) -- (136.8723pt,93.9688pt) -- (132.8723pt,97.9688pt) -- cycle;
\draw[pstyle3] (132.8723pt,240.3086pt) -- (132.8723pt,275.3086pt);
\draw[pstyle4] (128.8723pt,265.3086pt) -- (132.8723pt,275.3086pt) -- (136.8723pt,265.3086pt) -- (132.8723pt,269.3086pt) -- cycle;
\draw[pstyle3] (132.8723pt,181.9375pt) -- (132.8723pt,206.3398pt);
\draw[pstyle4] (128.8723pt,196.3398pt) -- (132.8723pt,206.3398pt) -- (136.8723pt,196.3398pt) -- (132.8723pt,200.3398pt) -- cycle;
\draw[pstyle3] (209.9535pt,169.9375pt) -- (265.7445pt,169.9375pt);

\begin{scope}[yshift = -1.8cm]

\begin{scope}[yshift=1cm]
\draw[color=plantucolor0002,fill=plantucolor0002,line width=1.0pt] (261.7445pt,247.8086pt) -- (265.7445pt,257.8086pt) -- (269.7445pt,247.8086pt) -- (265.7445pt,251.8086pt) -- cycle;
\end{scope}
\draw[pstyle3] (265.7445pt,220.5pt) -- (265.7445pt,341.2773pt);
\draw[pstyle3] (265.7445pt,341.2773pt) -- (144.8723pt,341.2773pt);
\draw[pstyle4] (154.8723pt,337.2773pt) -- (144.8723pt,341.2773pt) -- (154.8723pt,345.2773pt) -- (150.8723pt,341.2773pt) -- cycle;
\draw[pstyle3] (132.8723pt,353.2773pt) -- (132.8723pt,373.2773pt);
\draw[pstyle4] (128.8723pt,363.2773pt) -- (132.8723pt,373.2773pt) -- (136.8723pt,363.2773pt) -- (132.8723pt,367.2773pt) -- cycle;
\end{scope}

\draw[pstyle3] (132.8723pt,137.9375pt) -- (132.8723pt,157.9375pt);
\draw[pstyle4] (128.8723pt,147.9375pt) -- (132.8723pt,157.9375pt) -- (136.8723pt,147.9375pt) -- (132.8723pt,151.9375pt) -- cycle;

\end{scope}

\begin{scope}[yshift=-1.8cm]
\draw[color=black,line width=1.0pt] (132.8723pt,383.2773pt) ellipse (10pt and 10pt);
\draw[pstyle0] (132.8723pt,383.2773pt) ellipse (6pt and 6pt);
\draw[pstyle1] (132.8723pt,329.2773pt) -- (144.8723pt,341.2773pt) -- (132.8723pt,353.2773pt) -- (120.8723pt,341.2773pt) -- (132.8723pt,329.2773pt) -- cycle;
\end{scope}

\end{tikzpicture}
		}
		\label{fig:rel}
	}
	\subfigure[]{
	\scalebox{.82}{
\definecolor{plantucolor0000}{RGB}{0,0,0}
\definecolor{plantucolor0001}{RGB}{255,255,255}
\definecolor{plantucolor0002}{RGB}{0,0,0}
\begin{tikzpicture}[yscale=-.59, xscale=0.59
,pstyle0/.style={fill=black,line width=1.0pt}
,pstyle1/.style={color=plantucolor0002,fill=plantucolor0001,line width=1.0pt}
,pstyle3/.style={color=plantucolor0002,line width=1.0pt}
,pstyle4/.style={color=plantucolor0002,fill=plantucolor0002,line width=1.0pt}
]

\begin{scope}[yshift=1.9cm]
\draw[pstyle0] (163.6382pt,20pt) ellipse (10pt and 10pt);
\draw[pstyle1] (71.3524pt,66.9844pt) arc (180:270:16.9844pt) -- (88.3368pt,50pt) -- (238.9395pt,50pt) arc (270:360:16.9844pt) -- (255.9239pt,66.9844pt) -- (255.9239pt,66.9844pt) arc (0:90:16.9844pt) -- (238.9395pt,83.9688pt) -- (88.3368pt,83.9688pt) arc (90:180:16.9844pt) -- (71.3524pt,66.9844pt) -- cycle;
\draw[pstyle3] (163.6382pt,30pt) -- (163.6382pt,50pt);
\draw[pstyle4] (159.6382pt,40pt) -- (163.6382pt,50pt) -- (167.6382pt,40pt) -- (163.6382pt,44pt) -- cycle;
\end{scope}

\draw[pstyle1] (86.8632pt,157.9375pt) -- (240.4132pt,157.9375pt) -- (252.4132pt,177.1445pt) -- (240.4132pt,196.3516pt) -- (86.8632pt,196.3516pt) -- (74.8632pt,177.1445pt) -- (86.8632pt,157.9375pt) -- cycle;
\node at (86.8632pt,183.5469pt)[below right,color=black]{ };
\node at (48.4632pt,160.3398pt)[below right,color=black]{yes};
\node at (254.4132pt,160.3398pt)[below right,color=black]{no};
\begin{scope}[xshift=.6cm, xscale=1.2]

\draw[pstyle1] (10pt,223.3359pt) arc (180:270:16.9844pt) -- (26.9844pt,206.3516pt) -- (102.7419pt,206.3516pt) arc (270:360:16.9844pt) -- (119.7263pt,223.3359pt) -- (119.7263pt,223.3359pt) arc (0:90:16.9844pt) -- (102.7419pt,240.3203pt) -- (26.9844pt,240.3203pt) arc (90:180:16.9844pt) -- (10pt,223.3359pt) -- cycle;
\end{scope}
\draw[pstyle1] (225.5465pt,223.3359pt) arc (180:270:16.9844pt) -- (242.5309pt,206.3516pt) -- (282.2954pt,206.3516pt) arc (270:360:16.9844pt) -- (299.2798pt,223.3359pt) -- (299.2798pt,223.3359pt) arc (0:90:16.9844pt) -- (282.2954pt,240.3203pt) -- (242.5309pt,240.3203pt) arc (90:180:16.9844pt) -- (225.5465pt,223.3359pt) -- cycle;
\draw[pstyle1] (163.6382pt,246.3203pt) -- (175.6382pt,258.3203pt) -- (163.6382pt,270.3203pt) -- (151.6382pt,258.3203pt) -- (163.6382pt,246.3203pt) -- cycle;

\begin{scope}[yshift=-1.2cm]
\draw[pstyle1] (55.761pt,361.2734pt) arc (180:270:16.9844pt) -- (72.7454pt,344.2891pt) -- (254.5309pt,344.2891pt) arc (270:360:16.9844pt) -- (271.5153pt,361.2734pt) -- (271.5153pt,361.2734pt) arc (0:90:16.9844pt) -- (254.5309pt,378.2578pt) -- (72.7454pt,378.2578pt) arc (90:180:16.9844pt) -- (55.761pt,361.2734pt) -- cycle;
\begin{scope}[yshift=-.23cm, xshift=-.1cm]
\node at (65.761pt,354.2891pt)[below right,color=black]{generate next illness event $\illnessEvent(\p)$};
\end{scope}
\draw[color=black,line width=1.0pt] (163.6382pt,408.2578pt) ellipse (10pt and 10pt);
\draw[pstyle0] (163.6382pt,408.2578pt) ellipse (6pt and 6pt);
\draw[pstyle3] (163.6382pt,304.2891pt) -- (163.6382pt,344.2891pt);
\draw[pstyle4] (159.6382pt,334.2891pt) -- (163.6382pt,344.2891pt) -- (167.6382pt,334.2891pt) -- (163.6382pt,338.2891pt) -- cycle;
\draw[pstyle3] (163.6382pt,378.2578pt) -- (163.6382pt,398.2578pt);
\draw[pstyle4] (159.6382pt,388.2578pt) -- (163.6382pt,398.2578pt) -- (167.6382pt,388.2578pt) -- (163.6382pt,392.2578pt) -- cycle;
\end{scope}

\draw[pstyle3] (74.8632pt,177.1445pt) -- (64.8632pt,177.1445pt);
\draw[pstyle3] (64.8632pt,177.1445pt) -- (64.8632pt,206.3516pt);
\draw[pstyle4] (60.8632pt,196.3516pt) -- (64.8632pt,206.3516pt) -- (68.8632pt,196.3516pt) -- (64.8632pt,200.3516pt) -- cycle;
\draw[pstyle3] (252.4132pt,177.1445pt) -- (262.4132pt,177.1445pt);
\draw[pstyle3] (262.4132pt,177.1445pt) -- (262.4132pt,206.3516pt);
\draw[pstyle4] (258.4132pt,196.3516pt) -- (262.4132pt,206.3516pt) -- (266.4132pt,196.3516pt) -- (262.4132pt,200.3516pt) -- cycle;
\draw[pstyle3] (64.8632pt,240.3203pt) -- (64.8632pt,258.3203pt);
\draw[pstyle3] (64.8632pt,258.3203pt) -- (151.6382pt,258.3203pt);
\draw[pstyle4] (141.6382pt,254.3203pt) -- (151.6382pt,258.3203pt) -- (141.6382pt,262.3203pt) -- (145.6382pt,258.3203pt) -- cycle;
\draw[pstyle3] (262.4132pt,240.3203pt) -- (262.4132pt,258.3203pt);
\draw[pstyle3] (262.4132pt,258.3203pt) -- (175.6382pt,258.3203pt);
\draw[pstyle4] (185.6382pt,254.3203pt) -- (175.6382pt,258.3203pt) -- (185.6382pt,262.3203pt) -- (181.6382pt,258.3203pt) -- cycle;
\draw[pstyle3] (163.6382pt,137.9375pt) -- (163.6382pt,157.9375pt);
\draw[pstyle4] (159.6382pt,147.9375pt) -- (163.6382pt,157.9375pt) -- (167.6382pt,147.9375pt) -- (163.6382pt,151.9375pt) -- cycle;

\begin{scope}[yshift=1.8cm, xshift=-.1cm]
\node at (81.3524pt,60pt)[below right,color=black]{generate new illness $\illness$ for $\p$};
\end{scope}

\node at (86.8632pt,157.9375pt)[below right,color=black]{$\exists$ appointment satisfying };
\begin{scope}[xshift=.0cm, yshift=.1cm]
\node at (86.8632pt,170.7422pt)[below right,color=black]{patient $\p$'s requirements?};
\end{scope}

\begin{scope}[xshift=0.6cm, yshift=-.1cm]
\node at (20pt,216.3516pt)[below right,color=black]{make appointment};
\end{scope}
\begin{scope}[xshift=0.0cm, yshift=-.1cm]
\node at (235.5465pt,216.3516pt)[below right,color=black]{walk-in};
\end{scope}
\end{tikzpicture}
	}
	\label{fig:ill}
}
	\caption{Processing of (a) arrival events $\arrivalEvent(\gp,\p)$, (b) release events $\releaseEvent(\gp, \p)$, and (c) illness events $\illnessEvent(\p)$; $\p \in \SP$ and $\gp \in \SGP$.}
	\label{fig:p1}
\end{figure*}

\paragraph{Follow-up events} are indicated by $\followUpEvent(\gp,\p,i)$.
Some families of illnesses $\familyOfIllnesses \in \SFamiliesIllnesses$ cannot be treated via a single visit.
Instead, the related illnesses $\illness\in \Illnesses$ require follow-up treatments in intervals defined by the parameter $\tf\neq \emptyset$.
Ensuring continuous follow-up treatments, patients always try to arrange a follow-up appointment immediately after the treatment of illnesses requiring follow-up consultation. 
To account for the fact that no feasible follow-up appointment might be available, \SimulationModel generates a follow-up event $\followUpEvent(\gp, \p, \illness)$ at time $\pointInTime^{\text{treat}} + \tf$  every time illness $\illness\in \Illnesses$ with $\tf\neq \emptyset$ suffered by patient $\p \in \SP$ is treated by physician $\gp\in \SGP$ at time $\pointInTime^{\text{treat}}\in \PointsInTime$.
Follow-up events serve as the patient's reminder to actively re-pursue follow-up consultation for illness $\illness$ after the duration of the follow-up interval.
Triggered by a follow-up event $\followUpEvent(\gp,\p,i)$, patient $\p$ reattempts to arrange a follow-up appointment with physician $\gp$.
Should $\gp$ once again be unable to provide a suitable appointment, $\p$ seeks follow-up consultation as a walk-in patient.
Every follow-up treatment of an illness $\illness\in \Illnesses$ invalidates all associated existing follow-up events, as the follow-up interval is reset.
Therefore, \SimulationModel deletes all existing follow-up events $\followUpEvent(\gp, \p, \illness) \in \eventQueue$ associated with illnesses $\illness\in\Illnesses$ that were treated during a visit before the new follow-up events are generated.
As a result, follow-up events only trigger if an illness has not been treated for the duration of its follow-up interval $\tf\in \PointsInTime$.

\paragraph{Release events} are indicated by $\releaseEvent(\gp,\p)$.
As illustrated in Figure~\ref{fig:rel}, release events mark the event of physician $\gp$ releasing patient $\p$ after a treatment is performed.
Whenever a new treatment begins, the sampled service time determines the time of the subsequent release event $\releaseEvent(\gp, \p)$.
All treated illnesses $\illness\in \tmpIllnesses$ without duration $(\duration=\emptyset)$ are cured through a one-time treatment and thus removed from $\tmpIllnesses$.
Subsequently, all existing follow-up events corresponding to treated illnesses are deleted and new follow-up events are generated in the previously described manner.
The successful treatment revokes existing emergency flags, i.e., we set $\emergencyFlag=0$. 
If the patient's chronic illness $\chronicIllness \in \Illnesses$ was treated, the next recurrent regular appointment $\regappointment\in \Appointments$ with physician $\fgp$ is requested at time $\pointInTime^{\text{treat}} + \nu_\chronicIllness$.
Then, patients request an \stdApp appointment $\stdappointment\in\Appointments$ with physician $\gp$ for the follow-up treatment of the persisting \act illness $\illness^* = \text{argmin}_{\illness \in \tmpIllnesses : \tf\neq \emptyset} \,\tf$ with smallest follow-up interval.
The requested appointments ensure the follow-up treatment of all illnesses suffered by patient $\p$ and will preempt the previously generated follow-up events.
Finally, physicians implement their treatment strategy to select the next patient from the waiting room if the latter is non-empty.
Otherwise, physicians remains idle until the next arrival event triggers the treatment strategy. As a result of this behavior, physicians are never intentionally idle.

\paragraph{Illness events} are indicated by $\illnessEvent(\p)$.
As illustrated in Figure~\ref{fig:ill}, they describe that patient $\p$ starts to suffer from a new \act illness.
This means that the model generates a new \act illness $\illness\in\AllTmpIllnesses$ with stochastic qualities that depend on the patient's age and health condition and adds it to  their set of illnesses $\Illnesses$.
To treat the emerged illness, patients request an appointment from their preferred physicians or, in case this does not succeed, directly visit the preferred physician as a walk-in.
As a result, each illness event generates a corresponding arrival event $\arrivalEvent(\gp,\p)$ and adds it to the queue $\eventQueue$.
Finally, each illness event generates a future illness event $\illnessEvent(\p)$ for patient $\p$ and adds it to the queue $\eventQueue$ to  mark the next point in time patient $\p$ develops an \act illness.

\paragraph{Recovery events} are indicated by $\recoveryEvent(\p,i)$. They mark the event of patient $\p$ recovering from \act illness $i\in \tmpIllnesses$.
Whenever the model generates a new \act illness $\illness\in \AllTmpIllnesses$ with $\duration\neq \emptyset$, it also generates a corresponding recovery event $\recoveryEvent(\p,\illness)$ at time $\pointInTime^{\text{ill}} + \duration$, where $\pointInTime^{\text{ill}}\in \PointsInTime$ is the \PIT illness $\illness$ is developed.
Illnesses without duration $(\duration= \emptyset)$ do not require a recovery event as they are immediately cured through their initial treatment.
A recovery event removes illness $\illness$ from $\Illnesses$ and deletes any associated follow-up event $\followUpEvent(\gp, \p, \illness)\in \eventQueue$.
If patient $\p$ does not suffer from \act illnesses following the removal of illness $\illness$, i.e., $\tmpIllnesses = \emptyset$, the model revokes existing emergency flags by setting $\emergencyFlag=0$ and assumes that $\p$ may cancel scheduled \stdApp appointments.
Such cancellations occur with the patient's age-class-specific probability $\probToCancel\in [0,1]$ and consequently delete the associated arrival event $\arrivalEvent(\gp,\p)$.
As a result, some patients keep their existing \stdApp appointment for a final debriefing. 
Should patient $\p$ be currently seeking walk-in treatment due to persisting chronic illness $\chronicIllness\in \Illnesses$, this effort is continued.
Otherwise, current walk-in attempts are canceled and the associated arrival event $\arrivalEvent(\gp, \p)$ is deleted.

\paragraph{Open- and close events} are indicated by $\openEvent(\gp)$ and $\closeEvent(\gp)$, respectively. They mark the beginning and ending (including buffer) of a session $\session \in \Sessions$ operated by physician $\gp$. They ensure that treatment strategies become aware of a session's beginning, e.g., to allow for strategies that do not treat early-arriving patients before $\sessionBegin(\session)$, and that overtime is incurred for all treatments performed beyond the anticipated buffer time of $\session$.

\subsection{Modeling Variability}
\label{sec:stochasticity}
\SimulationModel relies on stochastic values to both approximate real-world variability and control the frequency of events. This applies to aspects of illnesses as well as to patient arrivals, appointment cancellations and service times. In consequence, every simulation experiment includes multiple stochastic repetitions of the modeled time period, termed \emph{simulation runs}. When examining simulation output, we account for the resulting variability through confidence intervals.

In the following, we highlight the aspects of the model that  are probabilistic rather than deterministic and discuss how the distributions underlying the random values are parameterized.

\paragraph{Frequency of Acute Illnesses.}
The occurrence of \act illnesses in \SimulationModel is modeled via a Poission process.
Patients develop \act illnesses at a frequency that depends on their age and health condition. For patients $\p \in \SP$ of age class $\age \in \SAgeClass$ with health condition $\condition\in [0,1]$, the expected  number of  \act illnesses per year is given by the parameter $\expAnnualIllnesses(\condition)$.
The intensity (or rate) of the Poission proccess is thus $\expAnnualIllnesses(\condition)/364$ per day.
Moreover, the duration between two consecutive illness events $\illnessEvent(\p)$ for patient $\p$ can be sampled from  an exponential distribution with rate $\expAnnualIllnesses(\condition) /364$; see~\cite[chapter 2]{Daley03}.

\paragraph{Type of Acute Illnesses.}
Whenever an illness event $\illnessEvent(\p)$ occurs and patient $\p \in \SP$ falls ill, the model generates an \act illness $\illness\in \AllTmpIllnesses$ according to the patients' age class $\age \in \SAgeClass$ and health condition $\condition\in [0,1]$.
The model assumes a probabilistic link between illness family $\familyOfIllnesses\in\TemporaryFamiliesIllnesses$ and the patient's age class $\age$ that is expressed via the age class-illness distribution $\illnessAgeclassDist$; see Section~\ref{illnessAgeDist}.
To that end, any emerging \act illness of patient $\p$ is randomly assigned to an illness family according to the discrete probability distribution  $f \mapsto \illnessAgeclassDist(\age, f)$ for $f\in\TemporaryFamiliesIllnesses$.

\paragraph{Qualities of Acute Illnesses.} For any new illness  $\illness\in \AllTmpIllnesses$ of family $\familyOfIllnesses\in \SFamiliesIllnesses$ generated through \SimulationModel, its seriousness $\seriousness\in [0,1]$ depends on a triangular distribution defined on the closed interval $[0,1]$.
The distribution's mode is the health condition $\condition \in [0,1]$ of the patient $\p \in \SP$ developing illness $\illness$.
Thus, patients with a bad health condition tend to develop more serious illnesses.

The duration $\duration\in \PointsInTime$ of illness $\illness$ depends on a log-normal distribution. Given $\illness$'s family of illnesses $\familyOfIllnesses\in \SFamiliesIllnesses$, seriousness $\seriousness\in [0,1]$, and the patient's age class $\age\in \SAgeClass$, we define the age-adjusted expected duration of illness $\illness$ as  $\ageAdjustedExpDuration\coloneqq \changeInDuration \cdot \nominalDuration{\familyOfIllnesses}(\seriousness)$. Therefore, \SimulationModel samples the illness' duration $\duration$ from a log-normal distribution with sdlog $\sigma= 0.3$ and meanlog $\mu = \log(\ageAdjustedExpDuration)- \sigma^2/2$.

Patient $\p$'s willingness to wait for the initial treatment of illness $\illness$ as specified by  $\wtwIllness \in \PointsInTime$ depends on a Weibull distribution.
Given $\illness$'s family of illness $\familyOfIllnesses\in \SFamiliesIllnesses$, seriousness $\seriousness\in [0,1]$, and the developing patient's age class $\age\in \SAgeClass$, the age-adjusted expected willingness to wait of illness $\illness$ is defined as  $\ageAdjustedExpWTW\coloneqq \changeInWTW \cdot \nominalWTW{\familyOfIllnesses}(\seriousness)$.
Analogous to Wiesche et.~al~\cite{Wiesche2017}, we sample $\wtwIllness$ from a Weibull distribution with shape parameter $p=2$ and derive the scale parameter from the age adjusted expected willingness to wait as $q= \ageAdjustedExpWTW / \Gamma(1 + (1/p))$ where $\Gamma$ denotes the gamma function.
Figure~\ref{fig:dist_WTW} visualizes the resulting density functions for various choices of the age-adjusted expected willingness to wait.

\begin{figure}
	\input{./graphics/dist_WTW.tex}
	\caption{Weibull distributions of $\wtwIllness\in \PointsInTime$ for different values of patient's age adjusted expected willingness to wait $\ageAdjustedExpWTW$.}
	\label{fig:dist_WTW}
\end{figure}

\paragraph{Patient Punctuality.} Patients do not always arrive on time for their scheduled appointments  $\appointment\in \Appointments$. Instead, \SimulationModel allows for patient arrivals to vary around the scheduled time $\pointInTime_\appointment\in \PointsInTime$ of the appointment by including an arrival deviation.
As suggested by Cayirli et al.~\cite{Cayirli2006}, the arrival deviation from $\pointInTime_\appointment$ depends on a normal distribution.
We choose a mean arrival deviation of $\mu = -5$ minutes and standard deviation of $\sigma = 6$ minutes such that roughly $20\%$ of all patients are expected to arrive late for their appointments which is consistent with the observations reported in~\cite{Fetter1966PatientsWT}.

\begin{figure}
	\centering
\begin{tikzpicture}[x=1pt,y=1pt,xscale=0.55, yscale=0.47]
\definecolor{fillColor}{RGB}{255,255,255}
\begin{scope}
\definecolor{drawColor}{RGB}{0,0,0}

\node[text=drawColor,anchor=base,inner sep=0pt, outer sep=0pt, scale=  1.00] at (192.68, 15.60) {Relative arrival};

\node[text=drawColor,rotate= 90.00,anchor=base,inner sep=0pt, outer sep=0pt, scale=  1.00] at ( 10.80,186.67) {Density};
\end{scope}
\begin{scope}
\definecolor{drawColor}{RGB}{0,0,0}

\path[draw=drawColor,line width= 0.4pt,line join=round,line cap=round] ( 59.83, 61.20) -- (325.52, 61.20);

\path[draw=drawColor,line width= 0.4pt,line join=round,line cap=round] ( 59.83, 61.20) -- ( 59.83, 55.20);

\path[draw=drawColor,line width= 0.4pt,line join=round,line cap=round] (112.97, 61.20) -- (112.97, 55.20);

\path[draw=drawColor,line width= 0.4pt,line join=round,line cap=round] (166.11, 61.20) -- (166.11, 55.20);

\path[draw=drawColor,line width= 0.4pt,line join=round,line cap=round] (219.24, 61.20) -- (219.24, 55.20);

\path[draw=drawColor,line width= 0.4pt,line join=round,line cap=round] (272.38, 61.20) -- (272.38, 55.20);

\path[draw=drawColor,line width= 0.4pt,line join=round,line cap=round] (325.52, 61.20) -- (325.52, 55.20);

\node[text=drawColor,anchor=base,inner sep=0pt, outer sep=0pt, scale=  1.00] at ( 59.83, 39.60) {$0.0$};

\node[text=drawColor,anchor=base,inner sep=0pt, outer sep=0pt, scale=  1.00] at (112.97, 39.60) {$0.2$};

\node[text=drawColor,anchor=base,inner sep=0pt, outer sep=0pt, scale=  1.00] at (166.11, 39.60) {$0.4$};

\node[text=drawColor,anchor=base,inner sep=0pt, outer sep=0pt, scale=  1.00] at (219.24, 39.60) {$0.6$};

\node[text=drawColor,anchor=base,inner sep=0pt, outer sep=0pt, scale=  1.00] at (272.38, 39.60) {$0.8$};

\node[text=drawColor,anchor=base,inner sep=0pt, outer sep=0pt, scale=  1.00] at (325.52, 39.60) {$1.0$};

\path[draw=drawColor,line width= 0.4pt,line join=round,line cap=round] ( 49.20, 70.49) -- ( 49.20,302.86);

\path[draw=drawColor,line width= 0.4pt,line join=round,line cap=round] ( 49.20, 70.49) -- ( 43.20, 70.49);

\path[draw=drawColor,line width= 0.4pt,line join=round,line cap=round] ( 49.20,128.58) -- ( 43.20,128.58);

\path[draw=drawColor,line width= 0.4pt,line join=round,line cap=round] ( 49.20,186.67) -- ( 43.20,186.67);

\path[draw=drawColor,line width= 0.4pt,line join=round,line cap=round] ( 49.20,244.77) -- ( 43.20,244.77);

\path[draw=drawColor,line width= 0.4pt,line join=round,line cap=round] ( 49.20,302.86) -- ( 43.20,302.86);

\node[text=drawColor,rotate= 90.00,anchor=base,inner sep=0pt, outer sep=0pt, scale=  1.00] at ( 34.80, 70.49) {$0.0$};

\node[text=drawColor,rotate= 90.00,anchor=base,inner sep=0pt, outer sep=0pt, scale=  1.00] at ( 34.80,128.58) {$0.5$};

\node[text=drawColor,rotate= 90.00,anchor=base,inner sep=0pt, outer sep=0pt, scale=  1.00] at ( 34.80,186.67) {$1.0$};

\node[text=drawColor,rotate= 90.00,anchor=base,inner sep=0pt, outer sep=0pt, scale=  1.00] at ( 34.80,244.77) {$1.5$};

\end{scope}
\begin{scope}
\definecolor{drawColor}{RGB}{0,0,0}

\path[draw=drawColor,line width= 0.4pt,line join=round,line cap=round] ( 59.83, 70.49) rectangle ( 89.35,148.92);

\path[draw=drawColor,line width= 0.4pt,line join=round,line cap=round] ( 89.35, 70.49) rectangle (118.87,227.34);

\path[draw=drawColor,line width= 0.4pt,line join=round,line cap=round] (118.87, 70.49) rectangle (148.39,260.01);

\path[draw=drawColor,line width= 0.4pt,line join=round,line cap=round] (148.39, 70.49) rectangle (177.91,255.11);

\path[draw=drawColor,line width= 0.4pt,line join=round,line cap=round] (177.91, 70.49) rectangle (207.44,245.31);

\path[draw=drawColor,line width= 0.4pt,line join=round,line cap=round] (207.44, 70.49) rectangle (236.96,217.54);

\path[draw=drawColor,line width= 0.4pt,line join=round,line cap=round] (236.96, 70.49) rectangle (266.48,166.89);

\path[draw=drawColor,line width= 0.4pt,line join=round,line cap=round] (266.48, 70.49) rectangle (296.00, 86.83);

\path[draw=drawColor,line width= 0.4pt,line join=round,line cap=round] (296.00, 70.49) rectangle (325.52, 72.13);

\begin{scope}[xshift=1.0cm, yshift=-.1cm]
\path[draw=drawColor,line width= 0.4pt,line join=round,line cap=round] (186.42,310.15) rectangle (335.15,262.15);
\definecolor{drawColor}{RGB}{0,0,0}

\path[draw=drawColor,line width= 1.0pt,line join=round,line cap=round] (194.92,294.15) -- (212.92,294.15);
\definecolor{drawColor}{RGB}{100,149,237}

\definecolor{drawColor}{RGB}{255,106,106}

\path[draw=drawColor,line width= 1.0pt,line join=round,line cap=round] (194.92,276.15) -- (212.92,276.15);
\definecolor{drawColor}{RGB}{0,0,0}

\node[text=drawColor,anchor=base west,inner sep=0pt, outer sep=0pt] at (222.42,290.98) {Empirical rates};

\node[text=drawColor,anchor=base west,inner sep=0pt, outer sep=0pt] at (222.42,270.98) {Beta$(1.93, 2.94)$};
\end{scope}

\definecolor{drawColor}{RGB}{255,106,106}
\path[draw=drawColor,line width= 1.0pt,line join=round,line cap=round] ( 59.83, 70.49) --
	( 62.51, 87.80) --
	( 65.20,102.72) --
	( 67.88,116.47) --
	( 70.56,129.30) --
	( 73.25,141.32) --
	( 75.93,152.62) --
	( 78.61,163.26) --
	( 81.30,173.26) --
	( 83.98,182.67) --
	( 86.67,191.51) --
	( 89.35,199.80) --
	( 92.03,207.58) --
	( 94.72,214.85) --
	( 97.40,221.63) --
	(100.08,227.94) --
	(102.77,233.79) --
	(105.45,239.20) --
	(108.14,244.18) --
	(110.82,248.74) --
	(113.50,252.90) --
	(116.19,256.67) --
	(118.87,260.05) --
	(121.55,263.07) --
	(124.24,265.72) --
	(126.92,268.03) --
	(129.61,269.99) --
	(132.29,271.62) --
	(134.97,272.94) --
	(137.66,273.94) --
	(140.34,274.64) --
	(143.03,275.06) --
	(145.71,275.18) --
	(148.39,275.04) --
	(151.08,274.63) --
	(153.76,273.96) --
	(156.44,273.05) --
	(159.13,271.90) --
	(161.81,270.52) --
	(164.50,268.91) --
	(167.18,267.10) --
	(169.86,265.08) --
	(172.55,262.86) --
	(175.23,260.45) --
	(177.91,257.87) --
	(180.60,255.11) --
	(183.28,252.19) --
	(185.97,249.11) --
	(188.65,245.89) --
	(191.33,242.53) --
	(194.02,239.03) --
	(196.70,235.41) --
	(199.38,231.68) --
	(202.07,227.84) --
	(204.75,223.89) --
	(207.44,219.86) --
	(210.12,215.74) --
	(212.80,211.55) --
	(215.49,207.28) --
	(218.17,202.96) --
	(220.85,198.58) --
	(223.54,194.16) --
	(226.22,189.70) --
	(228.91,185.21) --
	(231.59,180.70) --
	(234.27,176.17) --
	(236.96,171.64) --
	(239.64,167.11) --
	(242.32,162.59) --
	(245.01,158.08) --
	(247.69,153.60) --
	(250.38,149.15) --
	(253.06,144.75) --
	(255.74,140.39) --
	(258.43,136.08) --
	(261.11,131.84) --
	(263.80,127.67) --
	(266.48,123.58) --
	(269.16,119.58) --
	(271.85,115.67) --
	(274.53,111.87) --
	(277.21,108.18) --
	(279.90,104.60) --
	(282.58,101.16) --
	(285.27, 97.85) --
	(287.95, 94.68) --
	(290.63, 91.67) --
	(293.32, 88.81) --
	(296.00, 86.13) --
	(298.68, 83.63) --
	(301.37, 81.31) --
	(304.05, 79.18) --
	(306.74, 77.27) --
	(309.42, 75.57) --
	(312.10, 74.09) --
	(314.79, 72.85) --
	(317.47, 71.85) --
	(320.15, 71.12) --
	(322.84, 70.66) --
	(325.52, 70.49);
\definecolor{drawColor}{RGB}{255,106,106}
\end{scope}
\end{tikzpicture}
	\caption{Histogram and beta distributed maximum-likelihood fit for empirical walk-in arrival rates from~\cite{Shan18}.}
	\label{fig:arr_walk}
\end{figure}
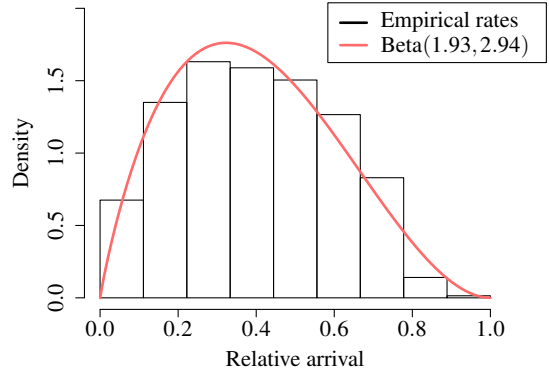

\paragraph{Walk-in Arrivals.} 
Walk-in patients have no prespecified time at which they are expected to arrive.
Instead, \SimulationModel defines for every walk-in patient an earliest arrival time $a \in \PointsInTime$ as well as a latest arrival time $b\in \PointsInTime$ which are both situational and thoroughly discussed in Section~\ref{Walk-in-Decision-Making}.
The walk-in patients' actual arrival within the given feasible arrival interval $[a,b]$ depends on a beta distribution.
Specifically, we fit a beta distribution using maximum likelihood estimation to the empirical arrival rates reported by Shan et al.~\cite{Shan18}.
As a result, we sample the arrival times of walk-in patients  from the interval $[a, b]$ of feasible arrival times according to a beta distribution with shape parameters $p = 1.93$ and $q = 2.94$; cf.~Figure~\ref{fig:arr_walk}. 

\begin{figure}
	\centering
	\input{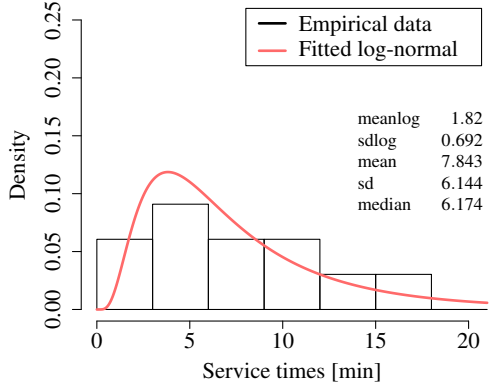}
	\caption{Histogram and log-normal maximum-likelihood fit for empirical service times of patients with appointment.}
	\label{fig:dista}
\end{figure}

\begin{figure}
	\centering
	\input{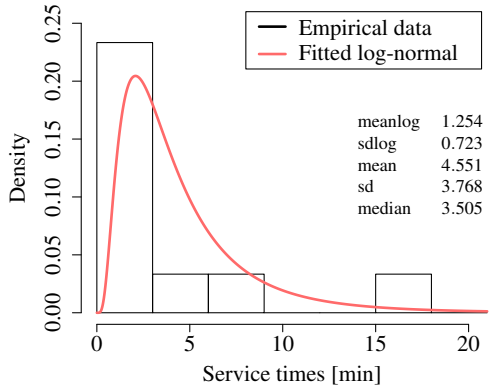}
	\caption{Histogram and log-normal maximum-likelihood fit for empirical service times of walk-in patients.}
	\label{fig:distw}
\end{figure}

\paragraph{Service Time.} \SimulationModel treats the service time per patient, i.e., the duration of treatments, as a random parameter.
To sample service times, we collected a set of $21$ service times in a local primary care practice.
As suggested in literature~\cite{Wiesche2017,POMS:POMS519}, we divide the sample into patients with and without appointment and apply a log-normal maximum likelihood fit.
Histograms of our empirical samples and the resulting distributions for walk-ins and patients with appointment are depicted in Figures~\ref{fig:dista} and~\ref{fig:distw}.
Based on the fitted distributions, we sample the service times of patients with appointment from a log-normal distribution with meanlog $\mu=1.82$ and sdlog $\sigma= 0.692$ and the service times for walk-in patients from a log-normal distribution with a meanlog $\mu = 1.254$ and sdlog $\sigma= 0.723$. As our collected data set does not incorporate transition times, we prolong all sampled service times by one minute.


\paragraph{Appointment Cancellations.} Patients that recover from all their current \act illnesses, i.e., $\tmpIllnesses = \emptyset$, cancel their existing \stdApp appointment $\stdappointment\in \Appointments$ with the age-class-specific probability $\probToCancel\in [0,1]$; compare Section~\ref{ageClass}.
As long as patients suffer from \act illnesses, they only cancel their \stdApp appointment if they require earlier treatment due to a newly emerged \act illness. All patients that have not canceled their appointment will arrive for it. As chronic illnesses are static within the model, regular appointments are never canceled.
\subsection{Emergence and Observation}
\label{emergenceObs}
\SimulationModel tracks key performance indicators from the point of view of patients, primary care physicians, and policy makers.
Thereby, it aims to illustrate the trade-offs between the stakeholders' objectives.
As these indicators emerge from agent interactions based on patients' evolving preferences and physicians' evolving strategies, they are difficult to predict in general. 

From the patients' point of view, key performance indicators include access time, travel distance, and waiting time.
\emph{Access time} measures the time a patient has to wait for an appointment, i.e, given the earliest acceptable appointment time $\pointInTime\in \PointsInTime$ and the time of the arranged appointment $\pointInTime_\appointment\in \PointsInTime$ it is defined as $\accessTime \coloneqq \pointInTime_\appointment - \pointInTime$.
The \emph{travel distance} measures the one-way distance patient $\p \in \SP$ has to travel when visiting physician $\gp \in \SGP$, i.e., $\travelDist\coloneqq \dist(\location_\p, \location_\gp)$, where $\dist(\location_1, \location_2)$ denotes the driving distance between locations $\location_1 \in \Locations$ and $\location_2 \in \Locations$ in kilometers.
The patient's \emph{waiting time} measures the time spent on-site before the actual treatment commences.
For walk-in patients, we define the waiting time for given walk-in arrival $\pointInTime^{\text{arr}}\in \PointsInTime$ and treatment commencement $\pointInTime^{\text{treat}} \in \PointsInTime$ as $\waitingTime \coloneqq \pointInTime^{\text{treat}} - \pointInTime^{\text{arr}}$.
For patients with appointment, we define the waiting time for given time of the appointment $\pointInTime_\appointment \in \PointsInTime$, patient's arrival at the practice $\pointInTime^{\text{arr}}\in \PointsInTime$,  and treatment commencement $\pointInTime^{\text{treat}} \in \PointsInTime$ as $\waitingTime \coloneqq \max \{ \pointInTime^{\text{treat}} - \max\{\pointInTime_\appointment,  \pointInTime^{\text{arr}}\} ,\, 0 \}$.
To evaluate patient's indicators, \SimulationModel keeps track of the total access time of arranging \stdApp and \regApp appointments, the total number of arranged \stdApp and \regApp appointments, the total number of attended appointments, the total number of walk-in patients, the total distance traveled by patients to access physicians, and the total waiting time for both patients with appointment and walk-ins.

From the physicians' point of view, key performance indicators include the utilization, overtime, number of treatments, and number of rejected patients with and without appointment.
A physician's \emph{utilization} describes the percentage of the available working time spent treating patients, i.e., for a session $\session\in \Sessions$ with total treatment duration $\pointInTime\in \PointsInTime$ it is defined as $\utilization\coloneqq \pointInTime / (\sessionEnd(\session) - \sessionBegin(\session) + \frac{1}{24})$.
Note that our definition of utilization clearly underestimates a physician's actual utilization as we do not account for additional tasks such as reporting, accounting, and answering phone calls that are not modeled in \SimulationModel.
\emph{Overtime} describes the physician's working time beyond the anticipated buffer, i.e., if the last patient in session $\session \in \Sessions$ is released at time $\pointInTime^{\text{rel}}\in \PointsInTime$ it is defined as $\overtime \coloneqq \max\{\pointInTime^{\text{rel}} - \sessionEnd(\session) - \frac{1}{24},\,0 \}$; see Figure~\ref{timeline}. 
To evaluate the  physician's indicators, \SimulationModel collects on physician level the total service time spent treating patients, the total number of performed treatments, the total overtime, and the total number of rejected patients with and without appointment.
The total available working time per PCP required to compute the utilization, can be derived from the opening hours $\openingHours$ and the modeled time horizon $\timeHorizon$.

\subsection{Input, Initialization, and Warm-Up}
\label{sec:init}

\SimulationModel codes a large number of values as flexible parameters.
Setting up a simulation experiment requires an input scenario to specify these parameter values.
Each simulation scenario represents a particular setting, in which a specific set of patients interacts with a specific set of physicians under specific circumstances.

As part of every simulation scenario, the \user specifies the families of illness $\SFamiliesIllnesses$, the age classes $\SAgeClass$, the age class-illness distribution $\illnessAgeclassDist$, and the set of physicians $\SGP$ with all their attributes.
The set of patients $\SP$ is only partially defined through the simulation scenario:
Each scenario specifies the total number of chronic and non-chronic patients.
Moreover, every patient's location $\location\in \Locations$, health condition $\condition\in [0,1]$, age class $\age\in \SAgeClass$, availabilities $\availabilities\colon \SessionOfWeek \to \{0,1\}$, and, for chronic patients, a chronic illness $\chronicIllness\in \AllChroIllnesses$ are given.
The remaining attributes of patients, e.g., ratings and illnesses, are initialized as described below.

At initialization, patients do not suffer from \act illnesses, i.e., $\tmpIllnesses= \emptyset$ and are not considered emergencies, i.e., $\emergencyFlag=0$.
Furthermore, all patients are initialized without scheduled appointments, i.e., $\stdappointment=\regappointment=\emptyset$.
The consideration set of physicians $\ScGP \subseteq \SGP$ per patient $\p \in \SP$ is determined according to Algorithm~\ref{consideredPCP} where $\text{rand}(x)$ for $x> 0$ denotes a uniformly distributed float from the half-closed interval $[0, x)$.
As a result, each patient considers all physicians within a $\SI{15}{\si{\km}}$ driving radius.
Physicians outside this radius are considered with a $\SI{5}{\percent}$ chance as some patients may choose their physician according to criteria other than proximity to their home, e.g., for historical reasons or personal recommendations. 

 \begin{algorithm}[t]
	\setlength{\abovedisplayskip}{0pt}%
	\setlength{\belowdisplayskip}{0pt}%
	\setlength{\abovedisplayshortskip}{0pt}%
	\setlength{\belowdisplayshortskip}{0pt}%
	\caption{\small Determine patient's considered PCPs.}
	\begin{algorithmic}[1]
		\REQUIRE{patient $\p\in \SP$, set of primary care physicians $\SGP$}
		\STATE set $\ScGP=\emptyset$
		\FOR {$\gp \in \SGP$}
		\IF{ $\dist(\location_\p, \location_\gp)< \SI{15}{\km}$}
		\STATE add $\gp$ to $\ScGP$
		\ELSE{
			\IF{rand$(20)<1$} 
				\STATE add $\gp$ to $\ScGP$
			\ENDIF
		}
		\ENDIF
		\ENDFOR		
		\STATE \textbf{return} $\ScGP$
	\end{algorithmic}
	\label{consideredPCP}
\end{algorithm}

To initialize patients' appointment ratings $\rApp$ and walk-in ratings $\rWalk$ for every considered physician $\gp \in \ScGP$ and weekly session $[\session] \in \SessionOfWeek$, we denote the number of matches between the physician's opening hours and patient $\p$'s availabilities by $\matches(\p, \gp)\coloneqq |\{[\session]\in \SessionOfWeek: \availabilities([\session])\land \openingHours([\session])\neq \emptyset\}|$ and the maximal shortest access distance by $\dist^{\max}\coloneqq \max_{\p\in \SP} \, \min_{\gp \in \SGP} \,\dist(\location_\p, \location_\gp)$. The model then initializes appointment ratings as
\begin{align*}
\rApp = \begin{cases}
\parbox{.46\columnwidth} {$3\, \matches(\p,\gp) - \dist(\location_\p,\location_\gp)\\
+ \text{rand}(2 \,\dist^{\max}) + 100$ }   &\text{if } \matches(\p,\gp) {>} 0\\
0   &\text{else.} 
\end{cases}
\end{align*}
Walk-in ratings $\rWalk$ are session specific as immediate care requires physicians to be in service. Thus, sessions in which a physician is closed are not feasible for walk-in visits which is encoded by an empty rating.
The model initializes walk-in ratings as
\begin{align*}
\rWalk = \begin{cases}
\parbox{.38\columnwidth} {$\text{rand}(\dist^{\max}) \\\; - \dist(\location_\p,\location_\gp) +100$} & \text{if } \openingHours([\session])\neq \emptyset\\
 \emptyset \; &\text{else.}
\end{cases}
\end{align*}

From the initialized ratings, \SimulationModel subsequently determines the family physician for chronic patients as the physician from the consideration set that has the highest appointment rating, i.e., $\fgp = \text{argmax}_{\gp\in\ScGP}\,\rApp$ which completes the setup of all simulation entities.

At this point in the initialization process, the global event queue $\eventQueue$ is still empty and therefore running a simulation experiment would result in no agent actions. 
To make physicians take up their work, the model generates open- and close events $\openEvent(\gp)$ and $\closeEvent(\gp)$ for every session operated by physician $\gp \in \SGP$  and adds these to $\eventQueue$.
To start the process of patients continuously developing \act illnesses, the model generates an initial illness event $\illnessEvent(\p)$ for every patient $\p \in \SP$ and adds it to $\eventQueue$.
Finally, to start the regular treatments of chronic illnesses $\chronicIllness\in \chroIllnesses$, an initial follow-up event $\followUpEvent(\fgp, \p, \chronicIllness)$ is generated at a randomly chosen \PIT within $\chronicIllness$'s follow-up interval $\nu_\chronicIllness\in \PointsInTime$ according to a uniform distribution and subsequently added to $\eventQueue$. 

As discussed previously, several aspects of the simulation model rely on emergent values that require a ``warm-up'' period before producing meaningful results for decision support.
Therefore, we precede every simulation experiment with a warm-up during which patients develop \act illnesses, physicians fill their appointment books, and patients adjust their ratings.
The duration of the warm-up and the length of the modeled time horizon are both variable and specified by the \user through the input scenario.
Note, that this does not correspond to solely analyzing a steady state, as the agents' emergent interactions can result in the development of meaningful trends in the data.
\subsection{Submodels}
\label{submodels}

We consider different aspects of \SimulationModel that rely on an internal logic as submodels.
One of the most basic submodels describes the logic of distances and travel times.
More complex examples include the logic underlying patients' behavior when requesting appointments and visiting practices as walk-ins, as well as the physician's strategies which are submodels by themselves.
As \SimulationModel allows for modular PCP strategies, we exemplify each strategy through the specific approach that is used in the case study.
Further submodels describe the consequences of rejecting patients, service time reductions, patients' rating adjustments, patients' choice of their family physician, and treatment effects.

\subsubsection{Distances and Travel Times}
\SimulationModel does not feature a road network to compute travel distances and travel times.
Instead, it approximates the driving distance $\dist \colon \Locations \times \Locations \to \mathbb{R}$ between two locations in kilometers using the great circle distance computed through the haversine formula with a detour factor of $\num{1.417}$ as determined by Boscoe et al.~\cite{doi:10.1080/00330124.2011.583586}.
These authors also point out, that driving distances provide good approximations for travel times in minutes, i.e., we compute travel times by assuming a constant driving speed of $\SI{60}{\si{\km} \per \si{\hour}}$.
As a result we define the travel time $\travelTime\colon \Locations \times \Locations \to \PointsInTime$ as $\travelTime(\location_1, \location_2)\coloneqq \frac{\dist(\location_1, \location_2)}{60\cdot 24}$.
\subsubsection{Patients Requesting Appointments}
\label{submodel:appointment_requests}
 Patient agents $\p \in \SP$ request an appointment with a physician $\gp \in \SGP$, by specifying the earliest acceptable appointment time $\pointInTime\in\PointsInTime$ and their willingness to wait for this appointment $\wtw\in \PointsInTime$.
 As a result, newly-arranged appointments are feasible, if and only if they are scheduled in the time interval $[\pointInTime, \pointInTime+\wtw]$.
 
 The earliest acceptable appointment time $\pointInTime \in \PointsInTime$ depends on the request.
 The initial treatment of \act illnesses $\illness\in\tmpIllnesses$ is urgent, so that patients seek to schedule an appointment as soon as possible.
 Thus, for these initial treatments, the earliest acceptable appointment time is the time of the request $\pointInTime^{\text{req}}\in \PointsInTime$ plus a $\num{30}$ minute buffer (corresponding to $\frac{1}{48}$ in decimal time) plus the direct travel time, i.e., $\pointInTime= \pointInTime^{\text{req}} + \frac{1}{48} + \travelTime(\location_\p, \location_\gp)$.
 Follow-up treatments are planned at regular intervals specified by the parameter $\tf\in \PointsInTime$.
 Patients request follow-up appointments in two ways:
 First, at the very beginning of the follow-up interval as every patient requests a follow-up appointment directly after the treatment of illnesses that require aftercare.
 Second, at the very end of the follow-up interval (triggered by a follow-up event) in case no feasible appointment was available at the time of the previous treatment.
 In the latter case, the request is urgent and therefore the earliest acceptable appointment time is defined as above, i.e., $\pointInTime= \pointInTime^{\text{req}} + \frac{1}{48} + \travelTime(\location_\p, \location_\gp)$.
If the follow-up appointment is requested at the beginning of the follow-up interval, the next follow-up appointment for illness $\illness \in \Illnesses$ should be scheduled after the follow-up interval has passed, i.e., we set $\pointInTime= \pointInTime^{\text{req}} + \tf$.
 
 The willingness to wait $\wtw\in \PointsInTime$ defines the maximum acceptable waiting period between the earliest appointment time and the actual time of the appointment.
 As a result, it serves as an upper bound to the patient's access time defined in Section~\ref{emergenceObs}.
 Patients' willingness to wait for the initial treatment of \act illness $\illness \in \tmpIllnesses$ is illness specific and given by $\wtw= \wtwIllness$. 
 Analogously, the maximum duration chronic patients are willing to wait for their regular appointment depends on their chronic illness $\chronicIllness\in \chroIllnesses$, i.e., $\wtw= \omega_\chronicIllness$.
 If patients request a follow-up appointment for \act illness $\illness \in \tmpIllnesses$, the willingness to wait is proportional to the length of the follow-up interval $\tf\in \PointsInTime$.
 To ensure that the follow-up interval is not exceeded by an excessive time span, the willingness to wait for follow-up appointments regarding $\illness\in\tmpIllnesses$ is $\wtw= \frac{\tf}{5} + 1$. 
 Finally, emergency patients who were denied treatment are exceptionally impatient and their willingness to wait is $\wtw=0$.

Algorithm~\ref{makeapp} describes how a patient requests an initial appointment for a newly emerged \act illness.
First, patients check whether they have a pre-existing appointment within the  acceptable time frame.
From the patients' point of view, pre-existing appointments are particularly convenient as they require no further actions.
Therefore, patients accept pre-existing appointments as feasible, even if they exceed their willingness to wait by up to $12$ hours (or $\frac{1}{2}$ in decimal time); see lines $1 - 2$.
If the patient's existing appointments are infeasible for the newly emerged illness, the existing \stdApp appointment is canceled to make room for a new, earlier, \stdApp appointment (compare line $4$ and $5$).

Patients $\p \in \SP$ request appointments from the two currently  highest rated physicians $\gp_1, \gp_2 \in \ScGP$ in their consideration set (compare line $7$).
Physicians  $\gp_1$ and  $\gp_2$ are queried in order of their rating, i.e., patients first request an appointment with the higher rated PCP $\gp_1$ and only resort to $\gp_2$ if the request is unsuccessful.
When a physician cannot offer a fitting slot, patients reduce their rating for the respective PCP.

When their willingness to wait is at least $3$ days (compare line $10$), patients only accept appointments that fit their personal availability $\availabilities\colon \SessionOfWeek \to \{0,1\}$ (cf. Section~\ref{epp}).
In case $\wtw \leq 3$, the request is so urgent that patients are always available.

If neither $\gp_1$ nor $\gp_2$ offer a feasible slot, the search for a feasible appointment is deemed unsuccessful and patients resort to a walk-in visit. 

 \begin{algorithm}[t]
 	\setlength{\abovedisplayskip}{0pt}%
 	\setlength{\belowdisplayskip}{0pt}%
 	\setlength{\abovedisplayshortskip}{0pt}%
 	\setlength{\belowdisplayshortskip}{0pt}%
 	\caption{ \small Arranging appointment for \act illness.}
 	\begin{algorithmic}[1]
 		\label{alg1}
 		\REQUIRE{patient $\p\in \SP$, willingness to wait $\omega \in \PointsInTime$, earliest appointment time $\pointInTime \in \PointsInTime$}
 		\IF{$\p$ has \stdApp or \regApp appointment before \\ time $\pointInTime +\omega+ \frac{1}{2}$}
	 		\STATE return true
 		\ELSE{
 			\STATE cancel \stdApp appointment
 			\STATE delete associated arrival event $\arrivalEvent(\gp,\p)$}
 		\ENDIF
 		\STATE determine preferred physicians $\gp_1,\gp_2 \in \ScGP$ such that $r^{\text{app}}_\p(\gp_1) \geq r^{\text{app}}_\p(\gp_2) \geq \rApp \;\forall \gp \in \ScGP \setminus \{\gp_1, \gp_2\}$.
 		\FOR {$j=1, 2$}
 		\STATE query $\gp_j$ for an appointment
 		\IF{physician $\gp_i$ offers appointment within $[\pointInTime, \pointInTime+\omega]$ $\land$ (satisfying $\p$'s availabilities $\availabilities \lor \wtw \leq 3$)}
 		\STATE accept appointment \hfill \# adapt $\ratingApppointment(\gp_j)$
 		\STATE add $\arrivalEvent(\gp_j,\p)$  to $\eventQueue$
 		\STATE return true
 		\ELSE{
 			\STATE refuse appointment \hfill \# adapt $\ratingApppointment(\gp_j)$}
 		\STATE continue
 		\ENDIF
 		\ENDFOR		
 		\STATE return false
 	\end{algorithmic}
 	\label{makeapp}
 \end{algorithm}

When patients request follow-up appointments, they mostly follow the steps outlined in Algorithm~\ref{makeapp}.
The main difference concerns the inquiry process (cf.~line $7$ and $8$), as new follow-up appointments are exclusively arranged with the physician that performed the previous treatment.
Only pre-existing appointments can be used for follow-up visits although they are not with the physician that performed the previous treatment; compare line~1.
If the follow-up appointment request is made at the end of the follow-up interval triggered by a follow-up event, a failure initiates a walk-in attempt to ensure the patient's aftercare.
If the follow-up appointment is requested immediately after treatment at the beginning of the follow-up interval, a failure does not lead to a walk-in attempt as the corresponding follow-up event will eventually lead to a reattempt at arranging a follow-up appointment.

Chronic patients' \regApp appointments are essentially follow-up appointments and thus arranged according to the same logic.
The only difference concerns the evaluation of pre-existing appointments:
As \regApp appointments are exclusively arranged with the patient's family physician $\fgp\in \ScGP$, pre-existing \stdApp appointments are only perceived as feasible if they are with the family physician $\fgp$ (cf.~line $1$ and $2$).
Infeasible pre-existing \stdApp appointments are not canceled but instead an additional \regApp appointment is arranged with the family physician $\fgp$ (cf.~line $4$ and $5$). Only if the newly arranged \regApp appointment is before or at most $12$ hours after an existing \stdApp appointment, i.e., $\pointInTime_{\regappointment} \leq \pointInTime_{\stdappointment} + \frac{1}{2}$, the latter is canceled as all acute illnesses will be treated at the \regApp appointment.

 \subsubsection{Walk-in Decision Making}
 \label{Walk-in-Decision-Making}
 Within \SimulationModel, all walk-in visits are preceded by an unsuccessful appointment request.
 As walk-in visits are per se urgent, the earliest possible time $\pointInTime\in \PointsInTime$ for a walk-in visit of patient $\p \in \SP$ at physician $\gp \in \ScGP$ is, analogous to Section~\ref{submodel:appointment_requests}, defined as the current time $\pointInTime^{\text{curr}}\in \PointsInTime$ plus a 30 minute buffer plus the direct travel time, i.e., $\pointInTime= \pointInTime^{\text{curr}} + \frac{1}{48} + \travelTime(\location_\p, \location_\gp)$.
 The patients' willing to wait for the walk-in visit is the willingness to wait $\wtw\in \PointsInTime$ of the preceding appointment request.
 As a result, the patient's walk-in visit takes place in the time interval $[\pointInTime, \pointInTime+\wtw]$, unless this is impossible due to the physicians' opening hours.
 
As part of the walk-in decision making, patients decide on a physician $\gp^* \in \ScGP$ and session $\session^* \in \Sessions$ for their walk-in visit. 
To that end, \SimulationModel computes all physician-session combinations $\feasibleWalkInVisits \subseteq \ScGP \times \Lambda$ that fall into the interval $[\pointInTime, \pointInTime+\wtw]$ and thus can be targeted for a walk-in visit.
If $\feasibleWalkInVisits= \emptyset$, the model gradually increases the willingness to wait $\wtw$ until $\feasibleWalkInVisits \neq \emptyset$.

Patients select the physician-session combination $(\gp^*, \session^*) \in \feasibleWalkInVisits$ targeted for their walk-in visit on the basis of their walk-in ratings $\ratingWalkIn$ via
\begin{align*}
(\gp^*, \session^*)=	\text{argmax}_{(\gp,\session) \in \feasibleWalkInVisits} \, 0.95^{\timeBetween(\session)} \cdot \rWalk,
\end{align*}
where $\timeBetween(\session) \coloneqq \sessionEnd(\session) - \pointInTime$ denotes the time difference between the earliest possible walk-in time $\pointInTime\in \PointsInTime$ and the end of session $\session \in \Lambda$.
This takes into account that walk-in patients urgently want to visit a physician by discounting the ratings based on the approximate access time $\timeBetween(\session)\geq 0$.
Note that this discounting model yields undesired results if we allow for negative ratings, motivating the models limitation to non-negative ratings. 

Given the targeted physician-session combination $(\gp^*, \session^*)\in \feasibleWalkInVisits$ for the walk-in visit, the time interval during which the actual visit at $\gp^*$ may take place is defined as follows:
The earliest time for walk-in patients to arrive in session $\session^* \in \Sessions$ is 15 minutes before its beginning $\sessionBegin(\session^*)\in \PointsInTime$, but obviously not before the earliest possible arrival $\pointInTime \in \PointsInTime$.
The latest possible arrival in session $\session^* \in \Sessions$ is its ending  $\sessionEnd(\session^*) \in \PointsInTime$, but not after the latest possible arrival $\pointInTime + \wtw$. The resulting time interval for the patient's walk-in arrival is
\begin{align*}
\textstyle
\left[\max( \sessionBegin(\session^*) - \frac{1}{96},\; \pointInTime),\; 
\min( \sessionEnd(\session^*), \pointInTime+\wtw )\right].
\end{align*}
The patient's actual arrival within the feasible time interval is stochastic and sampled according to the distribution specified in Section~\ref{sec:stochasticity}.

As long as patients actively pursue walk-in treatment, they never arrange new appointments.
That is if a walk-in patient develops a new acute illness or seeks an immediate follow-up appointment triggered by a follow-up event, their need for medical attention is met through the ongoing walk-in visit.

\subsubsection{Service Time Reduction}
\label{sub:str}
Physicians' treatment strategies let them reduce service times to prevent congestion and minimize overtime.
Within the model, the service time reduction operationalizes via a multiplicative factor $\speedup\in [0,1]$.
Thus, a treatment with an original service time of $10$ minutes (sampled from the log-normal distribution described in Section~\ref{sec:stochasticity}) takes only $8$ minutes when performed by a physician with current consultation speed $\speedup=0.8$.
When there is no effort to reduce service times, i.e., $\speedup = 1$, the actual services time coincide with the sampled original service times.

 \subsubsection{Consequences from Rejection of Patients}

 Whenever a patient visits a physician either with an appointment or as a walk-in, the physician's admission strategy determines whether the patient is admitted or rejected.
 Following a rejection, patients reduce their personal ratings $\ratingApppointment$ or $\ratingWalkIn$ depending on whether they arrived for an appointment or as a  walk-in.
 As rejected patients have been denied treatment, they are subsequently flagged as emergencies, i.e, $\emergencyFlag=1$.
 In order to be treated, rejected patients then start a walk-in attempt with reduced willingness to wait $\wtw=0$, i.e., they visit their preferred physician according to the updated walk-in preferences $\ratingWalkIn$ in the earliest possible session; compare Section~\ref{Walk-in-Decision-Making}.
A patient's emergency flag is only revoked after the next successful treatment or if the patient fully recovers from all \act illnesses.

\subsubsection{Rating Adjustments}

Throughout the simulation, patients adjust their ratings of physicians according to their experiences via additive factors.
To that end, patients increase ratings based on positive experiences and decrease ratings following negative experiences.
Thereby, patients with appointment update their appointment ratings $\ratingApppointment$ while walk-in patients update their walk-in ratings $\ratingWalkIn$.
Table~\ref{ratingUpdates} lists all events that trigger a rating adjustment.

\begin{table}
	 \begin{threeparttable}
	\centering
	\caption{Adaptation of patient ratings $\ratingApppointment$ and $\ratingWalkIn$.}
	\label{ratingUpdates}
	\begin{tabular*}{\linewidth}{@{}lr@{}}
		\toprule
		Positive Event                                               & Adjustment \\ \midrule
		waiting time $< \SI{7}{\minute}$ 															& $+5$			\\
		successful arrangement of appointment               & $+4$          \\
		successful treatment as walk-in                     & $+3  \speedup$          \\
		successful treatment with appointment               & $+2  \speedup$          \\
		\midrule
		Negative Event & Adjustment \\
		\midrule
		waiting time $> \SI{30}{\minute}$							    & $-10$		\\
		no appointment within willingness available         & $-\wtw$         \\
		rejected as walk-in                                 & $-10$       \\
		rejected with appointment                           & $-20$        \\		
		\bottomrule
	\end{tabular*}
    \begin{tablenotes}
    	\vspace*{.15cm}
	\small
	\item Parameter $\wtw\in\PointsInTime$ describes patient's willingness to wait and $\speedup\in [0,1]$ the physician's consultation speed.
\end{tablenotes}
 \end{threeparttable}
\end{table}

In \SimulationModel, only the effect of a failed appointment request and the effect of a successful treatment are parameterized.
All other event effects are hard-coded to represent the following intuition about patient perceptions:
Unanticipated events cause a stronger adjustment, while anticipated events only cause a slight adjustment.
For example, visiting a physician with an appointment and not being admitted is considered highly unlikely and therefore highly penalized. Furthermore, patients react more strongly to negative experiences, reflecting the so-called \emph{negativity bias} \cite{ae890d4ccb214c59ad442d152faa1bad}.

When a physician fails to offer a fitting appointment, the negative adjustment depends on the patient's associated willingness to wait $\wtw\in \PointsInTime$.
As $\wtw\geq 0$, the adjustment $-\wtw$ is always non-positive.
When the willingness to wait is high, the expectation of receiving a fitting slot is also high, so that the resulting disappointment leads to a stronger negative adjustment. 

When physicians reduce their service time as part of their treatment strategy, patients feel rushed. Therefore, the model scales the positive adjustment following a successful treatment as dependent on the physician's current consultation speed. For example, at a consultation speed of $\speedup=0.5$ a successful treatment with appointment increases $\ratingApppointment$ only by a value of $0.5 \cdot 2 = 1$.

To ensure the desired behavior of discounting ratings as described in Section~\ref{Walk-in-Decision-Making}, we bound all ratings from below by zero, i.e., we enforce
$\rApp\geq 0$ and  $\rWalk \geq 0$ for all $\p\in\SP$, $\gp \in \SGP$,  $[\session]\in \SessionOfWeek$.
As a result, negative adjustments have no effect on physicians with a rating of zero.

\subsubsection{Family Physician Adjustments}
Every time chronic patients adjust their appointment ratings $\ratingApppointment(\gp)$ for any $\gp \in \ScGP$, they simultaneously reevaluate their family physician $\fgp$ according to Algorithm~\ref{reeval_family_pcp}.
Thereby, chronic patients change their family physician when another physician from the consideration set has a rating that is at least $\SI{20}{\percent}$ higher than the current family physician's rating. 

 \begin{algorithm}[t]
	\setlength{\abovedisplayskip}{0pt}%
	\setlength{\belowdisplayskip}{0pt}%
	\setlength{\abovedisplayshortskip}{0pt}%
	\setlength{\belowdisplayshortskip}{0pt}%
	\caption{\small Reevaluation of family physician.}
	\begin{algorithmic}[1]
		\REQUIRE{chronic patient $\p\in \SP$}
		\STATE let $\gp^* =  \text{argmax}_{\gp\in\ScGP}\,\rApp$
		\IF{$\ratingApppointment(\gp^*) \geq 1.2 \cdot \ratingApppointment(\fgp)$ }
		\STATE $\fgp = \gp^*$
		\ENDIF	
	\end{algorithmic}
	\label{reeval_family_pcp}
\end{algorithm}

\subsubsection{Treatment Effects}

Physicians treat all of a patient's current \act illnesses $\illness \in \tmpIllnesses$ during the same appointment.
As a result, all scheduled follow-up events $\followUpEvent(\gp, \p, \illness)$ for $\illness \in \tmpIllnesses$ are deleted.
Moreover, all illnesses $\illness\in\tmpIllnesses$ that  require only a single treatment, as indicated by $\duration=\emptyset$, are cured and thus removed from $\tmpIllnesses$.
Finally, new follow-up events are scheduled for all illnesses $\illness \in \tmpIllnesses$ that still require follow-up consultation as indicated by a positive follow-up interval $\tf > 0$.

Chronic illnesses are only treated during the recurrent regular appointments or during walk-in visits triggered by the unavailability of a feasible regular appointment. If $\chronicIllness\in \chroIllnesses$ is treated, any existing follow-up event $\followUpEvent(\gp, \p, \chronicIllness)\in \eventQueue$ is deleted and replaced by a new, updated one.

Finally, the successful treatment revokes any emergency flag the patient may have.

\subsubsection{PCP Strategies}
\label{pcp_strategies}
PCP Strategies determine physicians' decision making through exchangeable submodels, that are defined as part of every scenario.
For illustration, we describe the exemplary strategies implemented and evaluated in our case study.

Appointment scheduling strategy:
\textit{Individual-block/ Fixed-interval} (IBFI) evenly spaces out appointments throughout each session; see \cite{POMS:POMS519,KLASSEN199683}.
To that end, it divides the opening hours of each session in a $\num{140}$ day rolling horizon into slots of $15$ minutes length.
Each slot can accommodate one appointment and slots are offered to patients on a first-come-first-served (FCFS) basis.
Thus, no appointments are withheld and every patient is offered the earliest feasible appointment at the time of inquiry.  

Treatment strategy:
\textit{Priority first come, first served} (PFCFS) is popularly used in studies of health systems~\cite{POMS:POMS519}.
In PFCFS, patients with appointment are prioritized over walk-ins and within their respective groups, patients are served in order of their arrival, i.e., FCFS; compare~\cite{doi:10.1287/opre.21.5.1030,doi:10.1080/02664768500000017}.
Patients that arrive before the beginning $\sessionBegin(\session) \in \PointsInTime$ of session $\session\in \Sessions$ have to wait and the physician does not start treatments until the session has officially begun. 
The PCP's standard consultation speed in PFCFS is $\speedup=1.0$, which is adjusted to $\speedup=0.8$ whenever more than $3$ patients await treatment; compare Section~\ref{sub:str}. 

Admission strategy:
\textit{Priority threshold} (PT) admits patients up to a certain utilization threshold; compare \cite{doi:10.1287/mnsc.2014.2057,QU2015303}. PT differentiates between appointment, walk-in, and emergency patients:
Emergency patients are always admitted, i.e., they have an infinite admission threshold.
Patients with an appointment in session $\session\in \Sessions$ are admitted as long as their time of arrival $\pointInTime^{\text{arr}}\in \PointsInTime$ is before the end of the session's buffer, i.e.,  $\pointInTime^{\text{arr}} \leq \sessionEnd(\session) + \frac{1}{24}$.
Appointment patients that arrive after the session's anticipated buffer are rejected.
For the admittance of walk-in patients, physicians predict their remaining workload by multiplying an expected service time with the number of currently waiting patients and upcoming scheduled appointments.
If this estimated workload is lower than the remaining duration of the current session including buffer, walk-in patients are admitted, otherwise rejected.
The expected service time is initialized to $7$ minutes and adjusted at the end of each session as follows: On the one hand, the expected service time is increased by one minute if three or more patients are awaiting treatment at the end of the anticipated buffer. On the other hand, the expected service time is reduced by $20$ seconds if the physician is idle at the end of the anticipated buffer although walk-ins were previously rejected.

\subsection{Structural Validation and Verification}
\label{sec:structuralValiVeri}
In \SimulationModel, validation and verification were carried out according to the best practices documented in the literature~\cite{KLEIJNEN1995145,Sargent2013}.
To ensure that our model implementation is correct (verification), we followed established good programming practices.
That is, we used object oriented programming to write modular code.
\SimulationModel is implemented in Java 8.
All random distributions are implemented using the Apache Common Math library~\cite{math2016apache}.
Each module is individually verified through unit testing.
Assertions ensure that variables remain within their specifications at runtime.
As an additional mean to detect undesired model behavior, \SimulationModel can trace the entire simulation process.
Traces are specialized logs that contain all information about the model's execution.
In \SimulationModel, traces are textual and comprehensible to  \users.
They enable the tracking of agents through the overall model and contain all the information that would be required to animate the model.
Analyzing traces and input output relationships, we performed dynamics tests for multiple simulation scenarios of various sizes with different system set-ups.

To ensure that our conceptional model serves as an adequate representation of real primary care systems (validation), we took several measures.
With regard to face validity, we presented the conceptual model to physicians and decision makers from health insurers as well as public authorities.
Furthermore, \SimulationModel builds on  data from the literature as well as empirical data collected on-site.
Moreover, we visited a primary care practice and interviewed staff to capture and understand the daily processes and routines of PCPs.
For the specific scenarios featured in the case study, we validated the simulation output with available empirical data.
Details on this historical validation can be found in the baseline analysis of the following case study.
\section{Case Study}
\label{sec:computationalExamples}
To demonstrate the potential of \SimulationModel, we present a case study evaluating the effects of changes in the population of a primary care system.
Specifically, we create a baseline scenario representing a real-world primary care system in the district of Aachen and investigate two possible changes in the primary care system's population from the status quo:
On the one hand, a decline in the number of PCPs as a result of a decreasing interest in opening a primary care practice in rural areas; see~\cite{MONITOR14}. 
On the other hand, an aging of the population causing a shift in the quality and intensity of illnesses and the resulting health care requirements.   
By considering both of these changes individually and in combination, we create three ``what-if'' scenarios that we compare to the baseline scenario.

Each scenario models a time period of one year preceded by warm-up period.
As \SimulationModel relies on stochastic values, every simulation experiment includes $20$ independent runs.
Section~\ref{sec:InstanceGeneration} details how the baseline scenario is derived from empirical data.
Section~\ref{baseline_analysis} documents the analysis and validation of the baseline scenario.
Sections~\ref{s1}, \ref{s2}, and \ref{s3} describe how the considered changes in the three ``what-if'' scenarios are implemented in \SimulationModel and subsequently benchmark these against the baseline scenario.
\subsection{Baseline Scenario}
\label{sec:InstanceGeneration}
The real-world primary care system that serves as the template for our study comprises three predominantly rural municipalities (Roetgen, Simmerath, and Monschau) in western Germany with a total population of approximately $\num{35000}$ inhabitants and $\num{20}$ primary care physicians.
In order to capture a real-world primary care system in the form of a simulation scenario, empirical data is required.
Most of this data is specific to a primary care system or its country of origin such that data collection has to be carried out for each system individually.
For the considered primary care system, empirical data concerning the physicians' distribution and opening hours was provided by the responsible department of public health or obtained from the responsible association of statutory health insurance physicians~\cite{KVN}.
The distribution of patients and their demographic composition is available from the national census~\cite{zensus2011} and official population projections by the federal state~\cite{itnrw}.
The distribution of illnesses and their characteristics can be estimated from publications of health insurances and federal government agencies~\cite{gekreport,GEDA12}.
All unavailable data was either empirically collected in a primary care practice or, where this was not possible, inferred.

In the following, we discuss how the available empirical data translates into a simulation scenario.
To that end, we detail the input parameter choices, i.e., the modeled physicians, patients, age classes, families of illnesses, and age class-illness distributions.

\subsubsection{Primary Care Physicians}
The population of primary care physicians $\SGP$ in our baseline scenario aims to model the actual primary care physicians in the considered primary care system.
According to data provided by the Aachen department of public health in $2017$, there are $20$ primary care physicians with health insurance accreditation in the three municipalities under consideration.
The physicians' exact locations are specified as part of the provided dataset (cf.~Figure~\ref{fig:data2}) and the physicians opening hours were obtained from the  Association of Statutory Health Insurance Physicians Nordrhein~\cite{KVN}.
Concerning the employed strategies, all physicians $\gp \in \SGP$ apply the individual-block/fixed-interval appointment scheduling strategy (IBFI), priority first come, first served treatment strategy (PFCFS), and priority threshold admission strategy (PT); cf.~Section~\ref{pcp_strategies}.

\subsubsection{Patients}
\label{study_patients}
\begin{figure}
	\centering
	\begin{tikzpicture}[scale=1.115]
	\node at (0,0) {	\includegraphics[width=.95\linewidth]{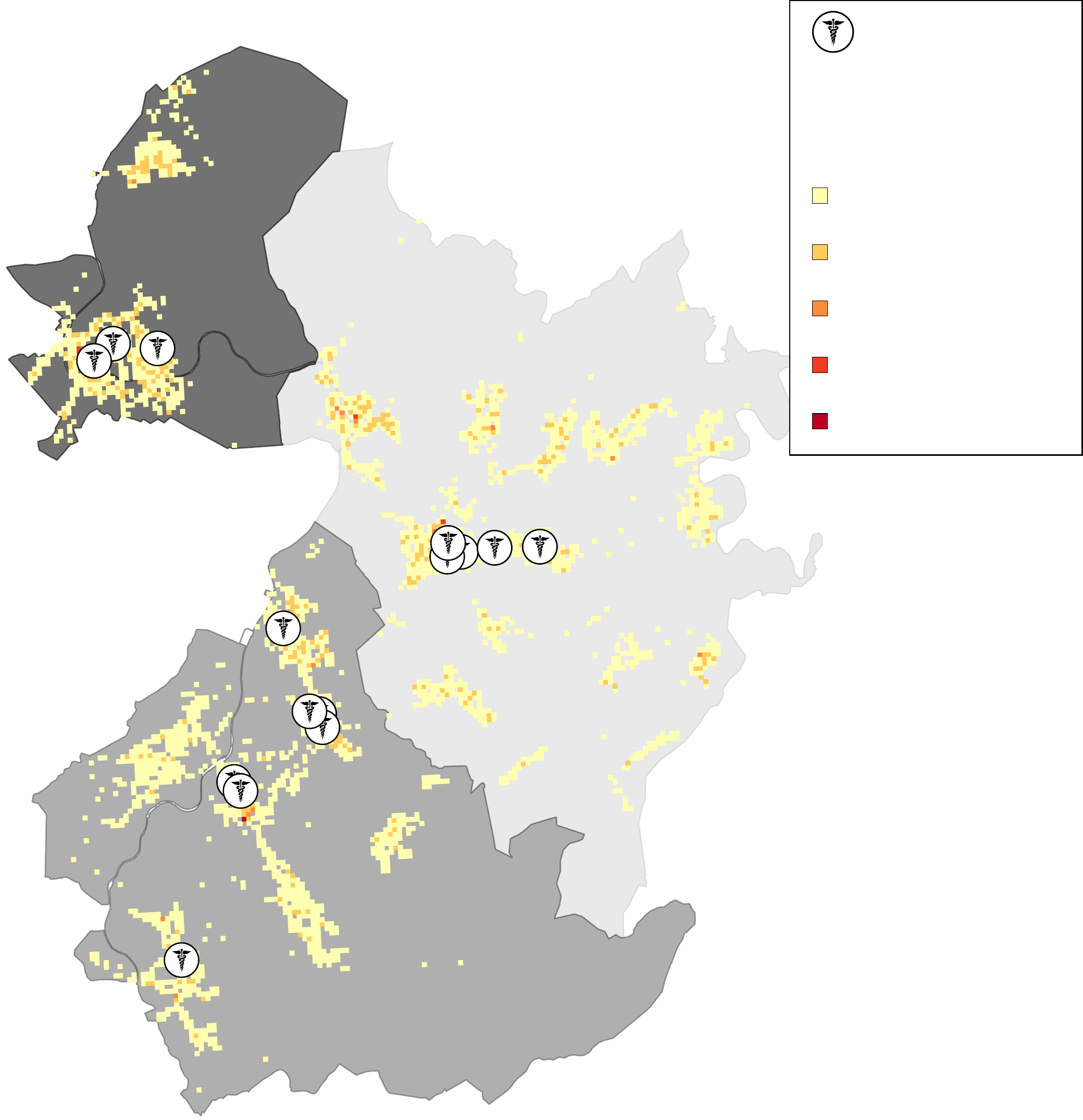}};
	\node[right] at (1.8,2.98) {\footnotesize PCPs};
	\node[right] at (1.5, 2.48) {\footnotesize Population};	
	\foreach \x [count=\y] in {$0 - 20$, $ 20 - 40$,  $ 40 - 60$, $  60 - 80$, $  80 - 100$ }
	\node[right] at (1.8, 2.38-\y*0.32) {{ \scriptsize \x}};
	\end{tikzpicture}

	\caption{Locations of PCPs with health insurance accreditation and population cells reported by the 2011 census~\cite{zensus2011} }
	\label{fig:data2}
\end{figure}
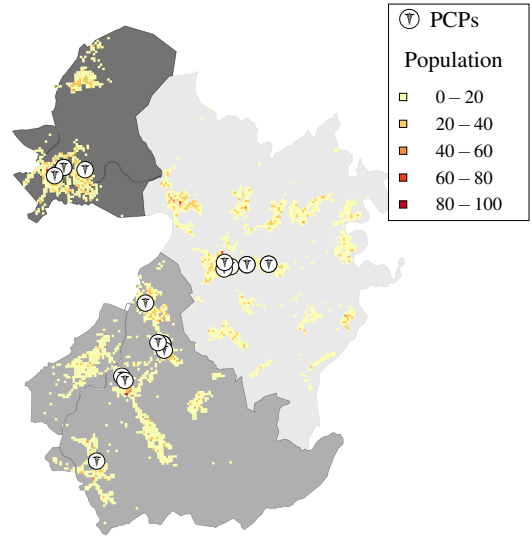

The population of patient agents $\SP$ in the baseline scenario aims to reflect the actual population in the considered primary care system.
The latest publicly available high resolution population data for the considered region is the German Census conducted in $2011$~\cite{zensus2011}.
At a resolution of $\num{2754}$ population cells measuring one hectare each, the $2011$ Census reports a total population of $\num{35542}$ for the three municipalities Roetgen, Simmerath, and Monschau; compare Figure~\ref{fig:data2}.
This population includes children under the age of $16$ who are excluded from our considerations, as children mainly consult pediatricians who are not modeled in this study.
Census data does not state the exact number of under $16$-year-olds in each population cell.
Instead, the Census reports the total number of under $16$-year-olds on municipality level: Roetgen $\num{1390}$, Simmerath $\num{2383}$, and Monschau $\num{1794}$.
To exclude children under the age of $16$, we proceed as follows: First, we fix one adult per population cell as we assume that children under the age of $16$ do not live on their own.
Then, we sample the number of under $16$-year-olds from the remaining population of each municipality according to a uniform distribution.
Exemplifying this procedure for Roetgen, the Census reports a total population of $\num{8288}$ distributed over $534$ population cells. We fix one adult per population cell, and uniformly distribute the $\num{1390}$ under $16$-year-olds among the $\num{7754}$ remaining inhabitants.
Performing this procedure for each municipality individually, we obtain the final patient population $\SP$ consisting of $\num{29975}$ patient agents distributed over $\num{2754}$ population cells.

The age-class independent attributes of each patient agent $\p \in \SP$ are determined as follows:
The location $\location\in\Locations$ for each patient is sampled from the associated population cell according to a uniform distribution.
Patients' health conditions $\condition \in [0,1]$ are sampled from a beta distribution with shape parameters $p=q=25$ such that all patients have an expected health condition of $\mathbb{E}(\condition)=0.5$.

\subsubsection{Age classes}
\label{study:age}
\SimulationModel accounts for the age dependency of various patient characteristics through the concept of age classes.
The baseline scenario differentiates three patient age classes: young (\youngClass), middle-aged (\midClass), and elderly (\oldClass). The characteristics of the modeled age classes $\SAgeClass$ are shown in Table~\ref{aclasses}.
Young patients (\youngClass) are, on average, the healthiest among all patients.
Thus, they are expected to develop the fewest \act illnesses per year from which they recover relatively quickly.
Their expected willingness to wait is prolonged and they are very unlikely to visit a PCP unless it is necessary.
Middle aged patients (\midClass) represent the working share of the population and we consider them to be our ``nominal'' patients.
They thus do not deviate from the expected illness duration and the expected willingness to wait as specified by families of illnesses.
On average, middle-aged patients (\midClass) develop more \act illnesses per year than young patients (\youngClass) while keeping slightly more appointments after recovery.
Elderly patients (\oldClass) are expected to develop the most annual \act illnesses and it takes them more time to recover from these.
Their expected willingness to wait is the lowest among all age classes and they are most likely to visit a PCP after all symptoms have subsided.

Based on Census data~\cite{zensus2011}, the age class $\age\in \SAgeClass$ of each patient depends on the discrete probability distribution shown in Table~\ref{asp}.
The age-class dependent attributes of each patient agent $\p \in \SP$ are subsequently determined as follows:
Each patient's session availabilities $\availabilities$ are determined by performing a Bernoulli trial for every session of the week $[\session] \in \SessionOfWeek$ based on the age-class dependent success probabilities from Table~\ref{asp}. 
To decide whether a patient is chronically ill, we perform a Bernoulli trial using the age class dependent success probabilities from Table~\ref{asp} that were estimated based on~\cite{GEDA12}. 

\begin{table}
	\centering
	\caption{Age classes $\SAgeClass$.}
	\label{aclasses}
	\begin{tabular*}{\columnwidth}{@{}llll@{}}
		\toprule
		& \youngClass & \midClass & \oldClass \\ \midrule
		exp.~illnesses  & $\expAnnualIllnesses(\condition){=} 6c$ & $\expAnnualIllnesses(\condition){=} 7c{+}1$ & $\expAnnualIllnesses(\condition){=} 9c{+}1$ \\
		dev.~duration  & $\changeInDuration{=} 0.8$ & $\changeInDuration{=}1.0$ & $\changeInDuration{=}1.2$ \\
		dev.~willingness & $\changeInWTW {=}1.2$ & $\changeInWTW {=}1.0$ & $\changeInWTW {=}0.8$ \\
		prob.~cancel  & $\probToCancel{=}0.95$ & $\probToCancel{=}0.8$ & $\probToCancel{=}0.7$ \\ \bottomrule
	\end{tabular*}
\end{table}

\begin{table}[t]
	\centering
	\caption{Age specific parameters for patient generation.}
	\label{asp}
	\begin{tabular*}{\linewidth}{@{}llll@{}}
		\toprule
		& \youngClass & \midClass & \oldClass \\ \midrule
		age class distribution & $0.1196$ & $0.6318$ & $0.2486$ \\
		availability probability & $0.85$ & $0.55$ & $0.95$ \\
		chronic illness probability & $0.12$ & $0.33$ & $0.52$ \\
		\bottomrule
	\end{tabular*}
\end{table}

\subsubsection{Families of Illnesses}
The most important classification system for illnesses world-wide is the International Classification of Diseases and Related health Problems (ICD) maintained by the World Health Organization.
In its current revision, ICD-10~\cite{ICD10} distinguishes more than $\num{14000}$ codes.
For the purpose of \SimulationModel, such a granular illness distinction is generally not necessary.
Thus, we can aggregate ICD-10 codes, e.g., using the $\num{22}$ chapters of ICD-10, or considering only a subset of all ICD-10 codes, e.g., the ones most frequently reported.
In the baseline scenario, we consider a subset of the $\num{100}$ ICD-10 codes most frequently reported to the Association of Statutory Health Insurance Physicians Nordrhein~\cite{KVICD16}.
The attributes of families of illnesses can be estimated based on historical treatment data which is commonly available to health insurers. 
Yet, such data is naturally protected by confidentiality and cannot be published.
Thus, we choose a less elaborate approach and only estimate all attributes which yields the families of illnesses $\SFamiliesIllnesses$ listed in Table~\ref{ci}.

\begin{table*}[t]
	\centering
	\caption{Characteristics of considered families of illnesses $f \in \SFamiliesIllnesses$.}
	\label{ci}
	\begin{tabular*}{ \linewidth }{llllll}
		\toprule
		ICD &Name & {{Exp.~willingness $W_f$}} &{{Exp.~duration $D_f$}} &{{Treatment frequency~$N_f$}} & Is chronic \\
		\midrule
		I10 & high blood pressure  &$W_f(s)=\num{-10}s+20$ & not applicable & $N_f(s)=-20s+100$ & true \\
		E11 & diabetes & $W_f(s)=-4s+14$ & not applicable & $N_f(s)=-10s+90$ & true \\
		I25 & ischemic heart disease & $W_f(s)=-4s+10$ & not applicable & $N_f(s)=-30s+100$ & true \\
		E78 & high cholesterol level & $W_f(s)=-5s+8$ & $D_f(s)=4s+8$ & $N_f(s)=-2s+11$ & false \\
		M54 & back pain & $W_f(s)=-3s+4$ & $D_f(s)=9s+5$ & $N_f(s)=-4s+11$ & false \\
		Z25 & vaccination & $W_f(s)=40$ & not applicable & not applicable & false \\
		J06 & cold & $W_f(s)=-2s+2$ & $D_f(s)=5s+4$ & $N_f(s)=-s+6$ & false \\ \bottomrule
	\end{tabular*}
\end{table*}

\vspace*{-.1cm}
\subsubsection{Age Class-Illness Distributions}
Age class-illness distributions define the expected occurrence of \act families of illnesses $\familyOfIllnesses \in \TemporaryFamiliesIllnesses$ per age class $\age \in \SAgeClass$.
For this distribution, the baseline scenario relies on the reported incidence rates of $\num{8.2}$ million customers of a large German health insurer published in~\cite{gekreport}.
We aggregate this data by gender and age to obtain the age class-illness distribution $\illnessAgeclassDist\colon \SAgeClass \times \TemporaryFamiliesIllnesses \to [0,1]$ shown in Table~\ref{iacd}.
Analogously, we determine the expected distribution of chronic families of illnesses $\ChronicFamiliesIllnesses$ among the modeled age-classes $\SAgeClass$ denoted by $\chroIllnessAgeclassDist\colon \SAgeClass \times \ChronicFamiliesIllnesses \to [0,1] $ shown in Table~\ref{iacd}.

The distribution $\chroIllnessAgeclassDist$ is not part of the baseline scenario itself.
Instead, it is only required to generate the unique chronic illness of chronically ill patients.
In the baseline scenario, we generate every chronic patient's chronic illness $\chronicIllness \in \AllChroIllnesses$ analogously to the process of generating \act illnesses as described in Section~\ref{sec:stochasticity}: Given the patient's age class $\age\in \SAgeClass$, the illness family $f_\chronicIllness \in \ChronicFamiliesIllnesses$ of $\chronicIllness$ depends on the discrete probability distribution $f \mapsto \chroIllnessAgeclassDist(\age, f)$ for $f\in\ChronicFamiliesIllnesses$.
The seriousness $s_\chronicIllness \in [0,1]$ of $\chronicIllness$ is sampled from a triangular distribution using the patient's health condition $\condition\in [0,1]$ as mode.
In turn, the seriousness defines $\chronicIllness$'s treatment frequency via $\nu_\chronicIllness=N_{f_\chronicIllness}(s_\chronicIllness)$ and willingness to wait as $\wtw_\chronicIllness= W_{f_\chronicIllness}(s_\chronicIllness)$. 
\begin{table}
	\caption{Age class-illness distributions $\illnessAgeclassDist$  and $\chroIllnessAgeclassDist$.}
\label{iacd}
	\begin{tabular*}{\linewidth}{@{}p{3cm}p{1cm}p{1cm}p{1cm}@{}}
		\toprule
		&\youngClass & \midClass & \oldClass \\ \midrule	
		high cholesterol level&\num{0.02} & 0.24 & 0.36 \\
		back pain&0.32 & 0.38 & 0.28 \\
		vaccination&0.14 & 0.14 & 0.27 \\
		cold&0.52 & 0.24 & 0.09 \\
		\midrule 
		high blood pressure&0.17 & 0.65 & 0.61 \\
		diabetes&0.33 & 0.16 & 0.2 \\
		ischemic heart disease&0.5 & 0.19 & 0.19 \\ \bottomrule
	\end{tabular*}
\end{table}

\vspace*{-.08cm}
\subsubsection{Duration of Warm-up}
\begin{figure}
\centering
\input{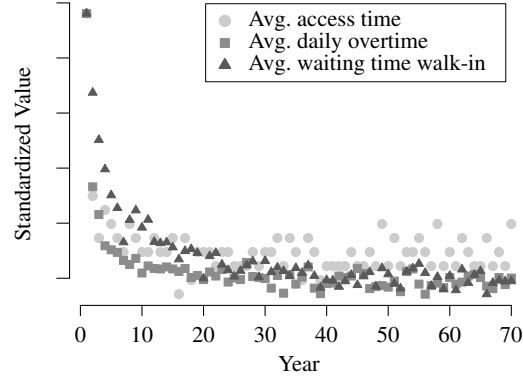}
\caption{Evolution of performance indicators in the baseline scenario for every year in a $70$ year time period.}
\label{fig:warm-up}
\end{figure}

\vspace*{-.08cm}
Every run of \SimulationModel contains a warm-up period; compare Section~\ref{sec:init}.
To determine an appropriate length for the warm-up period, we simulate the baseline scenario for a time period of $70$ years and track all performance indicator for each year individually.
Figure~\ref{fig:warm-up} shows the resulting evolution for the average access time of appointments, average daily overtime of physicians, and average waiting time of walk-in patients.
As we can see, all performance indicators are evolving in the first $30$ to $50$ years before they stabilize.
Similar behaviors can be observed for all other measured performance indicators.
Therefore, we set the duration of the warm-up period in each scenario to $60$ years.

\vspace*{-.2cm}
\subsection{Baseline Analysis}
\vspace*{-.14cm}
\label{baseline_analysis}
Table~\ref{t:allResults} reports the resulting expected key performance indicators as well as the associated exact $\SI{95}{\percent}$-confidence intervals for each tracked performance indicator; compare Section~\ref{emergenceObs}.
The results show that in the status quo, each physician in our primary care system performs, on average, $\num{10122.16}$ treatments per year.
This amounts to an average number of $\num{6.75}$ physician contacts per patient which is slightly above the $\num{6.6}$ annual PCP contacts reported back in $2006$~\cite{AP06}.
Roughly $\SI{47}{\percent}$ of patients visiting a physician in our baseline scenario are walk-in patients, which is consistent with the observed $\SI{48}{\percent}$ share of walk-ins in our collected empirical dataset of service times; compare Section~\ref{sec:stochasticity}.
Concerning overtimes, we were unable to obtain empirical data as most primary care physicians are self-employed and even the definition of overtime is unclear.
However, the estimated average daily overtime per physician (according to our definition) seems to be too low at just \num{0.8} minutes per day.
This can be explained by the fact that we incorporate buffers at the end of each session and do not include additional mandatory physician's activities such as reporting and accounting into our simulation model.

\begin{table*}[]
\setlength\tabcolsep{2.5pt}
\rotatebox{90}{
\begin{tabular}{@{}lllllllll@{}}
	\toprule
	& \multicolumn{2}{c}{Baseline Scenario}                               & \multicolumn{2}{c}{Decline in PCPs Short-term Shift}                             & \multicolumn{2}{c}{Decline in PCPs Medium-term Shift}                             & \multicolumn{2}{c}{Aging Patients Short-term Shift}                             \\ 
	\cmidrule(lr){2-3} \cmidrule(lr){4-5} \cmidrule(lr){6-7} \cmidrule(lr){8-9}
	& \multicolumn{1}{c}{Mean} & \multicolumn{1}{c}{\SI{95}{\percent}-CI} & \multicolumn{1}{c}{Mean} & \multicolumn{1}{c}{\SI{95}{\percent}-CI} & \multicolumn{1}{c}{Mean} & \multicolumn{1}{c}{\SI{95}{\percent}-CI} & \multicolumn{1}{c}{Mean} & \multicolumn{1}{c}{\SI{95}{\percent}-CI} \\ \midrule
	avg.~\# treatments                  & \num{10122.16} & [\num{10112.1}, \num{10132.22}]     & \num{12412.3}   & [\num{12399.87}, \num{ 12424.73}]   & \num{15006.28}  & [\num{14992.61}, \num{ 15019.95}]   & \num{10222.16}  & [\num{10211.54}, \num{ 10232.78}] \\
	avg.~\# walk-ins                    & \num{4731.53}  & [\num{4721.61}, \num{ 4741.46}]     & \num{6935.42}   & [\num{6923.02}, \num{ 6947.81}]     & \num{9361.44}   & [\num{9347.77}, \num{ 9375.11}]     & \num{4831.4}    & [\num{4820.79}, \num{ 4842.02}]   \\
	avg.~\# standard appts.~            & \num{3215.59}  & [\num{3214.14}, \num{ 3217.05}]     & \num{2766.77}   & [\num{2762.78}, \num{ 2770.77}]     & \num{2331.72}   & [\num{2325.66}, \num{ 2337.79}]     & \num{3194}      & [\num{3192.42}, \num{ 3195.58}]   \\
	avg.~\# regular appts.~             & \num{2175.03}  & [\num{2173.6}, \num{2176.47}]       & \num{2710.11}   & [\num{2705.98}, \num{ 2714.25}]     & \num{3313.12}   & [\num{3306.98}, \num{ 3319.25}]     & \num{2196.75}   & [\num{2195.35}, \num{ 2198.16}]   \\
	avg.~utilization {[}\%{]}           & \num{72.15}    & [\num{72.08}, \num{ 72.22}]         & \num{80.72}     & [\num{80.65}, \num{ 80.79}]         & \num{88.51}     & [\num{88.44}, \num{ 88.59}]         & \num{72.6}      & [\num{72.54}, \num{ 72.65}]       \\
	avg.~daily overtime {[}min{]}       & \num{0.8}      & [\num{0.74}, \num{ 0.86}]           & \num{2.89}      & [\num{2.75}, \num{ 3.02}]           & \num{10.03}     & [\num{9.79}, \num{10.27}]           & \num{0.77}      & [\num{0.73}, \num{ 0.81}]         \\
	avg.~\# rejected walk-ins           & \num{13.85}    & [\num{13.16}, \num{ 15.53}]         & \num{69.6}      & [\num{67.21}, \num{ 72.92}]         & \num{357.1}     & [\num{346.92}, \num{ 368.03}]       & \num{14.5}      & [\num{13.95}, \num{ 16.18}]       \\
	avg.~access time {[}d{]}            & \num{2.46}     & [\num{2.45}, \num{ 2.47}]           & \num{3.18}      & [\num{3.16}, \num{ 3.2}]           & \num{4.09}      & [\num{4.05}, \num{ 4.12}]           & \num{2.52}      & [\num{2.51}, \num{ 2.53}]         \\
	avg.~access time regular {[}d{]}    & \num{1.49}     & [\num{1.48}, \num{ 1.51}]           & \num{1.6}       & [\num{1.56}, \num{ 1.64}]           & \num{1.84}      & [\num{1.78}, \num{ 1.89}]           & \num{1.51}      & [\num{1.48}, \num{ 1.53}]         \\
	avg.~access distance {[}km{]}       & \num{4.95}     & [\num{4.94}, \num{ 4.96}]           & \num{6.66}      & [\num{6.65}, \num{ 6.66}]           & \num{7.51}      & [\num{7.5}, \num{7.52}]             & \num{5}         & [\num{5.0}, \num{5.01}]           \\
	avg.~waiting time appt.~{[}min{]}   & \num{2.09}     & [\num{2.08}, \num{ 2.1}]           & \num{2.22}      & [\num{2.2}, \num{2.23}]             & \num{2.18}      & [\num{2.16}, \num{ 2.2}]            & \num{2.11}      & [\num{2.1}, \num{2.12}]           \\
	avg.~waiting time walk-in {[}min{]} & \num{39.75}    & [\num{39.64}, \num{ 39.85}]         & \num{51.51}     & [\num{51.36}, \num{ 51.65}]         & \num{65.76}     & [\num{65.56}, \num{ 65.96}]         & \num{39.9}      & [\num{39.74}, \num{ 40.05}]       \\
	on-time appts.~{[}\%{]}             & \num{61.13}    & [\num{61.02}, \num{ 61.24}]         & \num{58.94}     & [\num{58.86}, \num{ 59.03}]         & \num{58.54}     & [\num{58.42}, \num{ 58.66}]         & \num{60.94}     & [\num{60.86}, \num{ 61.01}]       \\
	\# \act illnesses                   & \num{136454.2} & [\num{136283.89}, \num{ 136624.52}] & \num{136517.25} & [\num{136334.27}, \num{ 136700.23}] & \num{136499.55} & [\num{136348.12}, \num{ 136650.97}] & \num{137863.35} & [\num{137692.8}, \num{ 138033.9}] \\
	\# chronic patients                 & \num{10662}    & --                                  & \num{10662}     & --                                  & \num{10662}     & --                                  & \num{10776}     & --                                \\
	total PCP capacity {[}h{]}          & \num{32617}    & --                                  & \num{26455}     & --                                  & \num{22139}     & --                                  & \num{32617}     & --                                                            
	\\ \bottomrule
\end{tabular}
}
\hfill
\rotatebox{90}{
\begin{tabular}{@{}lllllll@{}}
	\toprule
	& \multicolumn{2}{c}{Aging Patients Medium-term Shift}                             & \multicolumn{2}{c}{Combined Effects Short-term Shift}                            & \multicolumn{2}{c}{Combined Effects Medium-term Shift}                            \\ 
	\cmidrule(lr){2-3} \cmidrule(lr){4-5} \cmidrule(lr){6-7} 
	& \multicolumn{1}{c}{Mean} & \multicolumn{1}{c}{\SI{95}{\percent}-CI} & \multicolumn{1}{c}{E} & \multicolumn{1}{c}{\SI{95}{\percent}-CI} & \multicolumn{1}{c}{E} & \multicolumn{1}{c}{\SI{95}{\percent}-CI} \\ \midrule
	avg.~\# treatments                  & \num{10300.37} & [\num{10288.82}, \num{ 10311.91}]   & \num{12536.05}  & [\num{12522.14}, \num{ 12549.95}] & \num{15269.52}  & [\num{15258.29}, \num{ 15280.75}]   \\
	avg.~\# walk-ins                    & \num{4909.14}  & [\num{4897.62}, \num{ 4920.66}]     & \num{7059.09}   & [\num{7045.14}, \num{ 7073.04}]   & \num{9624.34}   & [\num{9613.2}, \num{9635.49}]       \\
	avg.~\# standard appts.~            & \num{3162.45}  & [\num{3160.73}, \num{ 3164.17}]     & \num{2743.98}   & [\num{2739.24}, \num{ 2748.71}]   & \num{2257.32}   & [\num{2248.85}, \num{ 2265.78}]     \\
	avg.~\# regular appts.~             & \num{2228.78}  & [\num{2226.99}, \num{ 2230.56}]     & \num{2732.98}   & [\num{2728.25}, \num{ 2737.71}]   & \num{3387.86}   & [\num{3379.37}, \num{ 3396.35}]     \\
	avg.~utilization {[}\%{]}           & \num{72.95}    & [\num{72.88}, \num{ 73.03}]         & \num{81.11}     & [\num{81.03}, \num{ 81.2}]       & \num{89.35}     & [\num{89.27}, \num{ 89.43}]         \\
	avg.~daily overtime {[}min{]}       & \num{0.8}      & [\num{0.74}, \num{ 0.86}]           & \num{2.94}      & [\num{2.81}, \num{ 3.08}]         & \num{11.32}     & [\num{11.08}, \num{ 11.57}]         \\
	avg.~\# rejected walk-ins           & \num{16.4}     & [\num{15.89}, \num{ 17.63}]         & \num{75.15}     & [\num{72.01}, \num{ 79.12}]       & \num{428.9}     & [\num{421.41}, \num{ 437.2}]       \\
	avg.~access time {[}d{]}            & \num{2.58}     & [\num{2.57}, \num{ 2.58}]           & \num{3.3}       & [\num{3.27}, \num{ 3.32}]         & \num{4.34}      & [\num{4.3}, \num{4.38}]             \\
	avg.~access time regular {[}d{]}    & \num{1.51}     & [\num{1.49}, \num{ 1.54}]           & \num{1.68}      & [\num{1.63}, \num{ 1.73}]         & \num{1.94}      & [\num{1.88}, \num{ 2.01}]           \\
	avg.~access distance {[}km{]}       & \num{5.04}     & [\num{5.04}, \num{ 5.05}]           & \num{6.74}      & [\num{6.73}, \num{ 6.75}]         & \num{7.54}      & [\num{7.53}, \num{ 7.54}]           \\
	avg.~waiting time appt.~{[}min{]}   & \num{2.11}     & [\num{2.1}, \num{2.12}]             & \num{2.21}      & [\num{2.19}, \num{ 2.22}]         & \num{2.15}      & [\num{2.14}, \num{ 2.16}]           \\
	avg.~waiting time walk-in {[}min{]} & \num{40.11}    & [\num{39.97}, \num{ 40.24}]         & \num{52.07}     & [\num{51.91}, \num{ 52.23}]       & \num{67.2}      & [\num{67.01}, \num{ 67.39}]         \\
	on-time appts.~{[}\%{]}             & \num{60.85}    & [\num{60.76}, \num{ 60.93}]         & \num{58.98}     & [\num{58.89}, \num{ 59.07}]       & \num{58.68}     & [\num{58.6}, \num{58.77}]           \\
	\# \act illnesses                   & \num{138698.8} & [\num{138516.55}, \num{ 138881.06}] & \num{137830.15} & [\num{137657.8}, \num{138002.52}] & \num{138667.85} & [\num{138534.48}, \num{ 138801.22}] \\
	\# chronic patients                 & \num{10931}    & --                                  & \num{10776}     & --                                & \num{10931}     & --                                  \\
	total PCP capacity {[}h{]}          & \num{32617}    & --                                  & \num{26455}     & --                                & \num{22139}     & --                                                                    
	                 \\ \bottomrule
\end{tabular}
}
\vspace*{.3cm}
\caption{Mean performance indicators and $\SI{95}{\percent}$-confidence intervals obtained by repeating each simulation experiment $20$ times for each simulation scenario variant.}
\label{t:allResults}
\end{table*}

Patients in our baseline scenario are expected to travel almost $\SI{5}{\km}$ to visit a physician and have to wait an average number of $\num{2.46}$ days for their appointments.  
With regard to waiting times, we obtain an average expected waiting time of $\num{2.09}$ minutes for patients with appointment and $\num{39.75}$ minutes for walk-in patients. 
In comparison to the average waiting times observed when recording our service time dataset ($4$ minutes with appointment, $15$ minutes without appointment), our simulated waiting times are strikingly unfavorable for walk-in patients which suggests that physicians avoid excessive waiting times for walk-in patients through more sophisticated treatment strategies, e.g., accumulating priority queues~\cite{Stanford2014}.

\subsection{Scenario One: Decline in PCPs}
\label{s1}
Scenario one models a decline in the number of primary care physicians for a short- and a medium-term shift in time.
To that end, we exclude all primary care physicians from our baseline PCP population $\SGP$ that reached the statutory retirement age of $65$ by this point.
Specifically, we consider the year $2023$ by which $4$ out $20$ PCPs will have reached the  statutory retirement age as well as the year $2027$ by which $7$ out $20$ PCPs will have reached the statutory retirement age.
Assuming that none of the excluded physicians are replaced by a successor, we obtain our decimated population of primary care physicians $\SGPs$ for the short-term and $\SGPm$ for the medium-term shift.
By replacing the physician population $\SGP$ in our baseline scenario by $\SGPs$ and $\SGPm$, respectively, we obtain two scenarios variants for scenario one.
The patient and physician populations used in each scenario variant are summarized in Table~\ref{instances_in_detail}.

The simulation results for scenario one in Table~\ref{t:allResults} show a severe deterioration of all patient and physician indicators compared to the baseline scenario.
The physicians' expected workload measured through the average number of treatments increases by $\SI{23}{\percent}$ for the short-term and $\SI{48}{\percent}$ for the medium-term shift. 
Due to the increased scarcity of appointments, more and more patients are forced to visit physicians as walk-in patients ($\SI{56}{\percent}$ for short-term and $\SI{62}{\percent}$ for medium-term shift).
The average daily overtime for physicians (that neglects all the physicians' administrative and organizational tasks) increases by $2.09$ minutes for the short-term and $9.23$ minutes for the medium-term shift.
On average, patients wait $\SI{29}{\percent}$ longer for their appointments in the short-term and even $\SI{66}{\percent}$ longer in the medium-term shift scenario variant. 
Similar increases can be observed for the patients' average access distance, which increases by $\SI{35}{\percent}$  to $\SI{6.66}{\km}$ for the short-term and by $\SI{51}{\percent}$ to $\SI{7.51}{\km}$ for the medium-term shift.
The average waiting time for patients with appointment is almost unaffected by the decline in the number of physicians, which can be explained by the strict prioritization in PFCFS.
The average waiting time for walk-in patients increases by $\SI{30}{\percent}$ for the short-term and $\SI{65}{\percent}$ for the medium-term shift.

\begin{table}[]
	\begin{threeparttable}
		\caption{Populations in each simulation scenario variant.}
		\label{instances_in_detail}
		\setlength\tabcolsep{4.5pt}
		\begin{tabular*}{\linewidth}{@{}lcccccc@{}}
			\toprule
			& \multicolumn{2}{c}{Decl.~PCPs} & \multicolumn{2}{c}{Aging Patients} & \multicolumn{2}{c}{Comb.~Effects}\\
			\cmidrule(lr){2-3} \cmidrule(lr){4-5} \cmidrule(lr){6-7}
			&   s             & m            & s              & m       & s             & m            \\ \midrule
			patients   & $\SP$           & $\SP$           & $\SPs$ & $\SPm$ & $\SPs$ & $\SPm$       \\
			physicians & $\SGPs$ & $\SGPm$ & $\SGP$           & $\SGP$    & $\SGPs$  & $\SGPm$ \\ \bottomrule
		\end{tabular*}
		\begin{tablenotes}
			\vspace*{.15cm}
			\small
			\item  s = short-term shift, m = medium-term shift.
		\end{tablenotes}
	\end{threeparttable}
\end{table}
\subsection{Scenario Two: Aging Patients}
\label{s2}
Scenario two models the ongoing aging of the patient population for a short- and medium-term shift in time.
For this purpose, we adjust the discrete probability distribution determining the patients' age classes to generate two new patient populations. 
More precisely, we use current population projections~\cite{itnrw} for the years $2025$ and $2030$ to obtain the two adjusted discrete probability distributions for the patients' age classes shown in Table~\ref{t3}.
Using these distributions, we generate the aged patient population $\SPs$ for the short-term  and $\SPm$ for the medium-term shift. 
By replacing the patient population $\SP$ in our baseline scenario by $\SPs$ and $\SPm$, respectively, we obtain two scenario variants for scenario two; compare Table~\ref{instances_in_detail}.

\begin{table}
	\centering
	\caption{Age class distributions for aged patient population.}
	\smallskip
	\label{t3}
	\begin{tabular*}{\linewidth}{@{}p{2.5cm}p{1.1cm}p{1.1cm}p{1.1cm}@{}}
		\toprule
		\textbf{}  & \youngClass & \midClass & \oldClass \\ \midrule
		short-term shift       & 0.1051      & 0.6283    & 0.2666                  \\
		medium-term shift       & 0.1025         & 0.6033         & 0.2942                  \\ \bottomrule
	\end{tabular*}
\end{table}

The simulation results for scenario two (Table~\ref{t:allResults}) paint a similar picture as in scenario one, i.e., the majority of patient and physician indicators deteriorate, albeit far less severe.
As a result of the aging of the patient population, the average number of treatments per physician increases by $\SI{1}{\percent}$ and $\SI{2}{\percent}$ for the short-term and medium-term shift, respectively. 
However, in contrast to scenario one, additional treatments distribute more evenly between appointment and walk-in patients and thus the expected ratio of walk-in patients remains almost unchanged ($\SI{47}{\percent}$ for short-term and $\SI{48}{\percent}$ for medium-term shift).
Judging from the almost unaffected average overtime, physicians manage to accommodate the additional treatments mostly within their regular opening hours.
As a result of the increased treatment demand, patients wait on average $\SI{2}{\percent}$ longer for their appointments in the short-term and $\SI{5}{\percent}$ longer in the medium-term shift scenario variant. 
Moreover, they are willing to accept $\SI{1}{\percent}$ ($\SI{2}{\percent}$) longer average access distances in the short-term (medium-term) shift scenario to receive more timely treatment or avoid longer waiting times.
Patient waiting times with appointment are unaffected by the increased patient demand.
The average waiting times of walk-in patients in the short-term shift scenario variant remain almost unchanged, while they increase by $\SI{1}{\percent}$ for the medium-term shift.
\subsection{Scenario Three: Combined Effects}
\label{s3}
Scenario three models a combined decline in the number of primary care physicians and aging of the patient population for a short- and medium-term shift in time.
By replacing both, the patient and the physician population in our baseline scenario with the adjusted patient and physician populations from scenarios one and two, we obtain two scenario  variants for scenario three; compare Table~\ref{instances_in_detail}.

Analyzing our simulation results in Table~\ref{t:allResults} for scenario three, we can confirm that the combined effects of a decline in the number of PCPs and an aging population lead to the greatest deterioration of patient and physician indicators among all scenarios.
However, the effect of the combined changes compared to the combination of the individual effects from scenarios one and two varies between indicators:
For the average number of treatments and the ratio of walk-in patients, the effects of the combined changes correspond to the sum of the effects for the individual changes, e.g., a $\SI{24}{\percent}$ increase in the average number of treatments in short-term shift variant of scenario three versus a $\SI{23}{\percent}$ and $\SI{1}{\percent}$ increase in the respective variants of scenarios one and two. 
Concerning the physicians' average overtime, we can observe that a combined consideration of both changes has an amplifying effect.
For example, in the medium-term shift variants of scenarios one and two the average overtime increases by $9.23$ and $0$ minutes, respectively, while the combined changes in scenario three lead to an increase of $10.52$ minutes.
Similar amplifying effects can be observed for the patients' average access time and walk-in waiting time. 
Considering the patients' average access distance, the combination of both changes leads to different effects in the two scenario variants.
In the short-term shift variant, the effect of the combined changes corresponds to the sum of the effects for the individual changes.
In the medium-term shift variant, we can observe a slight dampening effect resulting from a combined consideration of both changes, i.e., while the individual changes lead to a respective $\SI{52}{\percent}$ and $\SI{2}{\percent}$ increase of the expected average access distance, the combined effects lead to an increase of $\SI{52}{\percent}$.    

\section{Discussion}
\label{sec:discussion}
The aim of \SimulationModel is to provide decision makers with a tool to analyze and optimize primary care systems.
\SimulationModel produces meaningful performance indicators that enable a far more detailed assessment of primary care systems compared to the current approaches based on patient-physician ratios.
Next to more accurate evaluation of the status quo, \SimulationModel can predict and quantify the influence of policy decisions and changes in the systems population, e.g., an aging of the population or a decline in the number of PCPs as illustrated in Section~\ref{sec:computationalExamples}. 
Thereby, the model can particularly take several system changes into account at the same time which enables the analysis of combined effects.
As all components of a simulation scenario can be easily adjusted, this opens up a broad field of potential applications ranging from physicians' location planning to the evaluation of specific PCP strategies, e.g., in the field of appointment scheduling.
Finally, the modular design of \SimulationModel perspectively allows for easy model extensions, e.g., to model prospective new supply concept such as mobile medical units or telemedicine.

The greatest entry requirement to using \SimulationModel is the complex and time consuming task of generating and validating the input scenarios.
As \SimulationModel models each agent individually, it requires detailed empirical data which has to be obtained from various parties or, even worse, could be unavailable.
Moreover, some model components such as the service time distributions are tailored to the German system and thus might have to be adjusted when using \SimulationModel to analyze, e.g., a primary care system in the United States.
Each of these model changes, potentially change the model's behavior and thus require a new validation process to ensure that insights derived from \SimulationModel are viable for the studied primary care system.

To help overcoming this entry requirement, we exemplified the scenario generation and validation process for a real-world primary care system in Germany.
Particularly, we detailed the generation process of all simulation entities and provided available empirical data sources.
Although data availability may vary for other primary care systems, this may hint at where the required empirical data can be obtained.
We validated our simulation scenario by comparing its output to available empirical data.
To show internal validity, we performed $20$ independent runs for each simulation experiment and captured the resulting model variability through confidence intervals.
However, we need to stress that additional validation should be performed before actual policy decisions are derived from the presented case study. 
Such validation measures should particularly include a sensitivity analysis as well as an expert validation which were out of scope for the purpose of this study.

In summary, \SimulationModel can serve as a versatile decision support tool in primary care planning when it is used with adequately validated simulation scenarios.
The process of generating and validating simulation scenarios is both challenging and time-consuming.
However, once this process is complete, \SimulationModel enables a detailed analysis and optimization of primary care systems that is superior to basic ratio-based approaches.
As a final motivation to use \SimulationModel, we present a potential real-world use-case from Germany:
At the beginning of $2019$, the German Bundestag passed the law for faster appointments and better care (TSVG)~\cite{TSVG} that increased the minimum weekly opening hours for physicians with statutory health insurance accreditation from $20$ to $25$ hours~\cite[Art.~$15$]{TSVG}.
The law is controversial, among other things, because there are doubts about the consequences of this policy decision~\cite{TSVG2}.
Using \SimulationModel, decision makers could have obtained insights into the effects of increased minimal opening hours from both the patients' and the physicians' perspective before its implementation.
\section{Conclusion and Further Steps}
\label{sec:conclusion}
In this paper, we presented the hybrid agent-based simulation model \SimulationModel that serves as a decision support tool in primary care.
\SimulationModel does not hard-code a specific primary care system, but requires an input scenario which encodes the speci\-fics of the primary system the \user wishes to evaluate.
The generation of a simulation scenario is non-trivial and we exemplified this process for real-world primary care system in Germany.
Based on this scenario, we performed a case study that serves as a proof of concept showcasing the capabilities of \SimulationModel.
Besides, it exemplifies the complex validation and calibration process that has to be carried out for each simulation scenario individually.

Future work will include further effort towards model validation and calibration as well as the implementation model extensions.
Currently, illness distributions are considered static by \SimulationModel.
By modeling dynamic illness distributions, we can make them dependent on seasonality or the patients' previous history of illnesses.
In the current model, the duration of an illnesses is independent of the actual treatment.
Interestingly, the results are convincing even without this causal link.
In the future, we want to compare whether implementing this link in the conceptual model significantly affects our findings.
A similar comparison shall investigate the influence of no show-patients, who introduce unexpected idle time into the physicians' schedules and require more sophisticated PCP strategies.
Yet other possible model extensions include: Illness specific appointments such that not all \act illnesses are treated during every appointment, intentional physicians' breaks, implementation of additional patient attributes such as gender, and mobile patient agents that move between different locations, e.g., their home and work.
Finally, we plan to release our implementation of \SimulationModel as open source software such that it can be easily accessed, studied, and adapted to the individual requirements of all \users.

\begin{acknowledgements}
During the development of \SimulationModel and the simulation scenarios in our case study, we received numerous support that we wish to acknowledge.
We thank the city region Aachen and the Actimonda health insurance who provided both data and the decision makers' expertise. 
We are grateful to all physicians, who provided their perspective and allowed us to collect empirical data in their practices.
We appreciate the help of  Marcia R\"uckbeil who advised us on statistics as well as our (former) students Tabea Krabs, Stephan Marnach, and  Mariia Anapolska who contributed towards the implementation of the simulation model.
Finally, we want to thank Sebastian Rachuba for his helpful remarks on an early version of this manuscript.
\end{acknowledgements}

\bibliographystyle{spmpsci}      
\bibliography{Reference}

\end{document}